\documentclass{aastex62}

\graphicspath{{./}{figures/}}
\received{June 17, 2020}
\revised{XX-XX-2020}
\accepted{XX-XX-2020}
\submitjournal{ApJ}

\shorttitle{Information in Cloudy Transit Spectra}
\shortauthors{Lacy et al.}

\begin{document}

\title{JWST Transit Spectra II: Constraining Aerosol Species, Particle-size Distributions, Temperature, and Metallicity for Cloudy Exoplanets}

\correspondingauthor{Brianna Lacy}
\email{blacy@princeton.astro.edu}

\author[0000-0002-0786-7307]{Brianna I. Lacy}
\affil{Princeton University \\
Peyton Hall, Ivy Lane \\
Princeton, NJ 08544, USA}

\author{Adam Burrows}
\affil{Princeton University \\
Peyton Hall, Ivy Lane \\
Princeton, NJ 08544, USA}

\begin{abstract}
JWST will provide moderate resolution transit spectra with continuous wavelength coverage from the optical to the mid-infrared for the first time. In this paper, we illustrate how different aerosol species, size-distributions, and spatial distributions encode information in JWST transit spectra. We use the transit spectral modeling code METIS, along with Mie theory and several flexible treatments of aerosol size and spatial distributions to perform parameter sensitivity studies, calculate transit contribution functions, compute Jacobians, and retrieve parameters with Markov Chain Monte Carlo. The broader wavelength coverage of JWST will likely encompass enough non-gray aerosol behavior to recover information about the species and size-distribution of particles, especially if distinct resonance features arising from the aerosols are present. Within the JWST wavelength range, the optical and mid-infrared typically provide information about 0.1-1 $\mu$m sized aerosols, while the near-infrared to mid-infrared wavelengths usually provide information about gaseous absorption, even if aerosols are present. Strong gaseous absorption features in the infrared often remain visible, even when clouds and hazes are flattening the optical and NIR portion of the spectrum that is currently observable. For some combinations of aerosol properties, temperature, and surface gravity, one can make a precise measure of metallicity despite the presence of aerosols, but more often the retrieved metallicity of a cloudy or hazy atmosphere has significantly lower precision than for a clear atmosphere with otherwise similar properties. Future efforts to securely link aerosol properties to atmospheric metallicity and temperature in a physically motivated manner will ultimately enable a robust physical understanding of the processes at play in cloudy, hazy exoplanet atmospheres. 
 \end{abstract}
 

\section{Introduction} \label{sec:intro}

Theories for utilizing the wavelength dependence of transit depths to learn about exoplanet atmospheres arose soon after the first detection of a transiting exoplanet (\citealt{Seager2000}; \citealt{Brown2001}; \citealt{Hubbard2001}), and it wasn't long before transit spectroscopy enabled the first detection of an exoplanet atmosphere \citep{Charbonneau2002}. Since then, many studies have applied Bayesian techniques to estimate atmospheric properties, successfully detecting individual molecules and estimating water abundances (see \citealt{Madhusudhan2018} and \citealt{Barstow2020a} for recent reviews). Today, sparse multi-wavelength transit measurements have been done for $\sim$100 planets, with more thorough wavelength coverage for a subset of only $\sim$30 or so\footnote{https://exoplanetarchive.ipac.caltech.edu}. Eventually, it is hoped that enough measurements can be made to search for overarching patterns in the abundances, metallicities, and C/O ratios of exoplanets which can be used to test theories of planet formation, migration, and subsequent evolution (\citealt{Oberg2011}; \citealt{Piso2016}).  

Some researchers have already started to use the small sample of transit spectra to discern patterns across planet mass and levels of stellar irradiation (\citealt{Iyer2016}; \citealt{Sing2016}; \citealt{Barstow2017}; \citealt{Fu2017}; \citealt{Fisher2018}; \citealt{Tsiaras2018}; \citealt{Pinhas2019}; \citealt{Wakeford2019}; \citealt{Welbanks2019}), but it is generally expected that more reliable results will come when the upcoming James Webb Space Telescope (JWST) and Atmospheric Remote-sensing Infrared Exoplanet Large-survey (ARIEL) expand the size and quality of available transit spectroscopy \citep{Burrows2014}. These two missions are complementary by design. JWST will observe tens of transiting exoplanets from 0.6 to 30 $\mu$m \citep{Stevenson2016}. It will have native spectral resolutions ranging from order R $\sim$ 100s to 1,000s which can be re-binned to trade off between $SNR$ and spectral information. This mission reaches out to much longer wavelengths than was previously possible with Hubble Space Telescope (HST) and Spitzer Space Telescope (Spitzer) and greatly improves the spectral resolution, especially at the longer wavelengths. ARIEL will have the ability to observe from 2.0 to 7.8 $\mu$m with a spectral resolution of R$\sim$100 \citep{Puig2016}. This satellite is dedicated solely to exoplanet science, so it will survey a much larger sample than JWST (around 1000 transiting planets, \citealt{Tinetti2018}). One core goal of ARIEL is to measure the mass-metallicity relationship of exoplanets (\citealt{Zellem2019}).

The future for transit spectroscopy looks very fruitful, but the method has some inherent challenges and limitations which we must work to overcome in order to realize the full promise of missions like JWST and ARIEL. Foremost among these is the reality that the slant geometry of transit spectroscopy makes this type of observation particularly sensitive to the presence of even trace amounts of aerosols in the upper atmosphere of the target exoplanet \citep{Fortney2005}. Note that, throughout this paper, we will adhere to the custom of referring to condensing species as \textit{clouds}, photochemically formed species as \textit{hazes}, and using the term \textit{aerosol} to encompass both. Initially, it was thought that the high temperatures of most exoplanets studied with transit spectroscopy would prevent clouds from forming. This notion turned out to be erroneous; the exoplanets observed so far exhibit a range of behavior from densely cloudy or hazy to completely clear (\citealt{Charbonneau2002}; \citealt{Lecavelier2008}; \citealt{Kreidberg2014b}; \citealt{Sing2016}; \citealt{Louden2017}). In fact, \citealt{Wakeford2019} used the statistics of aerosol effects in current transit spectroscopy to demonstrate that, if the currently available transit spectra are a representative sample, then observers proposing for time with JWST should anticipate signal sizes that are 30\% reduced from what one would see if atmospheres were clear.

When clouds or hazes are present, they make it more difficult to measure chemical abundances with current retrieval models. Clear atmospheres produce spectra with large variations in transit depth with wavelength, on the order several gas-pressure scale heights. Absorption and scattering by clouds and hazes can fill in the gaseous absorption windows, shrinking the size of the transit spectroscopy ``signal."  Furthermore, properties like metallicity are more easily inferred from chemical abundances if one only needs to account for gas-phase chemistry rather than coupling gas-phase chemistry to the microphysics and/or photochemistry of aerosols (\citealt{Woitke2018}; \citealt{Helling2019a}). For observations like those currently available (i.e. sparse coverage from optical to NIR), models accounting for a gray absorber and a varying optical slope have been adequate to marginalize over aerosol effects and obtain unbiased measurements of a planet's temperature and abundances (\citealt{Mai2019}; \citealt{Barstow2020c}). These measurements of abundances and temperature may be unbiased, but they are generally much less precise when thick hazes or clouds are present than for clear atmospheres (\citealt{Barstow2017}; \citealt{Fisher2018}; \citealt{Tsiaras2018}; \citealt{Wakeford2018}; \citealt{Pinhas2019}). 

It is hoped that the additional long-wavelength coverage of JWST and ARIEL will allow us to learn more about cloudy and hazy exoplanet atmospheres because there are stronger gaseous absorption features in the near-mid IR that may be visible above cloud or haze layers, and there may be distinctive spectral features in the mid IR arising from resonance modes within the aerosols themselves (\citealt{Budaj2015}; \citealt{Wakeford2015}; \citealt{Pinhas2017}; \citealt{Kitzmann2018}). If we can use transit spectra to identify which aerosol species are present and to constrain detailed size distributions, then this empirical information may even help refine efforts to model aerosol microphysics in detail (\citealt{Helling2019a} review the state of the art for exoplanet cloud modeling; \citealt{Kawashima2018b} provide an example of modeling photochemical haze formation in exoplanets). Eventually, if fast and accurate retrieval models can be developed which couple depletion and enrichment from clouds and hazes to gas-phase chemistry, then their presence need not hinder researchers from making accurate and precise measurements of chemical abundances, eventually uncovering chemical trends left by the processes of planet formation and evolution. Retrieved information from such transit spectra could even teach us surprising and interesting things about microphysics and photochemistry in alien environments (\citealt{Helling2019a}). This reasoning forms the basic motivation for our study.

In this work, we explore how well properties of the aerosols themselves can be constrained with JWST-like transit spectra, and to what extent adding longer wavelength coverage enables better measurements of metallicity and temperature, even in the presence of clouds or hazes. Our goal is to build an intuitive sense of how the temperature, mass, and metallicity of an exoplanet and the physical properties of any aerosols in its atmosphere translate into the shape of the full transit spectrum. We use the following questions to direct us: 
\begin{enumerate}
    \item Which JWST wavelengths contain the most information about aerosol properties and which provide information  about  gaseous  absorption? 
    \item How well can we recover atmospheric metallicities and temperatures, even when aerosols are present as we extend the wavelength coverage of transit spectra?
    \item Can we uniquely identify which dominant aerosol species are present in atmospheres using JWST transit spectroscopy? 
    \item Can we constrain the size-distribution of aerosols?
    \item How do these tasks differ for condensed clouds and photochemical hazes?
\end{enumerate}

Section \ref{sec:methods} describes our methods and discusses which aerosol species we consider. We use the recently developed code Multi-dimensional Exoplanet TransIt Spectroscopy (METIS, \citealt{Lacy2020}) to carry out Markov Chain Monte-Carlo (MCMC) retrievals, compute transit contribution functions, and conduct studies of model parameter sensitivity. Section \ref{sec:clear_fiducials} presents the four fiducial planets we use to anchor our study, which range in temperature from 700-1800 K in order to explore a range of possible condensate species. Section \ref{sec:wavelength_coverage} demonstrates that the additional long-wavelength coverage of JWST (and, to a lesser extent, ARIEL) will allow one to probe gas-phase molecules in the near-mid IR and aerosol properties at optical and mid-IR wavelengths. In \S \ref{sec:slab} and \S \ref{sec:equilibrium} we use MCMC experiments to test how well aerosol species can be distinguished and how well particle distributions can be recovered. In \S \ref{sec:slab} we incorporate aerosols as a uniform slab at an arbitrary pressure, a method suitable for either hazes or clouds. In \S \ref{sec:equilibrium}, we place a cloud base where the Clausius-Clapeyron line and the temperature-pressure profile intersect, and then have the cloud taper off. This phase equilibrium approach is suitable for condensing clouds. We summarize and draw our conclusions in \S \ref{sec:conclusions}.

\section{Methods} \label{sec:methods}

The code used throughout this work, METIS, is described in detail in \citealt{Lacy2020}, but we will review the important points here for the reader's convenience. The code takes in an arbitrary latitude-longitude-altitude grid of temperatures and pressures along with an atmospheric metallicity and returns the corresponding transit spectrum. It assumes thermochemical-equilibrium with a solar C/O ratio to assign the correct opacity and mean molecular weight to each grid point and assumes the ideal gas law to assign the appropriate densities. In this paper our focus is on exploring a wide variety of aerosol behaviors rather than 3D effects, so we simply use isothermal atmospheres as the input grids for METIS. When computing these isothermal structures we assume hydrostatic equilibrium and the ideal gas law. We use a reference pressure, $P_0$, and reference radius, $R_0$, a planet mass at that radius, $M_P$, a metallicity, $Z$, and a temperature, $T$ as input. The atmospheres have constant temperature with altitude but varying surface gravity and varying mean molecular weight in manner consistent with hydrostatic equilibrium and thermochemical equilibrium. 

We assume thermochemical equilibrium, so we can interpolate within pre-calculated tables of mixing ratios and matching pre-mixed tables of total opacity to assign the correct mean molecular weight and opacity for any combination of temperature and pressure. The tables are created using the thermochemical equilibrium calculations described in \citealt{Burrows1999} and \citealt{Sharp2007}, and opacity calculations as described in \citealt{Sharp2007}, with updated CH$_4$ opacities from \citealt{Yurchenko2014}. At times we also interpolate among several sets of tables with metallicities spanning 0.1 to 3.16, to allow us to perform MCMC retrievals of $Z$. We always assume a solar C/O ratio in this study.

The chemistry calculations minimize the Gibb's free energy for a network of hundreds of species and reactions,  recording mixing ratios for 30 important ions, atoms and molecules in the chemistry tables across a grid of temperatures and pressures. The corresponding opacity tables have a grid in temperature and density, assuming an ideal gas equation of state to convert pressure to density. The opacity tables sum together opacities from 26 ionic, atomic and molecular sources, as well as from H$_2$-H$_2$ and H$_2$-He collision induced absorption. Rayleigh scattering cross sections for the appropriate mixture of gases are provided in a separate table at the same temperatures and densities but only for a single reference wavelength of $\lambda_0=$1$\mu$m. This cross section is then scaled as $(\lambda/\lambda_0)^{-4}$. 

When aerosols are included, we do not account for any corresponding changes to the gas phase chemistry, so we are effectively assuming that the time scales for photochemical processes and condensation are long compared to gas-phase interactions and that replenishment of new material from deeper in the atmosphere keeps the gas phase unchanged. Alternatively, one could see this as an assumption that the amount of material tied up in aerosols is negligible compared to the gas-phase abundances of relevant atomic species.

\subsection{Aerosol Parameterizations}\label{sec:aerosol_parameterizations}

\begin{figure}
    \centering
    \includegraphics[width=0.475\textwidth]{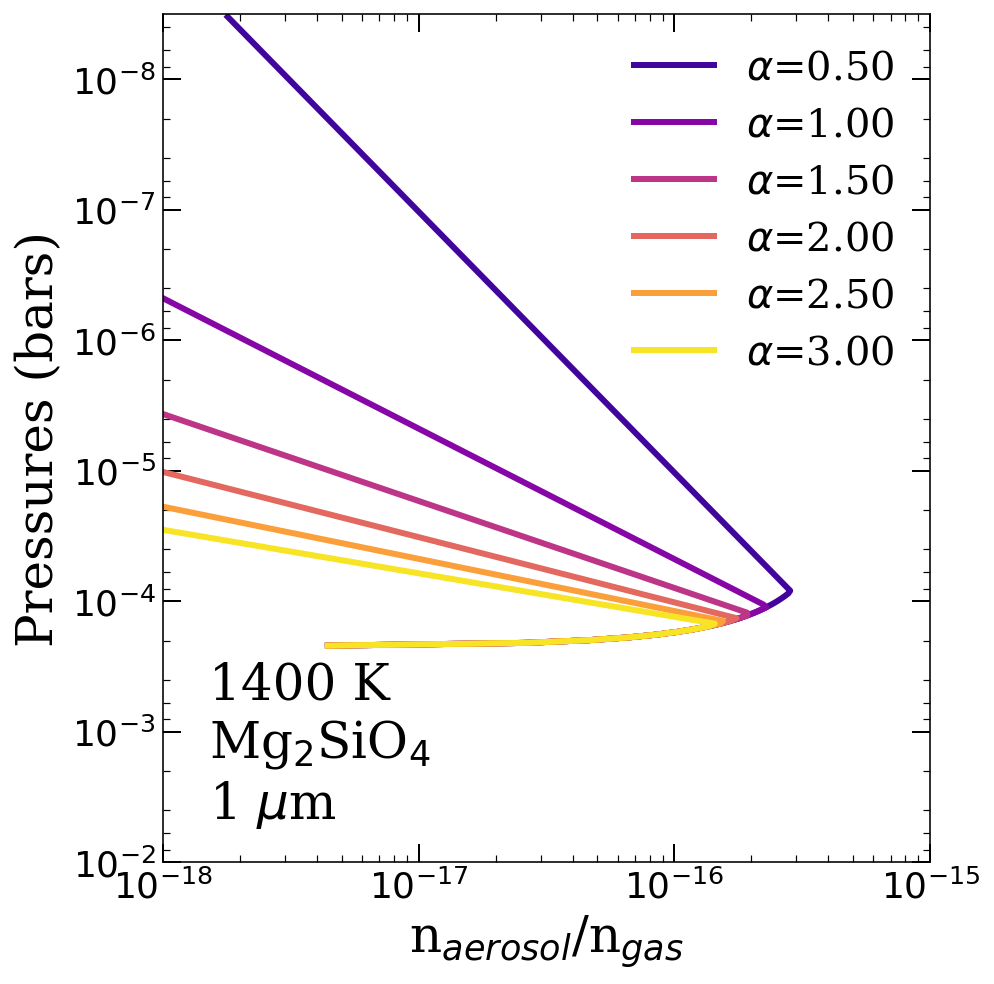}
    \includegraphics[width=0.475\textwidth]{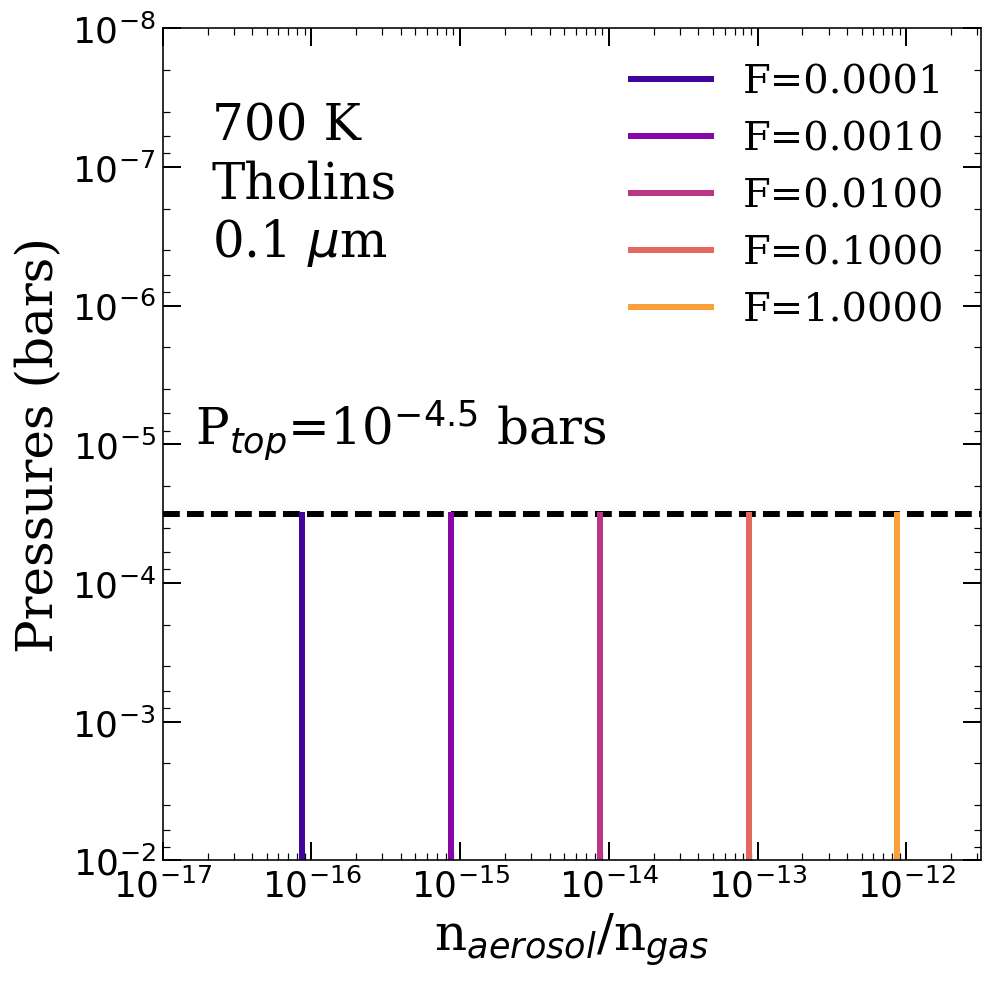}
    \caption{Demonstration of the meaning of parameters describing the spatial positions of aerosol particles. On the left we show the equilibrium cloud with varying values of $\alpha$ and on the right we show the slab with varying values of $F$. The equilibrium cloud examples assume a forsterite cloud with 1 $\mu$m particles in a 1400-K atmosphere. The slab examples assume a Tholin haze with 0.1 $\mu$m particles in a 700-K atmosphere and a top-pressure cut-off of 10$^{-4.5}$ bars. In both cases we assumed the atmosphere had solar metallicity. }
    \label{fig:aerosol_spatial_dists}
\end{figure}

We incorporate aerosol opacity using Mie theory, log-normal size distributions, and two options for specifying the spatial positions of particles in the atmosphere. We call these two forms of aerosol the ``slab" and the ``phase equilibrium cloud". Their parameters are summarized in Table \ref{tab:aerosol_spatial_parameters} and portrayed visually in Figure \ref{fig:aerosol_spatial_dists}. The table and figure are replicated from \citealt{Lacy2020} for the reader's convenience. 

\begin{table}[]\label{tab:aerosol_spatial_parameters}
    \centering
    \begin{tabular}{l|c|l|l}
       Name  & Parameters & Meaning & Intended Aerosol Type \\
    \hline
        Slab &P$_{top}$ & Top pressure cut-off & Condensing clouds \\
        (Fig. \ref{fig:aerosol_spatial_dists}, right)&F & Fraction of available material&or photochemical hazes\\
        &&that contributes to aerosol &\\
        &a$_m$& Modal particle radius&\\
        &$\sigma _a$& Size dispersion for log-normal &\\
    \hline
        Equilibrium Base&$\alpha$& Ratio of gas scale height to &Condensing clouds\\
        (Fig. \ref{fig:aerosol_spatial_dists}, left)&&aerosol scale height&\\
        &a$_m$& Modal particle radius&\\
        &$\sigma _a$& Size dispersion for log-normal &\\
    \end{tabular}
    \caption{Summary of aerosol spatial parameterizations. These can be paired with other size distributions besides the log normal parameters included here. Options to use any of the aerosol species listed in Table \ref{tab:aerosol_properties} are available. When the equilibrium base option is chosen, condensation curves shown in Figure \ref{fig:cclines} are used.}
\end{table}

These two forms of aerosol are complimentary. The slab aerosol is very flexible, and, as such, should be able to replicate a wide range of haze or cloud behavior seen in Nature, provided the correct aerosol species is used. It has one parameter, $F$, that sets what fraction of available material winds up bound into cloud or haze particles, and a second parameter, P$_{top}$, that can be invoked to set a top-pressure cut-off above which no particles form or remain for long, even if there is sufficient material. Flexibility is good for retrieval models, since one wants data to shape conclusions rather than pre-conceived notions. However, we do have many examples of clouds in our own solar system and ground-truth laboratory measurements of the temperatures and pressures at which substances can condense. The phase equilibrium cloud form we adopt makes some strong, but physically motivated, assumptions. Condensing clouds in the solar system tend to have their base around the intersection between the Clausius-Clapeyron line and the T-P profile, and they tend to have a smaller scale-height than the gas pressure scale height. This form sets the cloud base as such, and the amount of available material bound up in the cloud particles at the base is only that material which is in excess of super-saturation. From there, the cloud transitions to have the number of particles taper off according to $n_{aerosol}/n_{gas}(P)=n_{aerosol}/n_{gas}(P_{base}) \times (P/P_{base})^{\alpha}$. This $\alpha$ parameter thus captures the expectation that clouds will have a smaller scale height than the gas. If $\alpha$ is very high, your cloud falls off very quickly, while if $\alpha$ is 0 the ratio of cloud particle number density to gas particle number density remains constant with altitude. By exploring both these forms of aerosol, we can capture a wide range of likely behavior for photochemical hazes and condensing clouds, and we can demonstrate the advantages and disadvantages of applying retrieval models which favor more physical assumptions versus more flexibility as we seek to characterize transit spectra for cloudy hazy exoplanets.

There are a number of retrieval tools for transit spectra which account for clouds and hazes\footnote{some examples of transit retrieval codes which allow aerosols include: NEMESIS (\citealt{Irwin2008}), POSEIDON (\citealt{MacDonald2017}), PyRat-Bay (\citealt{Cubillos2017}), BART (\citealt{Blecic2017}), SCARLET (\citealt{Fraine2014}), CHIMERA (\citealt{Line2013}), $\tau$-REx (\citealt{Waldmann2015})}. \citealt{Barstow2020a} and \citealt{Barstow2020b} provide a recent review of retrievals and a direct comparison of results found with different aerosol parameterizations. Cloud and haze parameterizations usually involve some subset of the following: a single cloud opacity if the cloud is gray or an initial cloud opacity that will then be scaled with wavelength according to some rule if the cloud is non-gray, a specified range of pressures where the cloud will be present, a single particle size, and a scattering index (\citealt{Barstow2017}; \citealt{Fisher2018}; \citealt{Tsiaras2018}; \citealt{Pinhas2019}; \citealt{Mai2019}; \citealt{Ormel2019}). We take a slightly different approach in order to incorporate the actual complex indices of refraction from which ever aerosol species is used, along with a full log-normal size distribution not just a single particle size. These choices are necessary to answer one of our motivating questions: will JWST and ARIEL be able to distinguish between potential aerosol species? We also formulate our clouds and hazes such that we never put in a cloud or haze that has an impossibly large total mass of aerosol for a chosen metallicity and aerosol species. In the slab model, $F$ sets a fraction of \textit{available material} which will be incorporated into the haze or cloud, and the phase equilibrium assumption that only material in excess of super-saturation goes into cloud particles also scales with available material. As \citealt{Barstow2020b} point out, it will ultimately be most informative to conduct retrievals on transit spectra with several different aerosol parameterizations that have made different assumptions and then consider where the results agree and where they disagree and why. 

One limitation of our cloud and haze models is that they assume a single dominant species of aerosol, and they use a single log-normal particle-size distribution throughout the whole cloud or haze regardless of altitude. Detailed 1D microphysical models and models coupling microphysics to 3D GCMs indicate that condensed clouds likely have larger particles at their base and smaller particles near the upper boundary and that the chemical make-up of the clouds will vary with altitude and about the heterogeneous surface of highly irradiated exoplanets (\citealt{Parmentier2016}; \citealt{Powell2019}; \citealt{Helling2019a}; \citealt{Helling2019b}). If the cloud or haze in a target's atmosphere is not well-approximated by a single species and log-normal size distribution, then our cloud and haze models will likely fail to find a good fit or retrieve misleading results. Another limitation of our cloud and haze models is that we assume that the cloud or haze is uniformly present about the whole limb of the planet. This has been shown to return biased results (\citealt{Line2016}; \citealt{MacDonald2017}), so in order to apply our methods to real data we would need to adjust the model to account for patchiness. These limitations do not interfere with the purposes of this study because we are using simulated transit spectra and retrieval experiments on these synthetic observations to assess whether it is feasible that certain types of information could be embedded in JWST and ARIEL transit spectra rather than carrying out retrievals on actual data. This could be thought of as an upper limit of sorts. If information about aerosol species and particle-size distributions cannot be retrieved given our assumptions, then it will be even more difficult to do so when these assumptions are relaxed. If information can be retrieved, then this is a promising first step and further studies are warranted which include particle size-distributions and the full spectrum of the complex index of refraction in more complicated models. For example, one way to incorporate a varying size distribution with altitude is to apply the model suggested by \citealt{Ackerman2001}, which balances gravitational settling with a parameterized turbulent upwards mixing strength to determine the modal particle size at each level.

\subsection{Aerosol Species}\label{sec:aerosol_species}

The species of aerosols present in exoplanet atmospheres remains a mystery given our current observational capabilities. When fitting current data, it is generally sufficient to include an unidentified flat gray absorber and an optical slope along with gaseous opacity sources (\citealt{Mai2019}; \citealt{Barstow2020b}; \citealt{Barstow2020c}). This gives us a hint at the presence of both smaller and larger sized particles, but provides no smoking-gun signature of which set of species is present. In theory, transit spectra should be able to provide more information about aerosols because the effect of aerosols on a planet's transit spectrum depends on the optical properties of the aerosol species, the full size distribution of particles, where in the atmosphere the particles have formed, and how the presence of the aerosol affects the gaseous abundances of major absorbers (e.g. \citealt{Budaj2015}). However, within a narrow wavelength range, these different degrees of freedom can be degenerate and conspire to shape transit spectra in similar ways, even when species and size-distributions vary. Researchers have thus been left to make educated guesses as to what species are present. 

\begin{figure}
    \centering
    \includegraphics[width=0.5\textwidth]{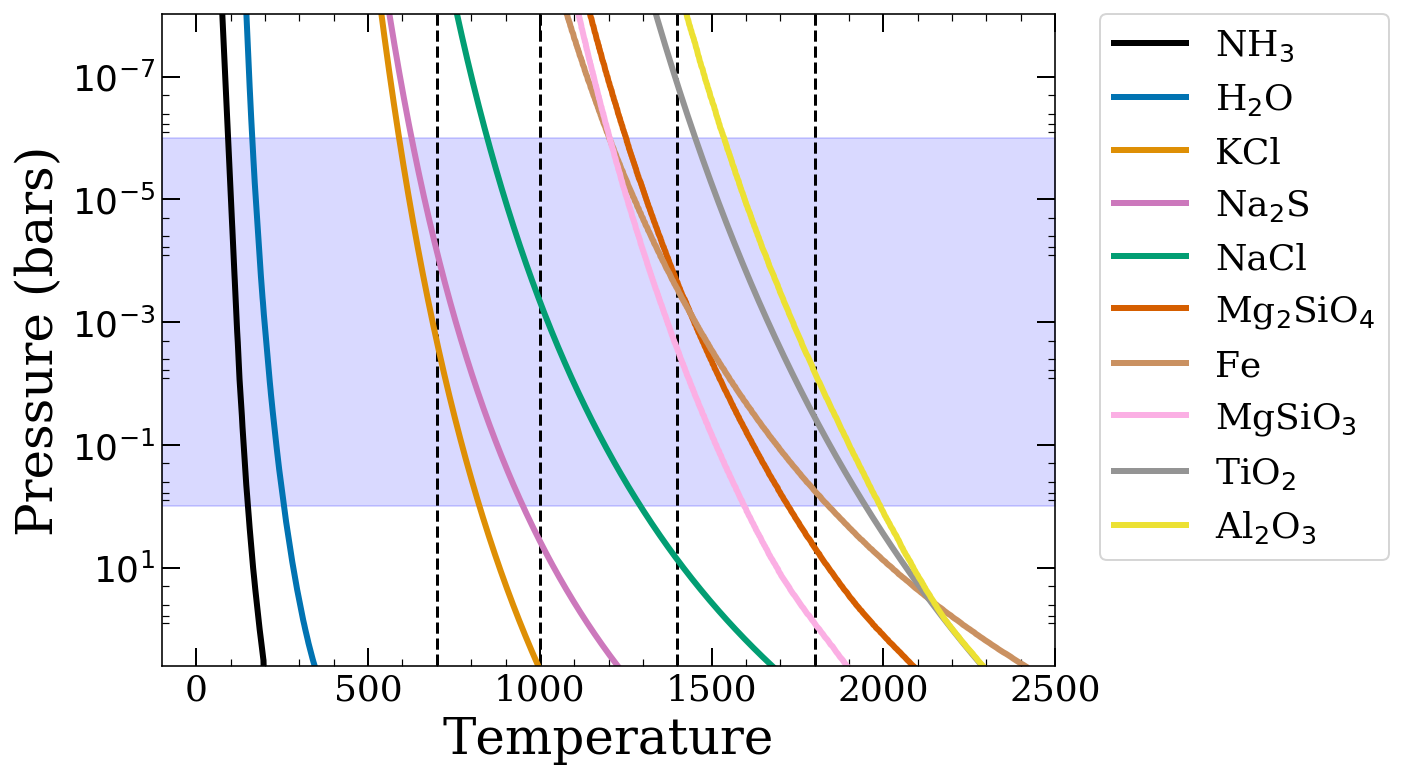}
    \caption{Claussius-Clapeyron lines for the condensed aerosol species included in this work (\citealt{Morley2012}; \citealt{Gao2018b}; \citealt{Sudarsky2003}). These curves are calculated for solar-metallicity atmospheres. The shaded blue region indicates the range of pressures probed by transit spectroscopy. The black dashed lines denotes four fiducial temperatures which we consider throughout this work. In our equilibrium cloud aerosol parameterization, we assume aerosols condense when the T-P profile and Claussius-Clapeyron line intersect.}
    \label{fig:cclines}
\end{figure}

\begin{table}
\begin{centering}
\begin{tabular}{c|c|c|c|c|c|c}
      Aerosol& Bulk density & Molar Mass  & $\sim$Condensation   & Solar Mixing Ratio& Source for Complex& \\
          Species  &  (g/cm$^3$)    & (g/mol)         & Temperature (K)& of Limiting Species&Indices of Refraction& Other names\\
\hline
\hline
Al$_2$O$_3$ & 4.02 & 101.96 & 1500-2000 &Al: 2.95$\times$10$^{-6}$ & (1) & Corundum \\
TiO$_2$ & 4.23 & 79.865& 1500-2000 &Ti: 9.77$\times$10$^{-8}$ & (1) &Anatase\\
Fe & 7.874& 55.845 & 1200-2500& Fe: 4.68e-5$\times$10$^{-5}$ & (1) & Iron \\
Mg$_2$SiO$_4$ & 3.25 & 140.69 &  1200-2000 &Mg: 3.8$\times$10$^{-5}$& (1) &Forsterite\\
MgFeSiO$_4$ & 3.25 & 153.31 &  1200-2000    & Fe: 4.68$\times$10$^{-5}$& (1) & Olivine\\
MgSiO$_3$ & 3.2 & 100.387 &  1200-2000    & Mg: 3.8$\times$10$^{-5}$& (1) & Enstatite \\ 
NaCl & 2.16 & 58.44& 900-1600&Cl: 3.16$\times$10$^{-7}$& (1) & Table Salt\\
Na$_2$S & 1.86& 78.0452& 700-1200 &S: 1.62$\times$10$^{-5}$& (1) & - \\
KCl & 1.98& 74.551&700-1000 &K: 1.32$\times$10$^{-7}$& (1) & Sylvite\\
H$_2$O & 0.997& 18.02& 200-300 &O: 8.51$\times$10$^{-4}$& (1) & Water Ice\\
NH$_3$& 0.8& 17.031&100-200 &N: 1.12$\times$10$^{-4}$& (2) & Ammonia Ice\\
\hline
- & $\sim$0.687& $\sim$27.0253 & - &N: 1.12$\times$10$^{-4}$& (3) & Titan Tholins\\
 (HCN)$_X$ & 0.687 & 27.0253 & - &N: 1.12$\times$10$^{-4}$& (4) & poly-HCN\\ 
C$_{14}$H$_{10}$ & 1.25&178.23 & - &C: 3.63$\times$10$^{-4}$& (5) & Carbonaceous Soot\\
C$_4$H$_8$& 0.588 & 56.106 & - &C: 3.63$\times$10$^{-4}$& (6) & Biomass Burning\\ &&&&&&(vegetation) \\
C$_3$H$_8$& 0.493& 44.1& - &C: 3.63$\times$10$^{-4}$& (7) & Biomass Burning\\
&&&&&&(propane) \\
\end{tabular}
    \caption{Summary of aerosol properties used in calculations. (1) \citealt{Kitzmann2018}; (2) \citealt{Robertson1975}; (3) \citealt{Khare1984}; (4) \citealt{Khare1994}; (5) \citealt{Chang1990}; (6) \citealt{Sutherland1991}; (7) \citealt{Wu2016}}
    \label{tab:aerosol_properties}
\end{centering}
\end{table}

After considering the large body of work positing which aerosol species are likely to be present, we have chosen to consider an extensive though not exhaustive list of 15 candidates in our study. These are listed in Table \ref{tab:aerosol_properties} and the corresponding indices of refraction and extinction efficiencies for 0.1, 1 and 10 micron particles are shown in Figures \ref{fig:indices} and \ref{fig:efficiencies} in the appendix. For condensing species, we include NH$_3$, H$_2$O,  KCl, Na$_2$S, NaCl, Mg$_2$SiO$_4$-Fe$_2$SiO$_4$ sequence, MgSiO$_3$-FeSiO$_3$ sequence, Fe, TiO$_2$, and Al$_2$O$_3$. This leaves out many of the species that have been considered, but includes those considered most likely to form based upon micro-physical modeling (\citealt{Gao2020}; \citealt{Helling2019a}), brown dwarf spectral modeling (\citealt{Leggett1998}; \citealt{Tsuji2002}; \citealt{Ackerman2001}), and the clouds and hazes seen on Solar System planets (\citealt{Robinson2015}). The condensation curves for these species are shown in Figure \ref{fig:cclines}. For photochemical hazes, we include Titan tholins, poly-HCN, a soot from propane burning, a PAH-dominated soot, and a soot resulting from burning vegetation. These are not predicted to be the exact Hydro-carbon hazes present in exoplanet atmospheres (\citealt{Horst2018a}; \citealt{Horst2018b}; \citealt{Horst2018c}), but they represent a range of plausible optical properties. These species are chosen mainly because they had readily available lab measurements of refractive indices across the wavelengths of interest. For the interested reader, we include a brief summary of the literature surrounding the likely make-up of exoplanet aerosols in the remainder of this section.

The natural place to start is to estimate the temperatures of objects and then consider which molecules made of available atomic species can exist at those temperatures. Looking purely at volatility, a number of studies have compiled lists of $\sim$30-40 candidate species that might exist at temperatures of 700-2500 K and have readily available lab measurements of complex indices of refraction (\citealt{Sudarsky2003}; \citealt{Budaj2015}; \citealt{Wakeford2015}; \citealt{Morley2012};  \citealt{Kitzmann2018}). Modeling transit spectra which include these aerosols shows that, for some species, relatively strong resonance features may show up in the continuous NIR-IR coverage of JWST (\citealt{Wakeford2015}; \citealt{Parmentier2016}; \citealt{Pinhas2017}; \citealt{Mai2019}). It will all depend on which species are present, their sizes and altitudes in atmospheres. Such studies have lead to the expectation that the higher SNR and broader wavelength coverage of future transit spectroscopy with JWST and ARIEL could allow us to identify which species are present in many cases. 
Other researchers have sought to winnow or rank this list of candidates by considering microphysical models of haze and cloud formation and the subsequent evolution of particle sizes and lifetimes in dynamic atmospheres (\citealt{Zahnle2016}; \citealt{Lavvas2017}; \citealt{Kawashima2018b}; \citealt{Kawashima2019a}; \citealt{Powell2019}; \citealt{Helling2019a}; \citealt{Helling2019b}; \citealt{Gao2020}). For condensate species, several studies find that TiO$_2$ is energetically the most likely to form via pure condensation (Helling et al. and Powell et al. Woitke Gao), while other species likely need seed particles. However, there is likely very little Ti in most atmospheres (\citealt{Anders1989}). It may be that TiO$_2$ forms seeds which other aerosols condense onto. In that case such aerosols would likely incorporate the optical properties of their outer layers rather than their tiny core of TiO$_2$ (\citealt{Powell2019}). \citealt{Gao2020} predict that planets with temperatures below 900 K will predominantly form photochemical hazes, planets with temperatures above 2000 K will be clear, and in between silicate clouds will dominate. When it comes to photochemical hazes, the list of possible species and mixes of species grows significantly! While the exact chemical mix of haze particles is hard to predict, it is widely agreed that any hazes will be dominated by hydrocarbons (\citealt{Kawashima2018b}; \citealt{Adams2019}; \citealt{Gao2020}). This is because the exoplanets probed by transit spectroscopy tend to be warm (typically 500-100 K), and receive a large amount of UV irradiation. If the atmospheres are CH$_4$ dominated rather than CO dominated, this is a perfect environment for hydrocarbon hazes to form easily through photochemical reactions triggered by photo-dissociation of methane (e.g. \citealt{Yung1984}). Theoretical studies have modeled haze production rates under different physical conditions and assumptions about metallicity and levels of UV flux, and assessing the effect on transit spectra (\citealt{Kawashima2018b}; \citealt{Kawashima2019a}; \citealt{Kawashima2019b}). Lab work is beginning to experiment with what types of hazes result as temperatures and input abundances vary (\citealt{Horst2018a}; \citealt{Horst2018b}; \citealt{Horst2018c}), however optical properties are not yet available for the resulting Hydro-carbon mixtures.

These types of detailed and approximate microphysical models have proven useful in studying clouds and hazes on earth and elsewhere in the solar system, but their adaptation to exoplanets is still in its infancy (\citealt{Powell2019}; \citealt{Helling2019a}; \citealt{Gao2020}). So far they have been mainly limited to making predictions (e.g. \citealt{Powell2019} predict observable effects from morning-evening asymmetries for hot Jupiters), but the data of JWST and ARIEL should start to test these predictions. If we can obtain credible empirical measures of aerosol properties from observations (species, size distributions, spatial positions), then detailed microphysical models can provide profound insight into the physical processes at work! 

Other hints at the make up of exoplanet clouds arise from inferences of condensates appearing and disappearing as temperatures change. Examples include the off-set peaks in some optical phase curves of hot Jupiters (\citealt{Parmentier2016}), the appearance and disappearance of reflective clouds on the elliptically orbiting Kepler-434b (\citealt{Dittmann2020}), and the strengthening and weakening of Fe lines in ultra high resolution spectroscopy WASP-76b (\citealt{Ehrenreich2020}). These works postulate MgS, KCl, NaCl, and condensates which contain Fe. 

\subsection{Simulated Data and Retrieval Frame Work}\label{sec:retrievals}
We simulate data reminiscent of JWST by binning our transit spectra to R$\sim$100 from 0.7 to 12 $\mu$m. This is recommended by \cite{Greene2016} as an optimal compromise between $SNR$ and spectral information. For noise, we take the Pandexo\footnote{https://exoctk.stsci.edu/pandexo/} errors for HD209458b with NIRISS, NIRCam I and II, MIRI to get the single-transit depth precision. These are then scaled by $\sqrt{N_{obs}}$ to represent the desired number of transit observations and added in quadrature with the systematic noise floors suggested in \cite{Greene2016}.

We use {\tt emcee}, a pure Python implementation of Goodman \& Weare’s affine invariant MCMC ensemble sampler, to carry out retrievals with METIS providing forward models and the aerosol parameterizations described in \S \ref{sec:aerosol_parameterizations}. Each chain is run to have a total length of 140,000 steps, which proved adequate for converged fits. During these retrievals we use the priors summarized in Table \ref{tab:priors}.

\begin{table}[]
    \centering
\begin{tabular}{c|c|l}
     Parameter& Prior &Description\\
     \hline
     P$_0$& 0.01 bars $<$ P$_0$ $<$ 10 bars & Reference pressure corresponding to known radius.\\
     Z & 0.1 $<$ Z/Z$_{\odot}$ $<$ 3.16 & Bulk metallicity \\
    a$_m$& 0.001 $\mu$m $<$ a$_m$ $<$ 100.0 $\mu$m &Modal particle size\\
    $\sigma _a$& 1.0 $<$ $\sigma _a$ $<$ 50.0 &Width of log-normal particle-size distribution\\
    $\alpha$ & 0.001 $<$ $\alpha$ $<$ 100 & Ratio of aerosol scale height to gaseous scale height \\
    P$_{top}$&   10$^{-7}$bars $<$ P$_{top}$ $<$ P$_0$& Top pressure cut-off of aerosol\\
    $F$&0 $<$ $F$ $<$ 1& Fraction of available material bound up in aerosol\\
\end{tabular}
    \caption{Priors used in MCMC retrievals.}
    \label{tab:priors}
\end{table}

\section{Fiducial Atmospheres when Clear} \label{sec:clear_fiducials}
 
 Before we present the main findings of this study (focused on the promising outlook for using JWST to study transit spectra of cloudy and hazy atmospheres), we first examine several fiducial atmospheres when they are clear. This section will provide useful context to help readers interpret later results by showing how information about atmospheric properties is embedded in transit spectra when clouds and hazes are absent. The parameters for four fiducial atmospheres are summarized in Table \ref{tab:fiducials_uniform}. The range of temperatures is chosen to hit each condensate that may form in the warm-hot exoplanets most suitable for study with transit spectroscopy (see black dashed lines in Figure \ref{fig:cclines}). We assume isothermal structures with temperatures of 700 K, 1000 K, 1400 K, and 1800 K respectively. The 1000-K, 1400-K, and 1800-K planets have masses and radii chosen to represent hot-Jupiters around sun-like stars. All are assumed to have the same mass but have increasing reference radii with temperature, so the planets have different surface gravities. The 700-K planet has a mass, radius, and stellar radius chosen to approximate a warmer version of the mini-Neptune GJ1214b. Along with temperatures, the surface gravities, and chemical abundances vary between the objects. 

\begin{table}[]
    \centering
    \begin{tabular}{c|c|c|c|c|c|l}
          T & M & P$_{base}$ & R$_{base}$ & Z &R$_{*}$&Possible Aerosols\\
          (K) & (M$_{J}$) & (bars) & (R$_{J}$)  &(Z$_{\odot}$) &(R$_{\odot}$) & \\
         \hline
        1800 &1.0&1.0&1.5&3.0&1.0&Fe, TiO$_2$, Al$_2$O$_3$, hazes\\
         1400 & 1.0 & 1.0 & 1.2 &3.0 &1.0&Mg$_2$SiO$_4$, MgSiO$_3$, Fe, hazes\\
         1000 & 1.0 & 1.0 & 1.0 &3.0 &1.0&Na$_2$S, NaCl, hazes\\
         700 &0.0203 &1.0 &0.2389&3.0 &0.2064&Na$_2$S, KCl, hazes\\
    \end{tabular}
    \caption{Model parameters and possible aerosols for our four fiducial atmospheres.}
    \label{tab:fiducials_uniform}
\end{table}

\begin{figure}
    \centering
    \includegraphics[width=0.45\textwidth]{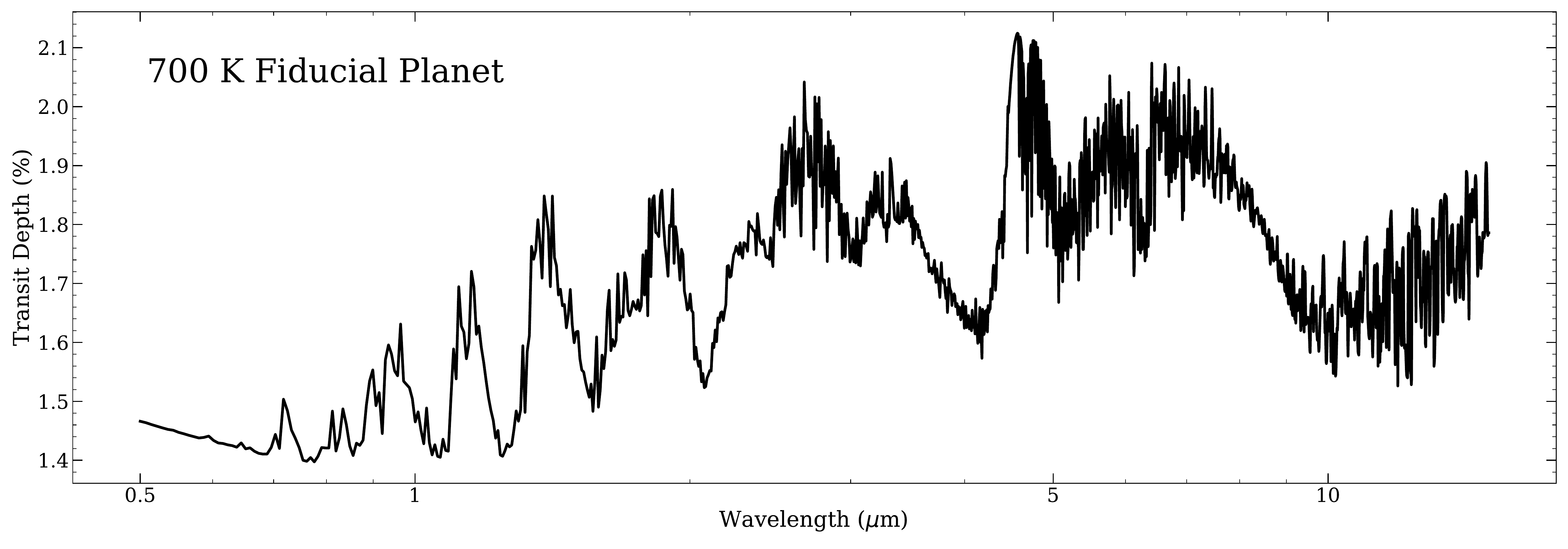}
    \includegraphics[width=0.425\textwidth]{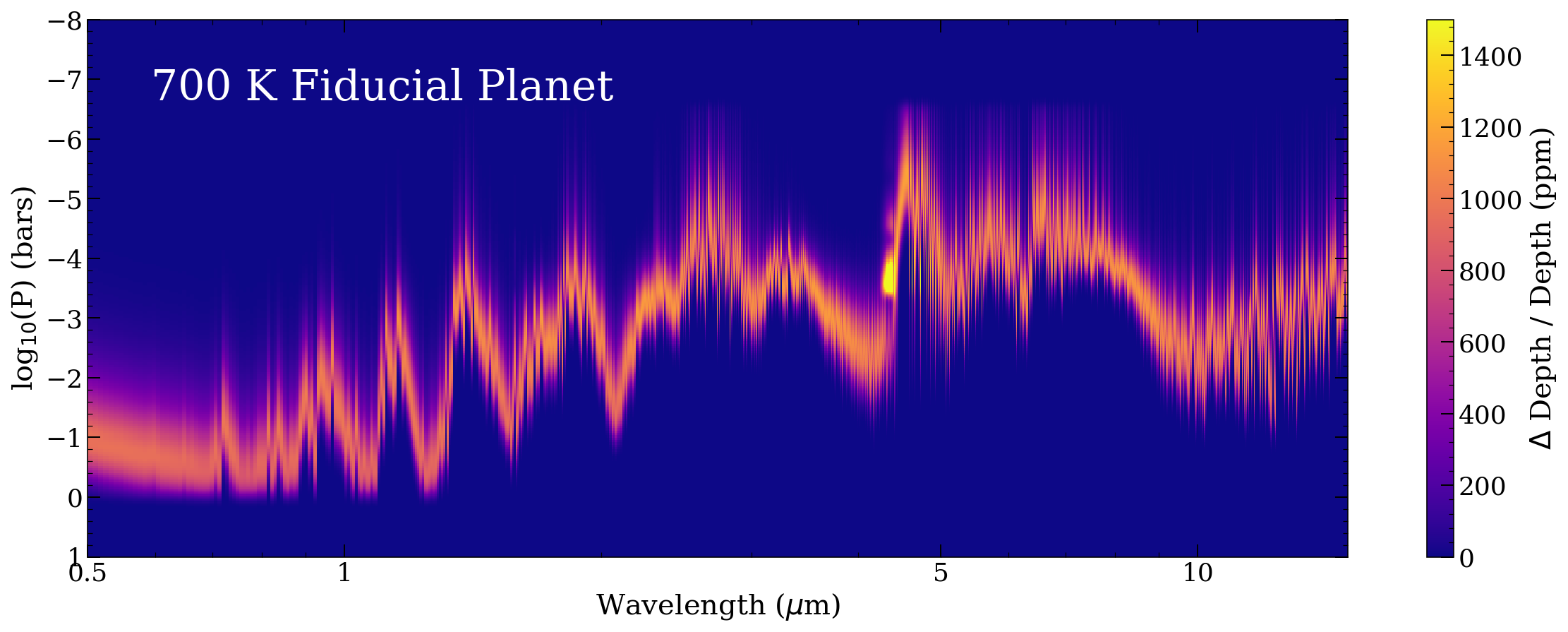}
    \includegraphics[width=0.45\textwidth]{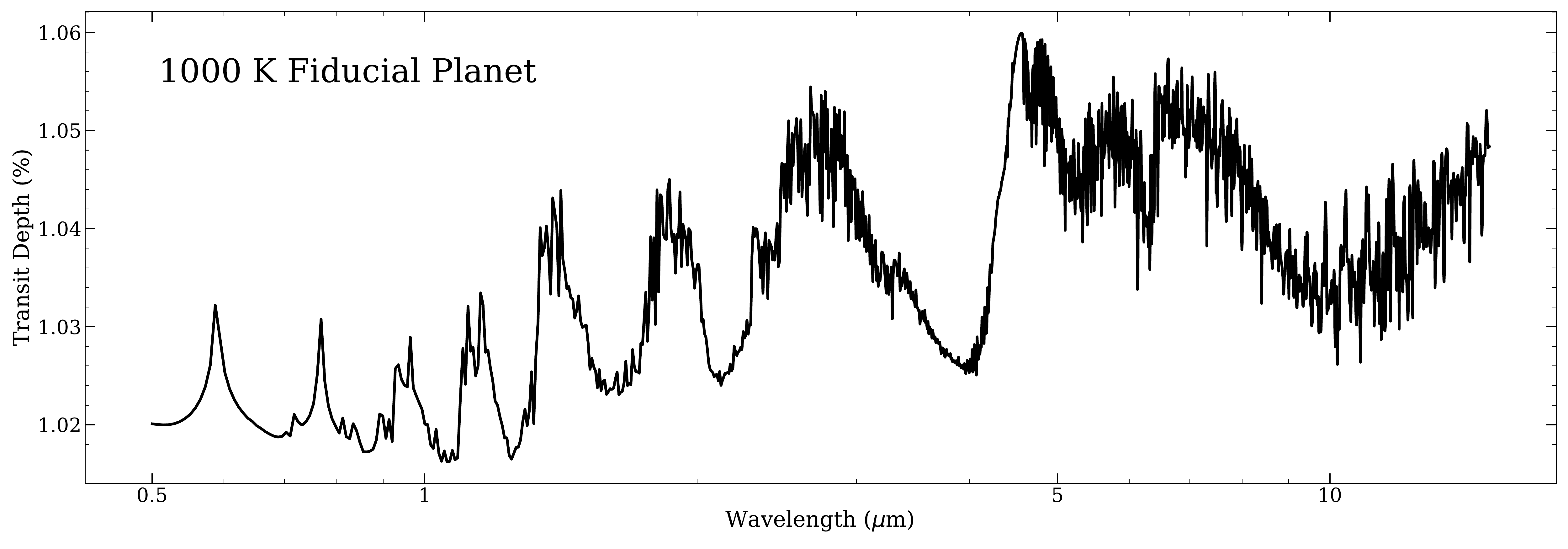}
    \includegraphics[width=0.425\textwidth]{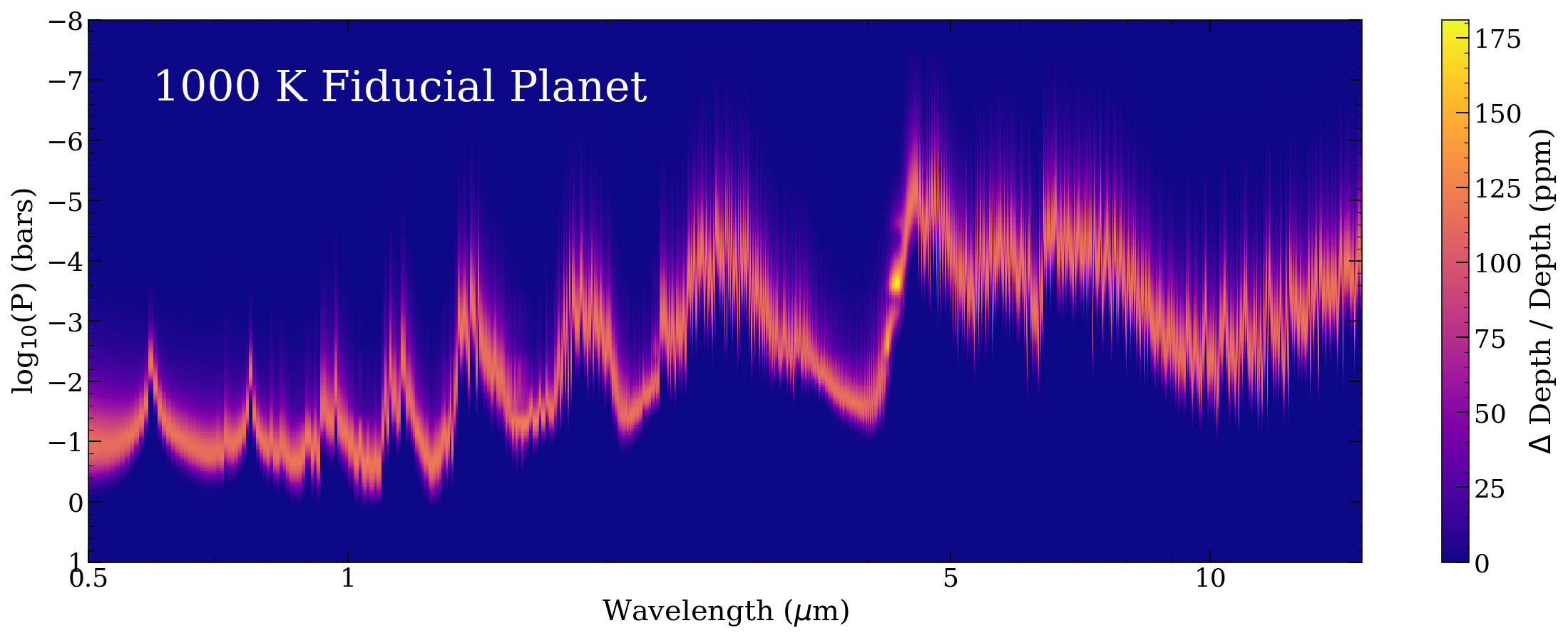}
    \includegraphics[width=0.45\textwidth]{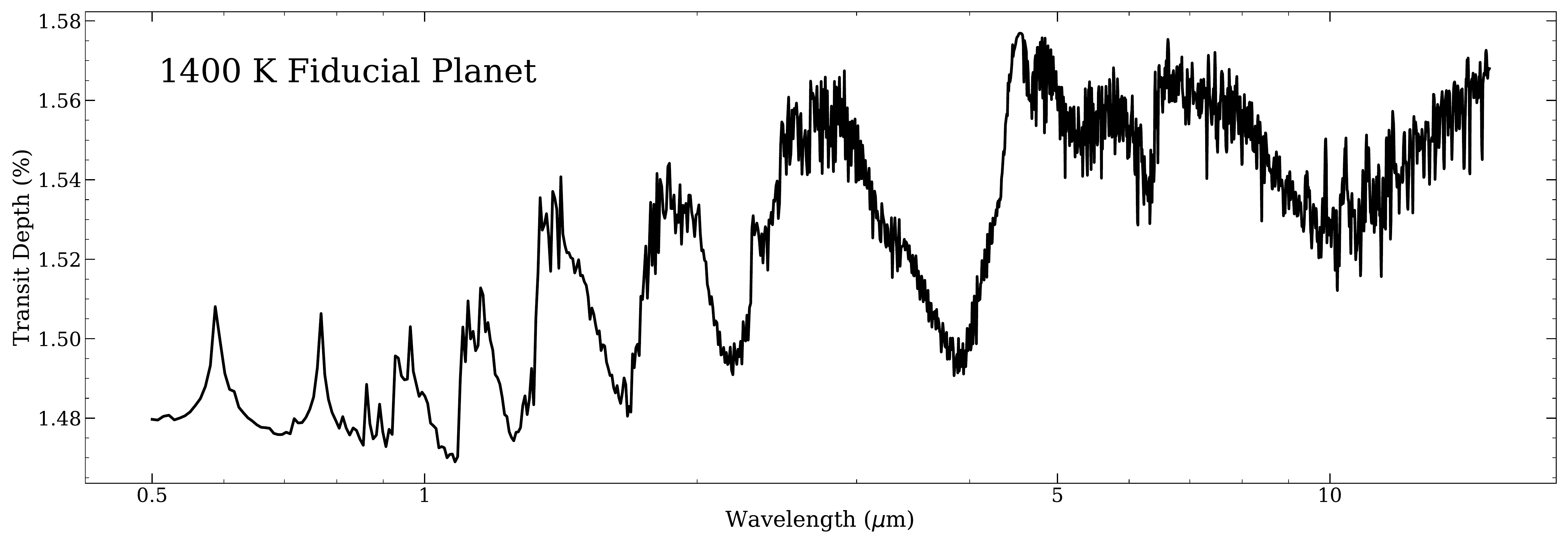}
    \includegraphics[width=0.425\textwidth]{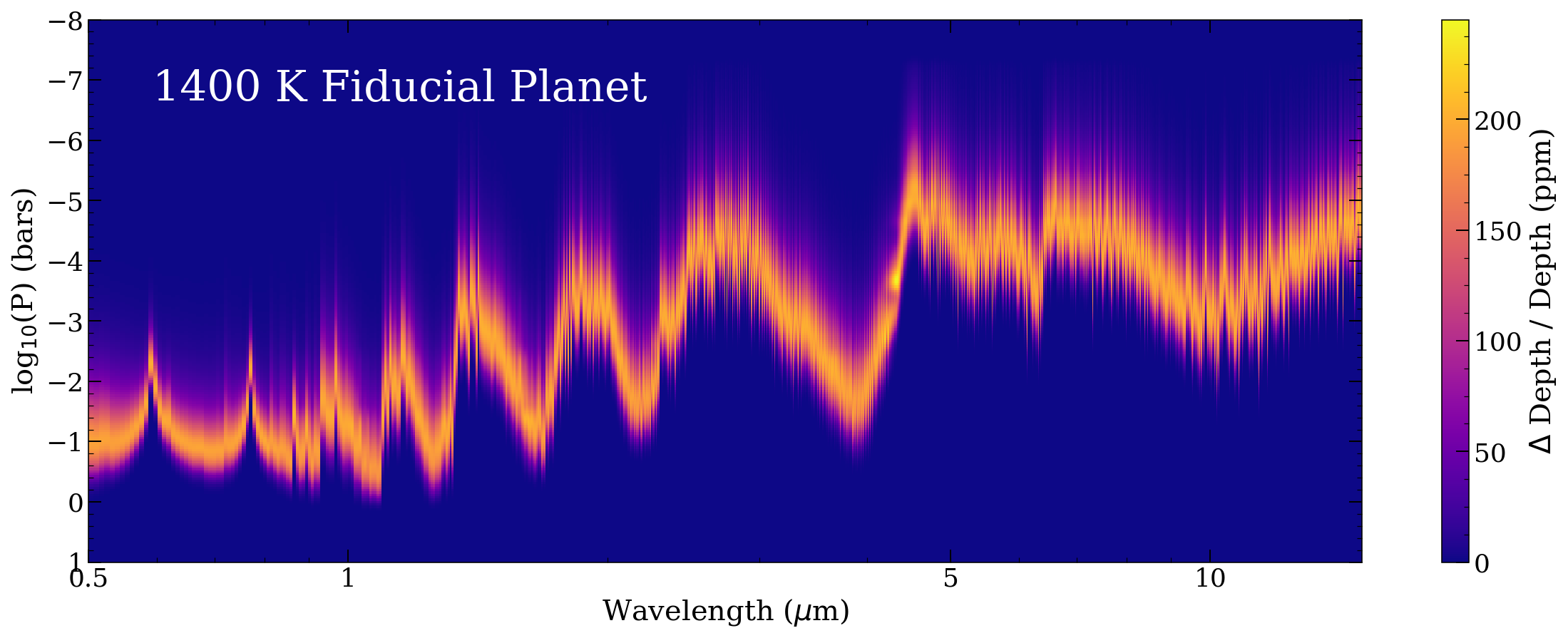}
    \includegraphics[width=0.45\textwidth]{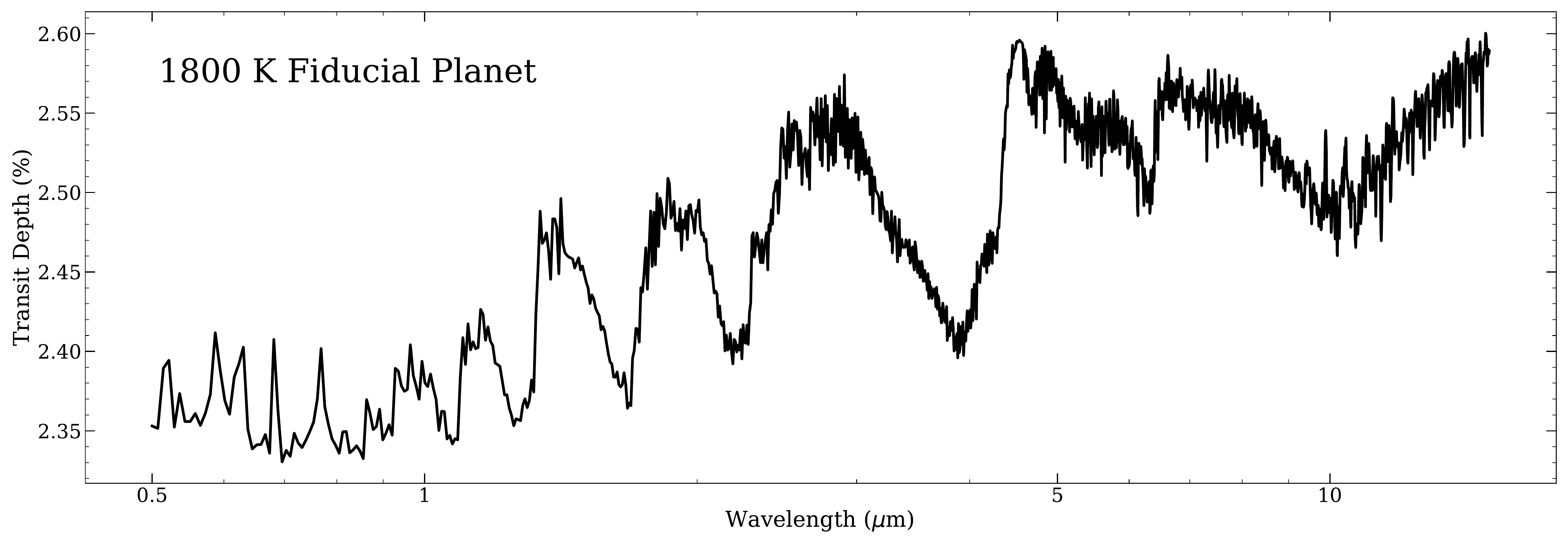} 
    \includegraphics[width=0.425\textwidth]{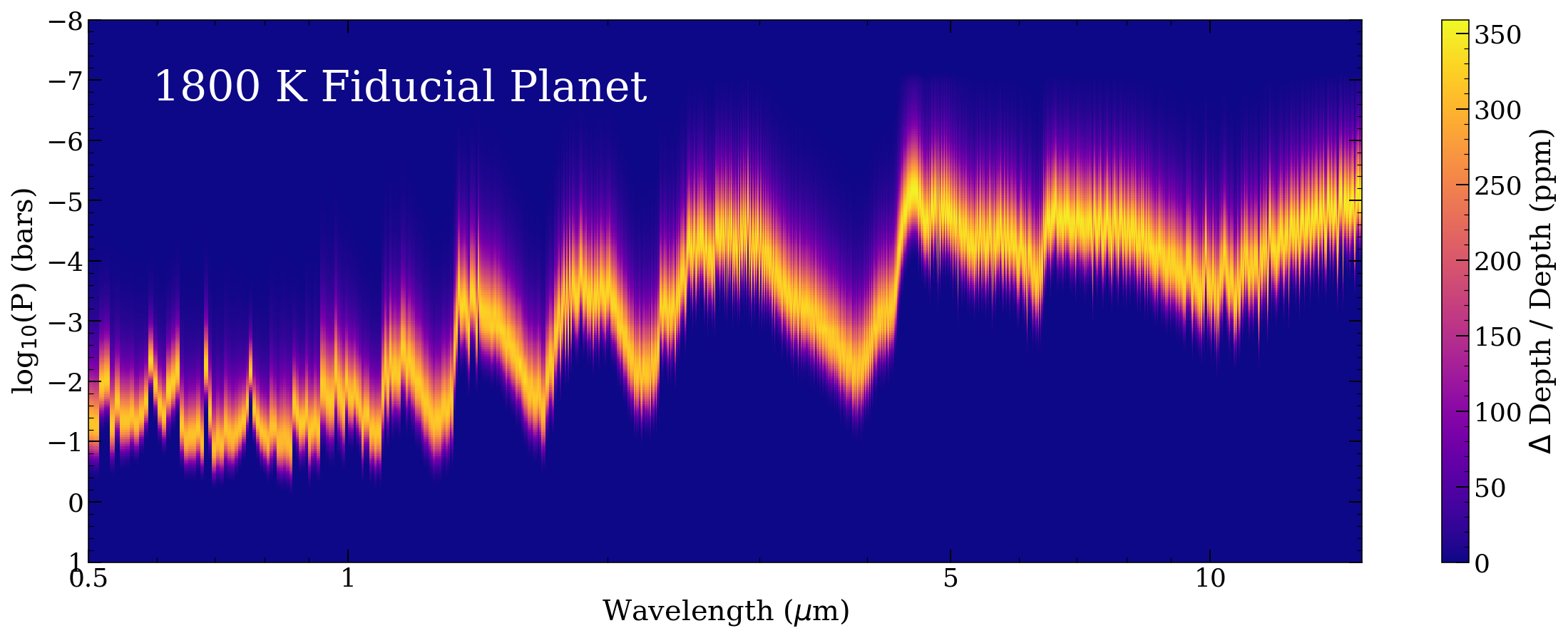} 
    \caption{The left column shows transit spectra for the four fiducial atmospheres with parameters listed in Table \ref{tab:fiducials_uniform}. The right column shows a calculation of which pressure levels are contributing most to the transit spectrum at each wavelength. Lighter colored, yellow and orange portions of the Figure represent the parts of the atmosphere that are shaping the transit spectrum, while darker blue portions are pressure levels that do not contribute to the transit spectrum.}
    \label{fig:clear_contributions}
\end{figure}

 Figure \ref{fig:clear_contributions} shows the fiducial clear transit spectra in the left-hand column and a measure of which pressure levels are shaping the transit spectra at each wavelength in the right-hand column. This calculation is done by setting the opacity in a given pressure layer to zero and then computing the resulting transit spectrum without that layer's contribution. This is then compared to the full transit spectrum including the opacity from all layers. If a layer is contributing to the transit spectrum, then setting it's opacity to zero will result in a large difference between the full spectrum and the spectrum missing one layer. We will refer to this calculation hereafter as the ``transit contribution function". In the figure, lighter colored, yellow and orange portions represent the parts of the atmosphere that are shaping the transit spectrum, while darker blue portions are pressure levels that do not contribute much to the transit spectrum. One can see that, at wavelengths where the low pressures (high altitudes) are shaping the transit spectrum, a correspondingly larger transit depth is seen in the transit spectrum on the left. 

In the 0.5-1.0 $\mu$m range, all four transit spectra are probing 10$^{-1}$ - 10$^{-3}$ bars. The 700-K case has Rayleigh scattering blueward of 0.7 $\mu$m. The 1000-K and 1400-K planets have prominent sodium and potassium doublets at 0.66 and 0.77 $\mu$m. In the 1800-K  atmosphere, metal hydrides like FeH, CrH, MgH and CaH begin to show up and the optical spectrum takes on a jagged flat shape. All temperatures have prominent water absorption features across the 1-15 $\mu$m range, augmented by CH$_4$ at the edges and in between water features at ~1.5, 2.25, and 3.5 $\mu$m. There is a strong CO absorption feature at 4.7 $\mu$m, and another smaller one at $\sim$2.3 $\mu$m. As the temperature lowers from 1800 K down to 700 K, CH$_4$ abundances increase and CO abundances decrease \citep{Sharp2007}. In the 1-2 $\mu$m range peaks of absorption features in the transit spectra probe around 10$^{-3}$ - 10$^{-4}$ bars, while between absorption features transit spectra probe pressures of around 10$^{-2}$ bars.  Longward of 2 $\mu$m, the peaks of absorption features probe around 10$^{-4.5}$ - 10$^{-5.5}$ bars. The windows between absorption features reach down to only 10$^{-4}$ bars. In the CO feature around 4.7 $\mu$m, a very high altitude/low pressure of around 10$^{-6}$ bars is shaping the transit spectrum.  

\begin{figure}
    \centering
    \includegraphics[width=\textwidth]{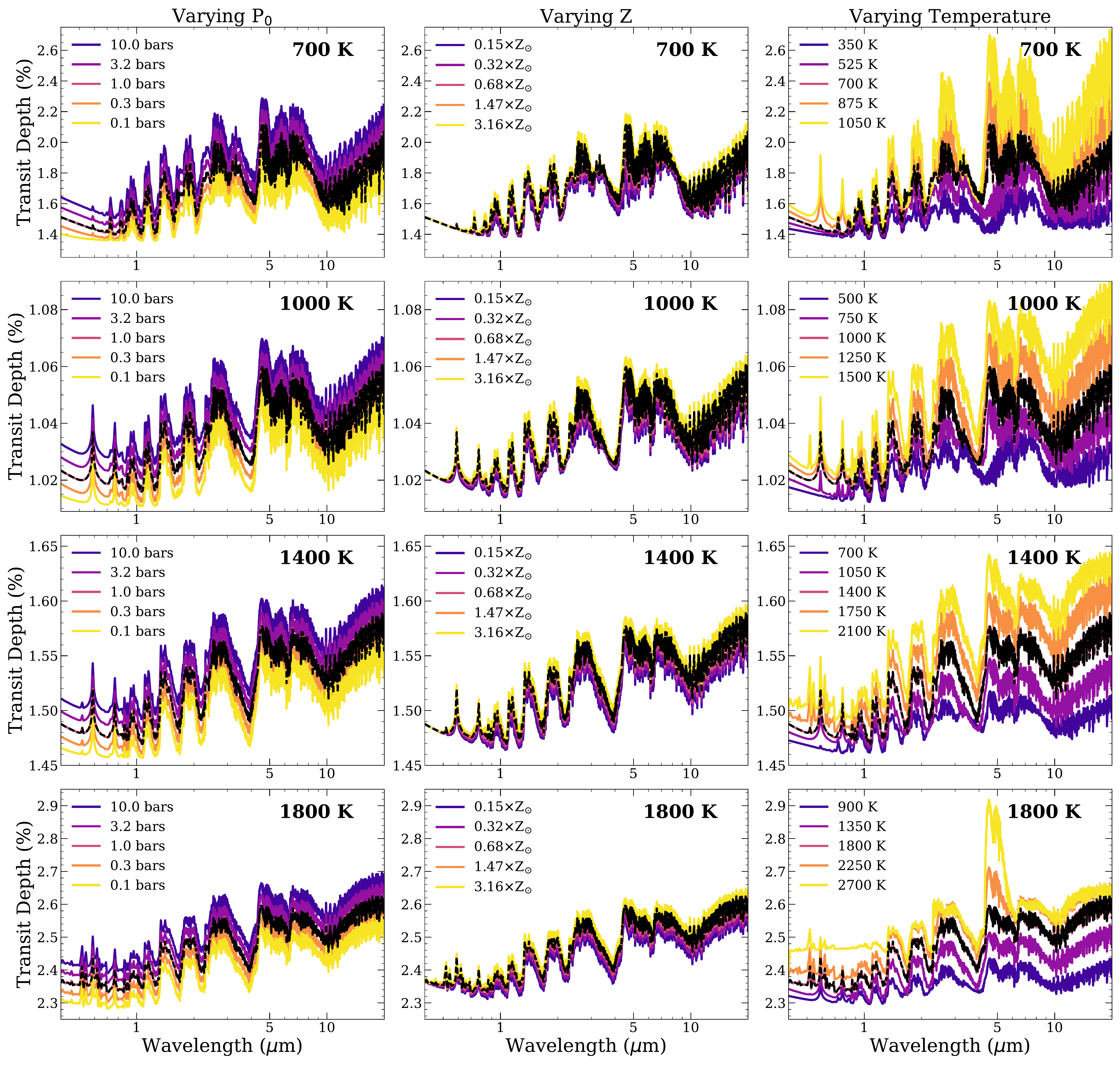}
    \caption{Demonstration of the model's sensitivity to the reference pressure $P_0$, the metallicity $Z$, and the temperature. We vary each parameter about the fiducial values for each of our four fiducial temperatures (black dashed lines show fiducial transit spectra). See Table \ref{tab:fiducials_uniform} for a summary. Each row varies a different parameter and each column shows a different fiducial temperature atmosphere. Within each column the y-scale is kept the same for easier comparison.}
    \label{fig:clear_1D_pstudy}
\end{figure}

Figure \ref{fig:clear_1D_pstudy} demonstrates how perturbing the non-aerosol parameters in our model (reference pressure, temperature, and metallicity) changes the resulting transit spectra. Each column contains transit spectra perturbing a different parameter, and each row shows a different one of the fiducial planets. 

\begin{figure}
    \centering
    \includegraphics[width=\textwidth]{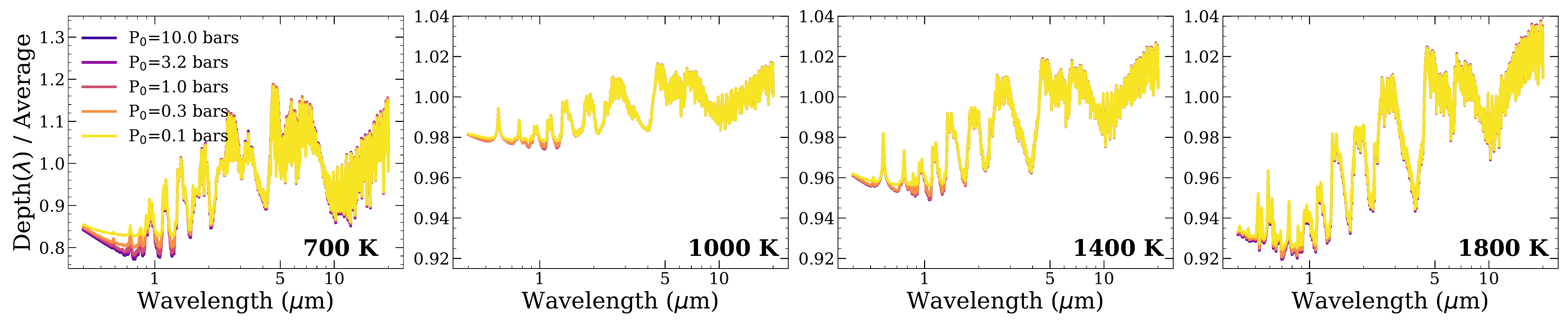}
    \caption{This Figure shows the wavelength-dependent transit-depth divided by the average transit depth across the wavelengths shown as the reference pressure changes. This is meant to isolate how changing the reference pressure alters the shape of the transit spectrum, not just the baseline. Each panel shows a different fiducial temperature atmosphere. In this case we have a different y-axis scale for the 700-K atmosphere but keep the same scale for the 1000, 1400, and 1800-K atmospheres.}
    \label{fig:clear_1D_p0_pstudy}
\end{figure}

Altering the reference pressure, P$_0$, keeps the relative shape of the transit spectra almost the same, mostly shifting the average transit depth up or down (see the left column of Figure \ref{fig:clear_1D_pstudy}). To demonstrate the subtle change in shape that results from changing P$_0$, Figure \ref{fig:clear_1D_p0_pstudy} shows the transit depth at each wavelength divided by the average depth over all wavelengths. The shape only changes for wavelengths around 1.5 $\mu$m and shorter. This indicates that changing P$_0$ has a different effect on Rayleigh scattering than it does on the gaseous absorption and scattering. The effect is barely perceivable in the higher surface gravity 1000-K, 1400-K, and 1800-K planets, but for the lower surface gravity 700-K planet, there is a non-negligible change in shape.

\begin{figure}
    \centering
    \includegraphics[width=\textwidth]{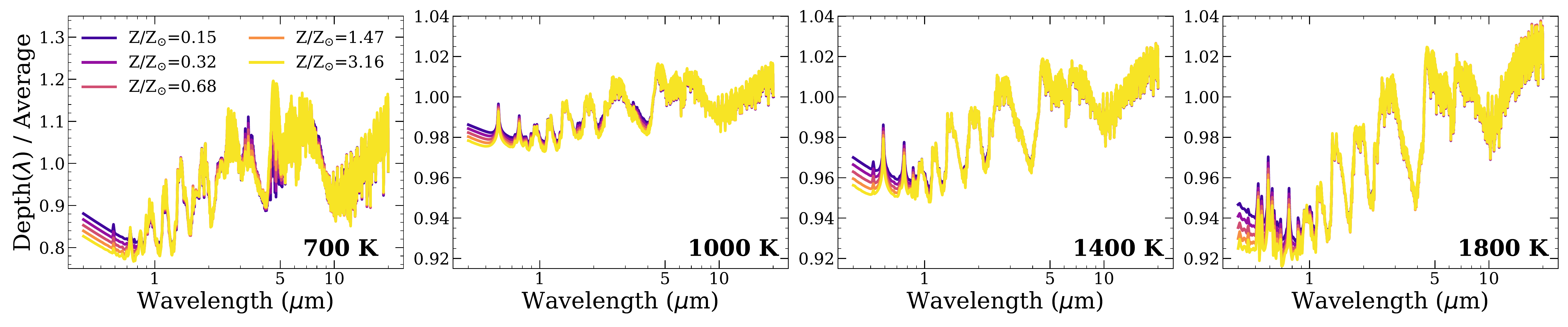}
    \caption{This Figure shows the wavelength dependent transit-depth divided by the average transit depth across the wavelengths shown as you vary the metallicity. This is meant to isolate how changing the metallicity alters the shape of the transit spectrum, not just the overall opacity. Each panel shows a different fiducial temperature atmosphere. One can see clearly that, within the metallicity range considered, the change in mean molecular weight due to the change in metallicity is negligible. Changing metallicity only influences the balance of CH$_4$ and CO and the amount of Rayleigh scattering relative to the other gaseous absorption.}
    \label{fig:clear_1D_z_pstudy}
\end{figure}

Changing the metallicity, $Z$, has a more subtle effect (see the center column of Figure \ref{fig:clear_1D_pstudy}). It systematically increases or decreases abundances of most opacity sources across the board, but also alters the relative thermochemical equilibrium abundances of a few important opacity sources. Finally, it can also change the scale height of the atmosphere by changing the mean molecular weight. Figure \ref{fig:clear_1D_z_pstudy} isolates the change in shape as metallicity varies. In order to get a handle on metallicity, one must have measurements that include some of the wavelengths that do not lie directly on top of each other in this figure, and measurements at wavelengths that do. Otherwise, there is either no change with metallicity, and/or a degeneracy between changing the reference pressure and the metallicity. For all four temperatures, the relative differences between wavelengths blueward of 0.75 $\mu$m and wavelengths redward of 0.75 $\mu$m can show metallicity changes. Unfortunately, these short wavelengths are very prone to being covered by aerosols since they are reaching deeper into the atmosphere to pressures of around 10$^{-2}$ bars. When the temperature is 700 K, as metallicity varies some longer wavelengths around  3.5 $\mu$m, 5 $\mu$m, and 8 $\mu$m change relative to the rest of the transit spectrum. These changes occur around pressures of 10$^{-4}$ bars, so they are more likely to be detectable above a cloud or haze, but still in danger of being obscured by a high-altitude aerosol. When the temperature is 1000 K, the depth and breadth of water absorption features vary slightly with metallicity, and changes are apparent in the windows at 1.8, 2.25, and 4.25 to 4.75 $\mu$m (i.e. where CH$_4$ is peaking through). These wavelengths are probing around 10$^{-2.5}$- 10$^{-3}$ bars. For 1400 K and 1800 K, there is really not much change in shape with metallicity that is probed above a pressure of 10$^{-2}$ bars. This hints that if any aerosols are present at altitude, it may be difficult to make metallicity measurements of exoplanets hotter than 1000 K.

\begin{figure}
    \centering
    \includegraphics[width=\textwidth]{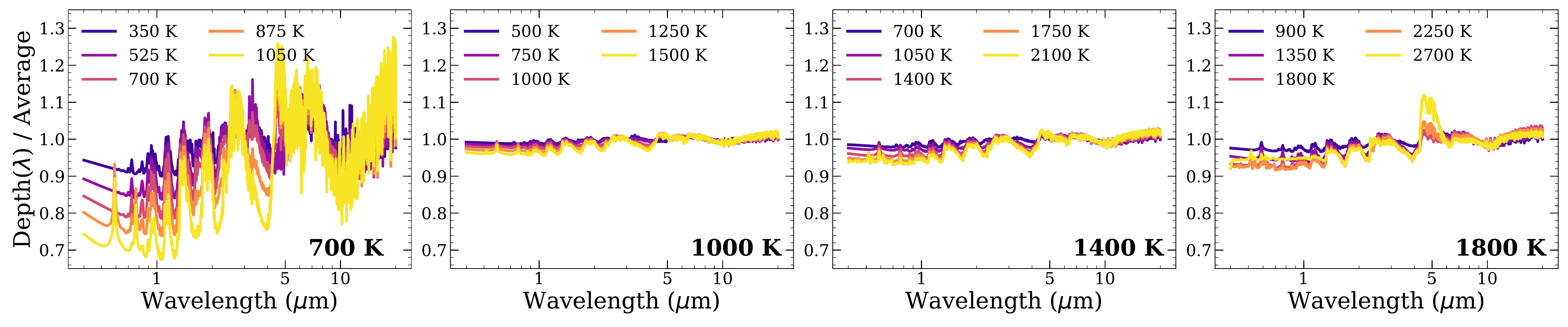}
    \caption{This Figure shows the wavelength dependent transit-depth divided by the average transit depth across the wavelengths shown as you vary the temperature. This is meant to isolate how changing the temperature alters the shape of the transit spectrum from any shifts upwards or downwards that are constant for all wavelengths. Each panel shows a different fiducial temperature atmosphere and varies the temperature about it. We keep the y-axis the same for all four planets. For the 1000, 1400 and 1800-K planets, the mass is always one Jupiter mass. This means that increasing temperature leads to increased scale heights, so for the 1800 K object we see larger variation across wavelengths. The 700 K object has a much smaller mass, so it's scale height is an order-of-magnitude larger than the others even though it has a cooler temperature. Changing the temperature has a very different affect than changing the metallicity or the reference pressure.}
    \label{fig:clear_1D_T_pstudy}
\end{figure}

Tweaking the temperature, $T$, has a significant effect on both the shape and baseline depth of the transit spectra, see Figure \ref{fig:clear_1D_T_pstudy} and the right-most column of Figure \ref{fig:clear_1D_pstudy}. First of all, the temperature sets the equilibrium chemistry, so changing the temperature alters which features are present and their relative strengths. At certain junctures, the change in temperature can dramatically change the shape: for example, between 700-800 K when the Na doublet becomes prominent, between 1500 and 1700 K when metal hydrides start shaping the optical, and above 2000 K when the CO feature at 4-5 $\mu$m grows extremely prominent. Around 2500 K, H$^{-}$ opacity starts to kick in blueward of 2 $\mu$m. Changing the temperature also changes the scale height, either stretching or squashing the features of the transit spectra. This effect is especially clear from looking at  Figure \ref{fig:clear_1D_T_pstudy}. Note that the 700-K planet has the smallest surface gravity, then the 1800-K planet, then the 1400-K planet, then the 1000-K planet.

Figures \ref{fig:clear_1D_pstudy} - \ref{fig:clear_1D_T_pstudy} show that changes to the temperature and the reference pressure have larger effects than changes to the metallicity (at least within the range of $Z$ = 0.1 - 3.16 $\times$ Z$_{\odot}$). They also show that changing the reference pressure and the metallicity can have similar effects on the transit spectrum if only limited wavelength coverage or low precision measurements are available. We see the impact of these trends play out in retrievals, as one would expect. Constraints on temperature and reference pressure are generally very tight, while constraints on metallicity are a bit looser. When thick clouds or hazes overpower the gaseous opacity in the optical wavelength range, the degree of degeneracy between reference pressure and metallicity tends to increase. 

\section{Gas versus Aerosol Opacity from Optical to IR} \label{sec:wavelength_coverage}

\begin{figure}
    \centering
    \includegraphics[width=0.9\textwidth]{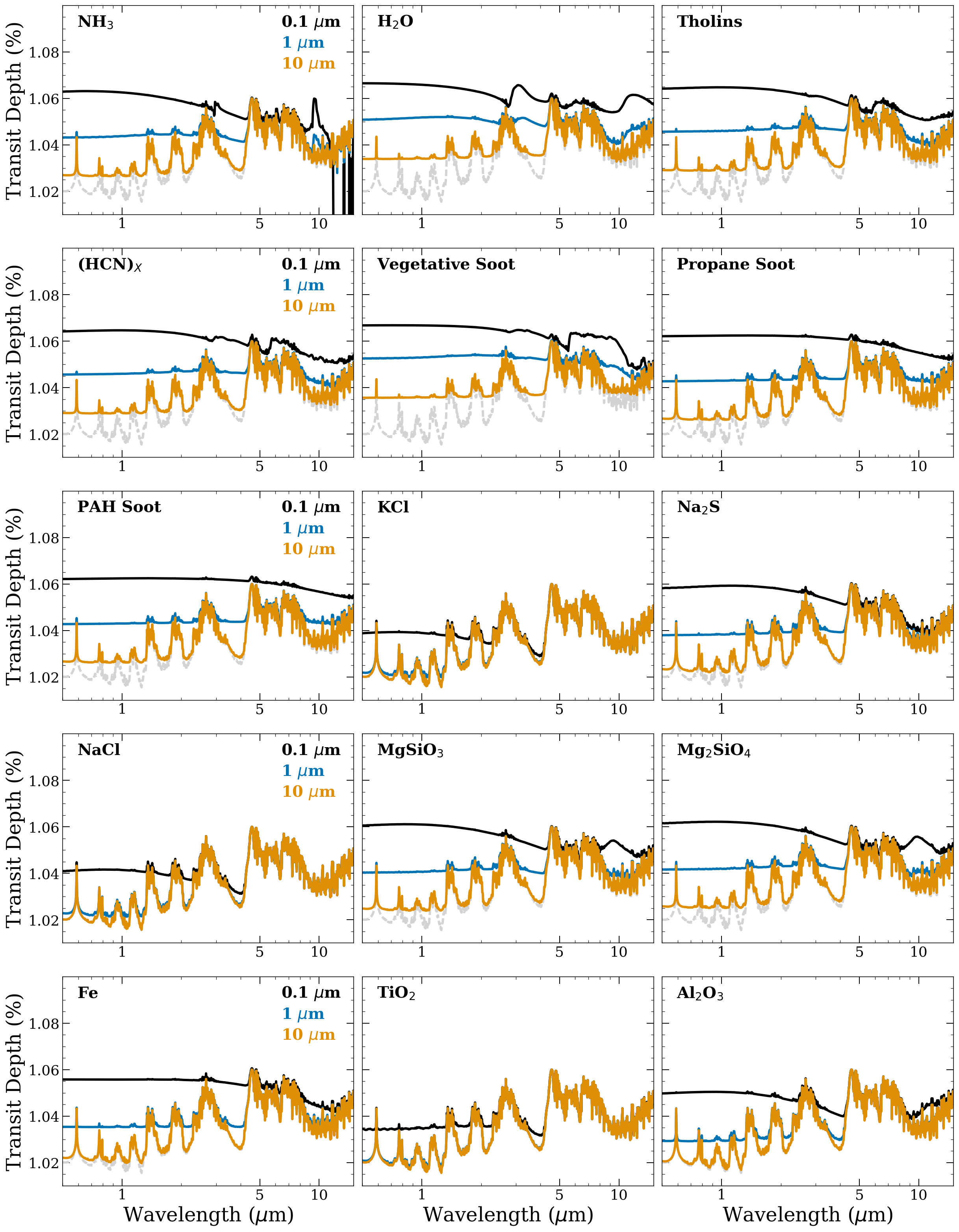}
    \caption{Transit spectra for all our aerosols species added to the 1000-K fiducial atmosphere as a slab aerosol with Z=1.05$\times$Z$_{\odot}$, $F$=0.25, and P$_{top}$ set to be at the top of the atmosphere. Modal particle sizes of 0.1, 1.0, and 10.0 $\mu$m are compared, always with a log-normal size dispersion of $\sigma _a$=2.5. Some of the condensing species would certainly not form at 1000 K, but we have kept a constant temperature for easier comparisons.}
    \label{fig:all_specs_diff_sizes}
\end{figure}

Now that we have looked at the transit spectra of our fiducial atmospheres when they are clear, we will move on to explore the effects of adding in different species and sizes of aerosols. It is widely hoped that the broader wavelength coverage of JWST will enable us to identify which aerosol species are present in exoplanet atmospheres and to access the stronger gaseous absorption features at longer wavelengths, even if clouds and hazes are diminishing the signal. Example transit spectra containing each species are shown in Figure \ref{fig:all_specs_diff_sizes} and compared to the corresponding clear transit spectrum (light gray dashed line). The spectra shown in Figure \ref{fig:all_specs_diff_sizes} assume our fidcuial 1000-K hot Jupiter around a sun-like star, a log-normal particle-size distribution with a dispersion of 2.5 and the three modal particle sizes indicated, and finally that a quarter of the available material went into forming the aerosols ($F$=0.25). Particles are allowed to form as high up into the atmosphere as there is sufficient material. This type of calculation has formed the basis for the community's hopes that JWST will provide a smoking gun revealing which aerosols are present (\citealt{Wakeford2015}; \citealt{Pinhas2017}; \citealt{Kitzmann2018}). 

Looking at Figure \ref{fig:all_specs_diff_sizes}, one can see that the contributions from aerosols in the 0.4-2 $\mu$m wavelength range are generally flat or consist of a smooth monotonic slope for a variety of species and particle sizes. These are the wavelength ranges readily observed to date, so it is no wonder we have not been able to identify which species are present. Species start to look more distinct from 3-15 $\mu$m as spectral features arising from resonant bending/vibrational/rotational modes may be present (e.g. in NH$_3$, H$_2$O, Titan tholins, (HCN)$_X$, vegetative soot, MgSiO$_3$, Mg$_2$SiO$_4$, and Al$_2$O$_3$). In the absence of such distinct features, there may still be a unique, non-gray shape to the smooth aerosol opacity (e.g. for KCl, Na$_2$S, NaCl, or Fe). Figure \ref{fig:all_specs_diff_sizes} hints at the advantage of broad wavelength coverage for both recognizing aerosols and for measuring gaseous properties when aerosols are present. It also demonstrates a point that we will reinforce in later results: there is a tension between obtaining strong constraints on aerosol properties and obtaining strong constraints on gaseous absorption.

Whether or not aerosol spectral features are actually observable depends on the relative strengths of aerosol opacities and gaseous opacities at wavelengths where aerosol spectral features peak, as well as whether particle-size distributions are such that these features are present at all. With the assumptions made here (no top-pressure cut-off, Z=1.05, and $F$=0.25), the smaller modal particle size of 0.1 $\mu$m forms many more aerosol particles and typically overpowers gaseous absorption at most wavelengths (black lines). These type of spectra, dominated by aerosol opacity across most wavelengths, would make it easy to identify which aerosol species is present and what the particle-size distribution is, but won't provide much of the information about the gaseous absorption. On the opposite end, the 10-$\mu$m modal particle distribution formed from the same available mass of material makes many less aerosol particles, and we see that aerosols fill in the deepest troughs in gaseous absorption features, but don't overwhelm the gaseous absorption peaks (orange lines). In these examples, the spectra won't provide much information about the aerosol species or properties, but you could expect to get strong constraints on things like temperature, reference pressure and metallicity. The spectra with a 1-$\mu$m modal particle size fall in between (blue lines). Aerosols tend to overpower gaseous opacities at shorter wavelengths, where water absorption is weaker, but not at longer wavelengths. The windows between water features at 4 and 10 $\mu$m provide the best chance of detecting aerosol features directly. For some species, the 1-$\mu$m spectra look like they could provide a smoking gun signature of which species is present, and also allow constraints on the gas-phase abundances. 

As we have seen from the spectra in Figure \ref{fig:all_specs_diff_sizes}, and might have expected intuitively, more wavelength coverage provides more information. However, one might ask: are some of these wavelengths more information dense than others? One way to assess this is to compare the Jacobians of our models for transit depth across wavelengths. We computed the partial derivatives of transit depth with modal particle size, log-normal size dispersion, and metallicity for transit spectra with a variety of particle sizes and aerosol species (Figure \ref{fig:700K_jacobian} here and Figures \ref{fig:1000K_jacobian} - \ref{fig:1800K_jacobian} in the appendix). The aerosols are roughly grouped by the temperatures at which they may condense, with species meant to represent hydrocarbon hazes added in where there are less than 3 candidate condensate species. If the \textit{absolute value} of the Jacobian is large, then it indicates that the measurement at that wavelength is sensitive to the parameter in question. 

\begin{figure}
    \centering
    \includegraphics[width=\textwidth]{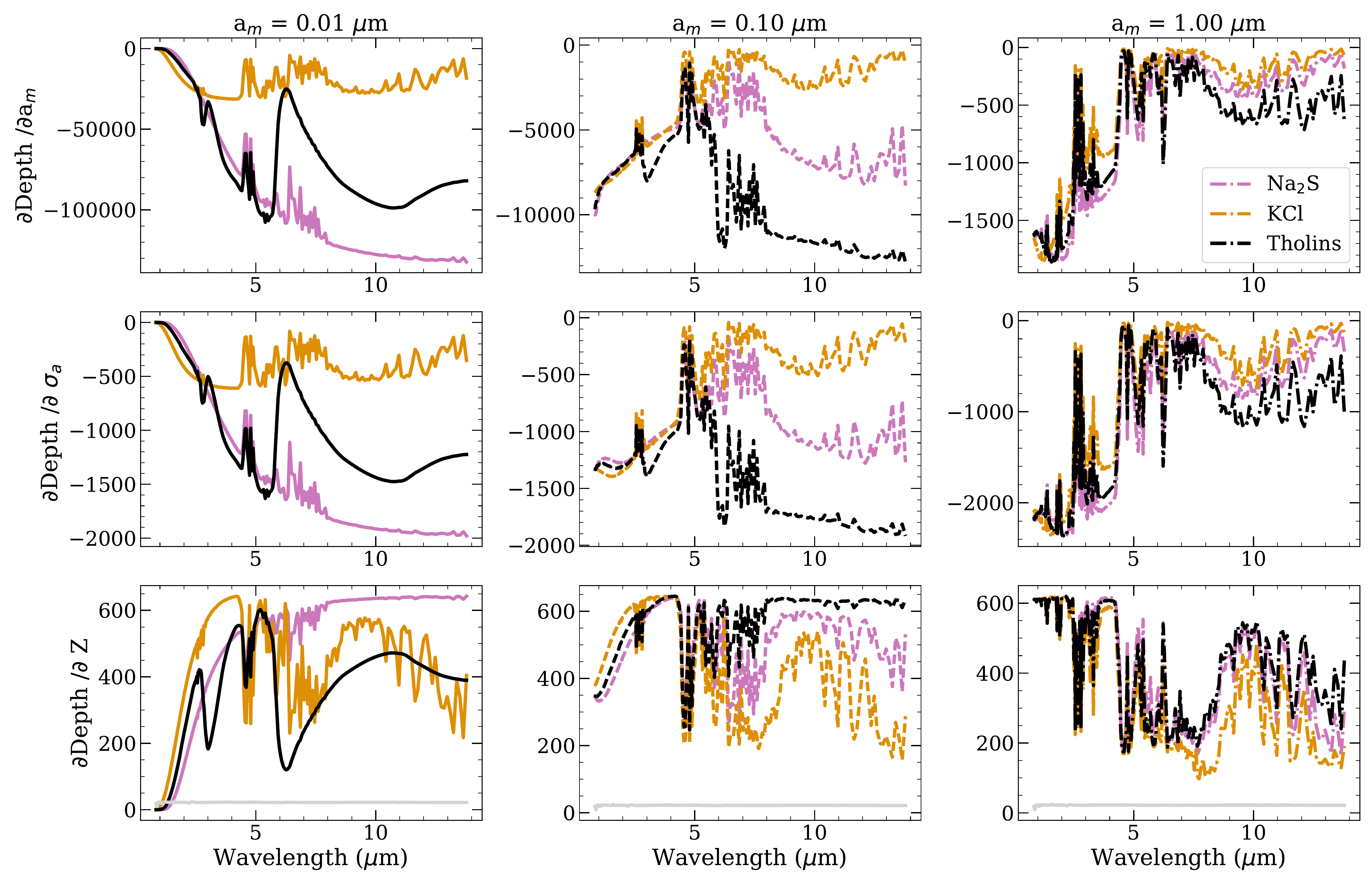}
    \caption{Transit depth Jacobians for our 700-K fiducial atmosphere when Na$_2$S (pink), KCl (orange), and Titan tholins (black) are included as slab aerosols. The top row is the partial derivative of the transit depth with modal particle size, the second row is the partial derivative with the size-dispersion, and the bottom row is the partial derivative with metallicity. In the metallicity panels we also include a light gray line showing the Jacobian for a clear atmosphere. Each column shows the Jacobians calculated for a different modal particle size as labeled above. We set $F$ for each aerosol species such that $F\times$the solar abundance of the limiting atomic species equal to 3$\times$10$^{-6}$. This was arbitrarily chosen such that the aerosols would not behave as a simple gray opacity source. The top-pressure cut-off for the slab aerosol was set to 10$^{-8}$ bars, which is well above where there is no longer enough material to form many aerosol particles. When Jacobians are further from zero, it means that the transit depth for that wavelength is more steeply dependent on whichever parameter was used for the partial derivative.}
    \label{fig:700K_jacobian}
\end{figure}

Figure \ref{fig:700K_jacobian} shows Na$_2$S, KCl, and Titan tholins, some aerosol species that may be present in a 700-K atmosphere. Figure \ref{fig:1000K_jacobian} in the appendix shows Na$_2$S, NaCl, and Titan tholins, some species that may be present in a 1000-K atmosphere. Figure \ref{fig:1400K_jacobian} in the appendix shows Mg$_2$SiO$_4$, MgSiO$_3$, and Fe, some species that may be present in a 1400-K atmosphere. Figure \ref{fig:1800K_jacobian} in the appendix shows Fe, Al$_2$O$_3$, and TiO$_2$, some species that may be present in a 1800-K atmosphere. For the top and middle row of each Figure, the partial derivatives are negative, so the lower, more negative values, represent more information/sensitivity to particle size and the spread in the size distribution, respectively. For the bottom row, the partial derivatives are positive, so higher more positive values represent more information/sensitivity to the metallicity. In computing these Jacobians, we have used the four fiducial atmospheres of Table \ref{tab:fiducials_uniform}, log-normal size distributions with a dispersion of 2.5 and the modal sizes listed at the top of each column, assumed the slab type aerosol with a very high top pressure, and varied the value of $F$ depending on the solar abundance of the limiting atomic species for each aerosol species. In some Jacobians, the jagged patterns of gaseous absorption are still clearly visible, while in others the Jacobians look smoother indicating that aerosol opacity is totally dominating. Where gaseous absorption patterns are imprinted in the Jacobians, one sees local minima of sensitivity to aerosol properties and to metallicity. There is a lot wrapped into these figures, so we break the implications into four main themes:

\begin{enumerate}
    \item Some wavelengths tend to be dominated by aerosol opacity while others tend to be dominated by gaseous opacity. For 0.1-$\mu$m and 1.0-$\mu$m modal particle sizes, the shortest wavelengths (under 2 $\mu$m) and the longer wavelengths (over 8 $\mu$m) tend to be most sensitive to particle size and the breadth of the particle-size distribution for a variety of species (though not all). In the 4- to 8-$\mu$m range, the Jacobians for modal particle size and size-dispersion tend to show less sensitivity, indicating that these wavelengths are still dominated by gaseous opacity, even when aerosols are present. In particular the profile of the CO feature at 4.5-$\mu$m is always apparent as a local minimum in sensitivity to modal particle size and breadth of the particle-size distribution. We attribute this trend to two things. First, the gaseous absorption is strong at these wavelengths. We saw in the previous section that the transit spectrum at these wavelengths is formed up at low pressures of 10$^{-6}$ bars for clear atmospheres (see Figure \ref{fig:clear_contributions}). Second, many aerosol species have a drop in their extinction efficiences around this wavelength range, at least for small particles (see Figure \ref{fig:efficiencies}). For a 0.01-$\mu$m modal particle size, the shortest wavelengths are less sensitive to modal particle size and the dispersion of the particle-size distribution compared to the 0.1-$\mu$m and 1.0-$\mu$m cases.  
    \item An important consequence of item (1) is that the gaseous CO and water absorption from $\sim$4 to 8 $\mu$m may remain detectable even if the optical and NIR portion of a transit spectrum is flattened or smoothed by aerosol extinction. Targets for which no gaseous absorption features are currently detected may prove more forthcoming with JWST.
    \item The newly available continuous long wavelength coverage of JWST incorporates wavelengths where different aerosol species look most distinct. The Jacobians for different species are more similar to each other at optical wavelengths than the infrared wavelengths for modal particle sizes of 0.1 $\mu$m and 1 $\mu$m (look particularly at Figures \ref{fig:1000K_jacobian}-\ref{fig:1800K_jacobian} rather than Figure \ref{fig:700K_jacobian}). The exceptions are 0.1-$\mu$m Iron which has enhanced sensitivity in the optical, and MgSiO$_3$ and Mg$_2$SiO$_4$ which tend to have very similar Jacobians across all wavelengths not just in the optical. They only have significant differences in their Jacobians when the resonance feature around 10 $\mu$m is visible (see column for modal particle size 0.01 $\mu$m in Figure \ref{fig:1400K_jacobian}). 
    \item These Jacobian calculations reinforce the idea that there is a tension between learning about aerosol properties and learning about gaseous atmospheric properties, but also indicate a possible sweet spot where we can have our cake and eat it too. For the 700-K, 1400-K, and 1800-K fiducial atmospheres, transit spectra with aerosols included are much more sensitive to metallicity than transit spectra for clear atmospheres (shown as a gray line in the panels in the bottom row of each Figure). For the 1.0-$\mu$m case, the Jacobians for modal particle size are small and the gaseous absorption peaks are almost all apparent. For the 0.01-$\mu$m case, the Jacobians for modal particle size are largest by an order of magntiude and, at many wavelengths, no gaseous absorptions effects are discernable at all. Our scaled $F$ was chosen such that a size distribution with mode 0.75 would show some gas absorption, but also non-gray aerosol effects. This means that there is plenty of aerosol material for the 0.01-$\mu$m modal size to overpower gaseous opacities, and that the 1-$\mu$m modal size tends to just fill in the windows between absorption features in a slightly non-gray manner. 
\end{enumerate}


These example transit spectra and Jacobian calculations lend further credence the community's hope that the extensive wavelength coverage and high precision of JWST and ARIEL will enable unambiguous measurements of both the gas phase and the properties of any aerosols present in exoplanet atmospheres. Different wavelengths of light probe different pressure layers in the atmosphere as they encounter a different combination of aerosol opacity and gaseous opacity. Smooth spectra in the sparse optical-NIR range currently available do not obviate the possibility of information-filled NIR-midIR JWST transit spectra. In the remainder of the paper, we will see how these results bear out in MCMC retrievals with slab hazes and clouds and phase equilibrium clouds, and how the picture changes as we alter the spatial positions of particles in a wider variety of ways than shown in this section.

\section{Slab-type Hazes and Clouds}\label{sec:slab}

In this section we will demonstrate the range of behavior that can be produced by the slab aerosol, and then we will test how well aerosol species and properties can be retrieved (along with other atmospheric parameters). The slab aerosol is a useful model to consider because it does not make strong assumptions about where in the atmosphere photochemical hazes and condensate clouds will form. Its only assumptions are that the same aerosol size distribution is present at all pressures which contain aerosol, and that the base of the haze or cloud will extend down to pressures too high to be probed by transit spectroscopy. 

\begin{figure}
    \includegraphics[width=0.499\textwidth]{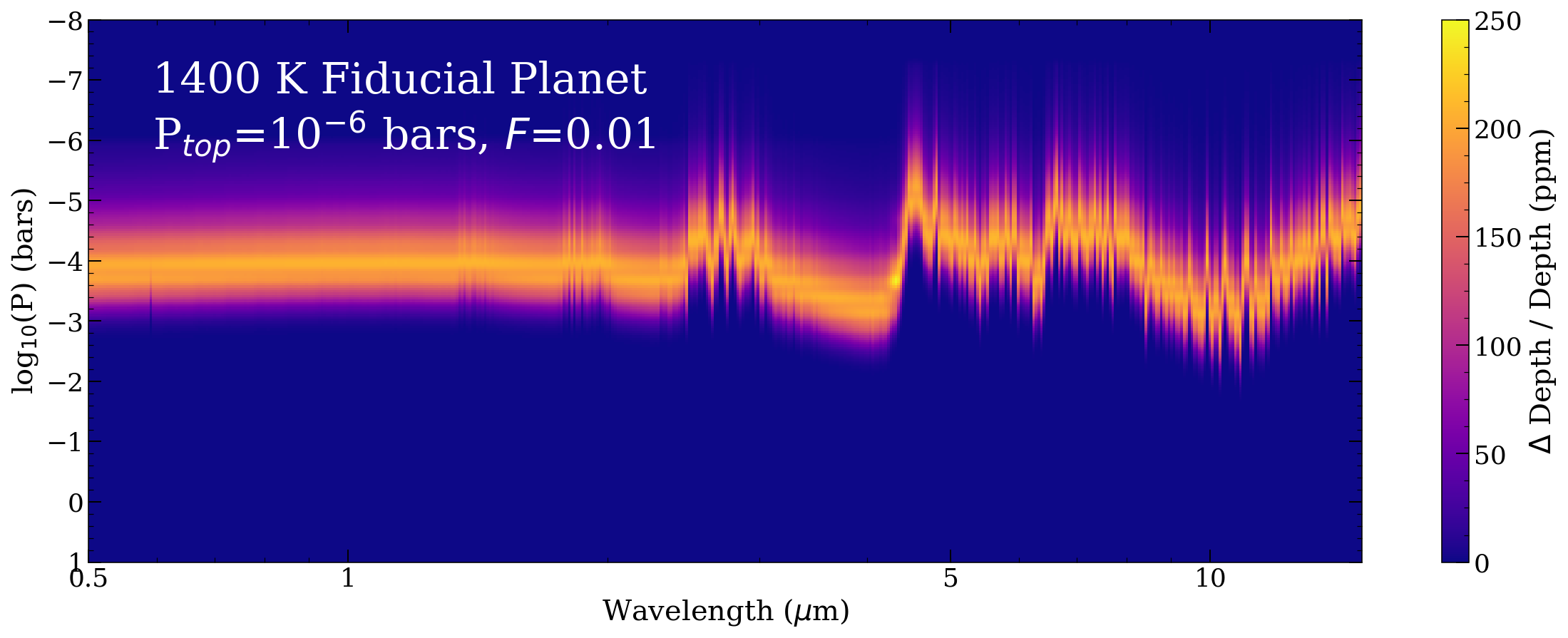}
    \includegraphics[width=0.499\textwidth]{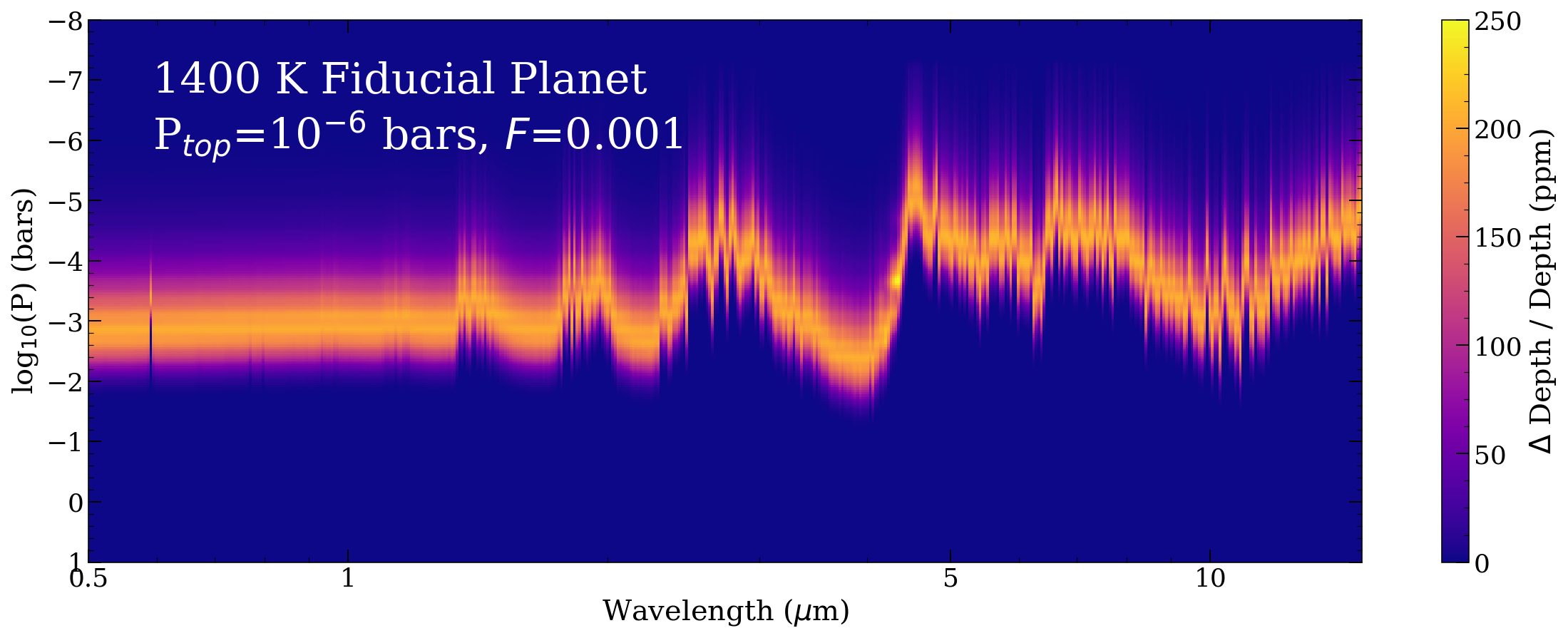}
    \includegraphics[width=0.499\textwidth]{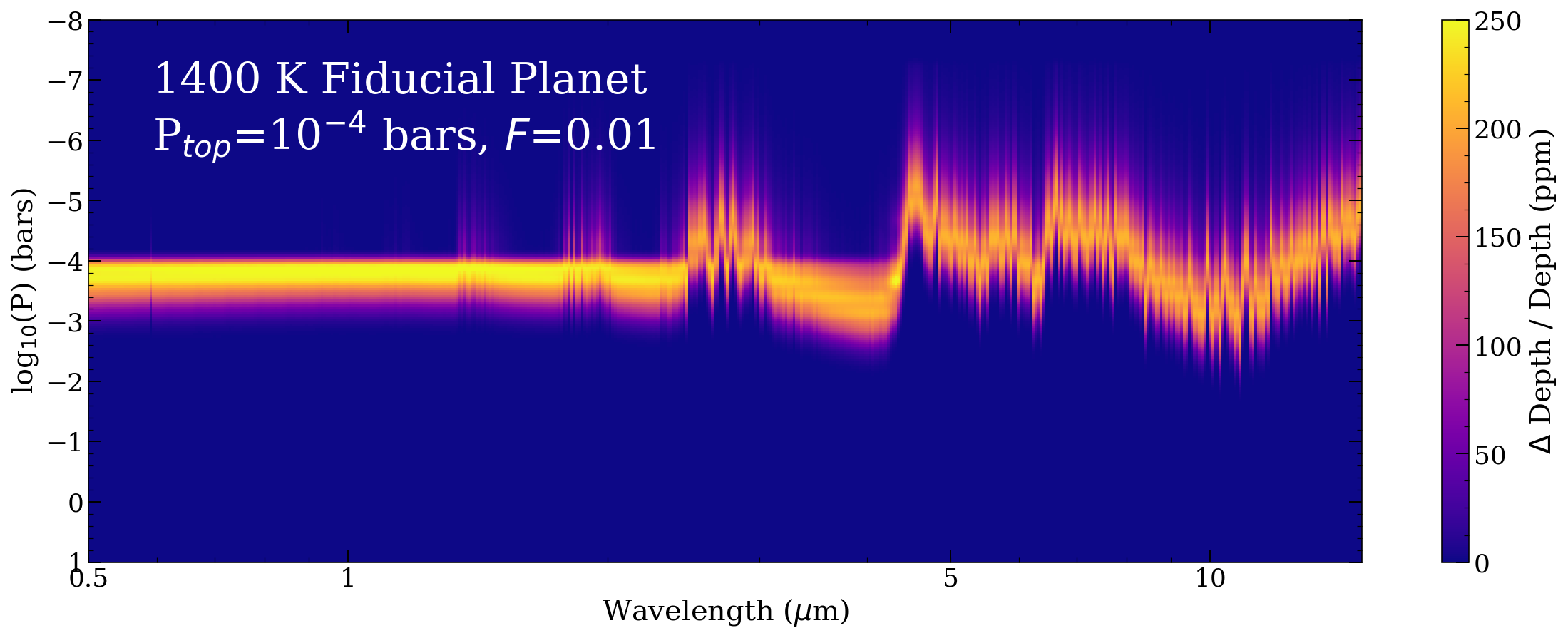}
    \includegraphics[width=0.499\textwidth]{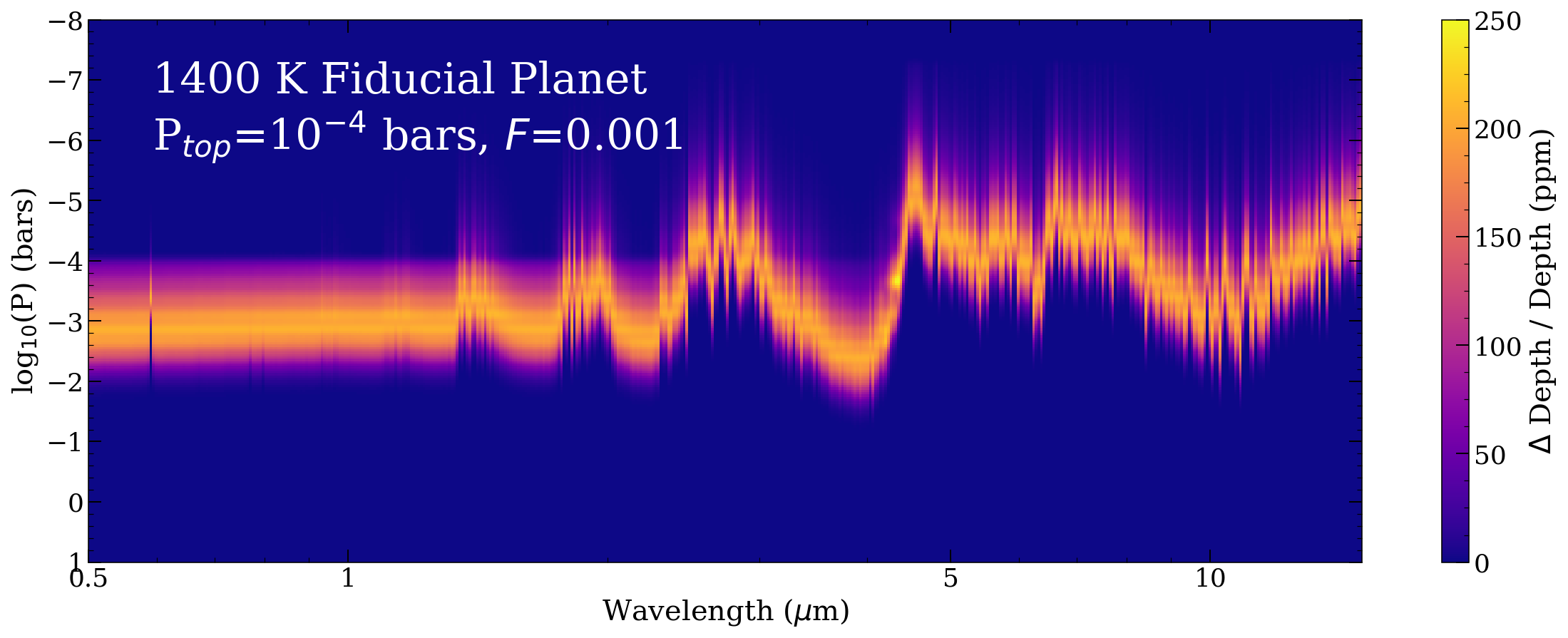}
    \includegraphics[width=0.499\textwidth]{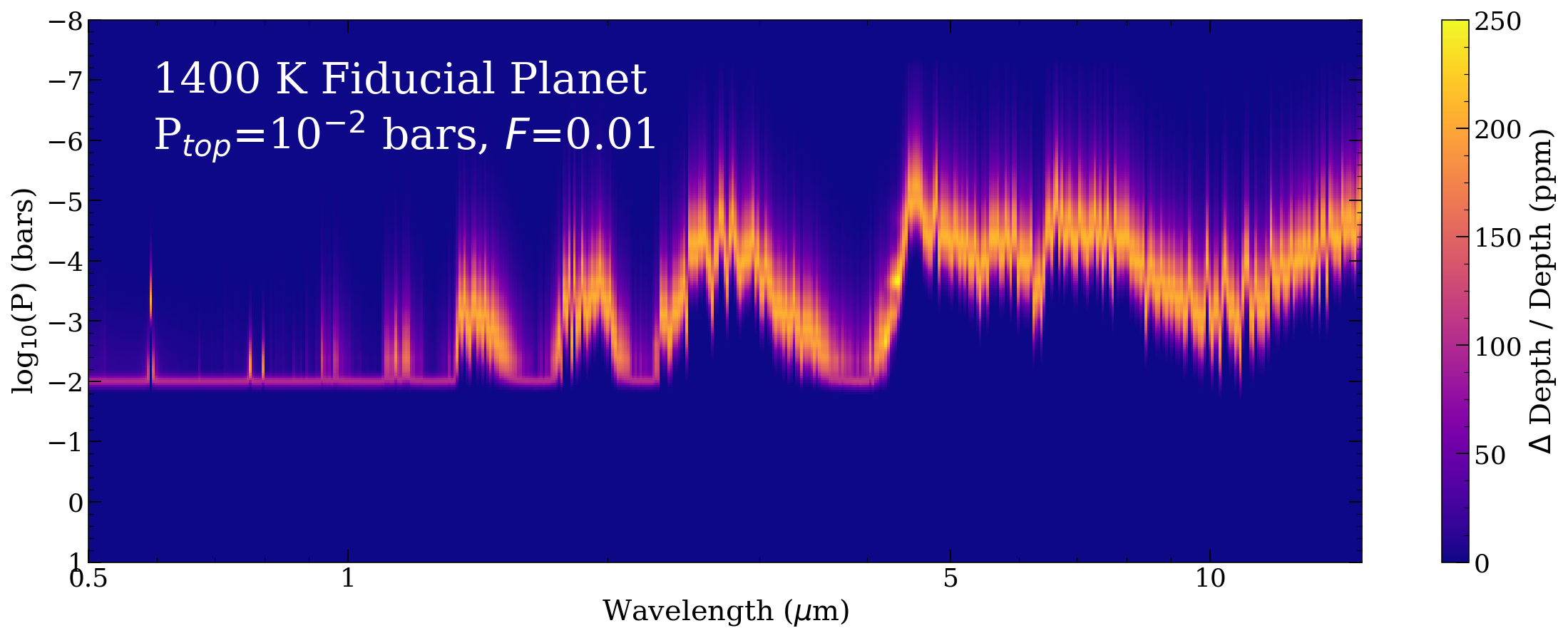}
    \includegraphics[width=0.499\textwidth]{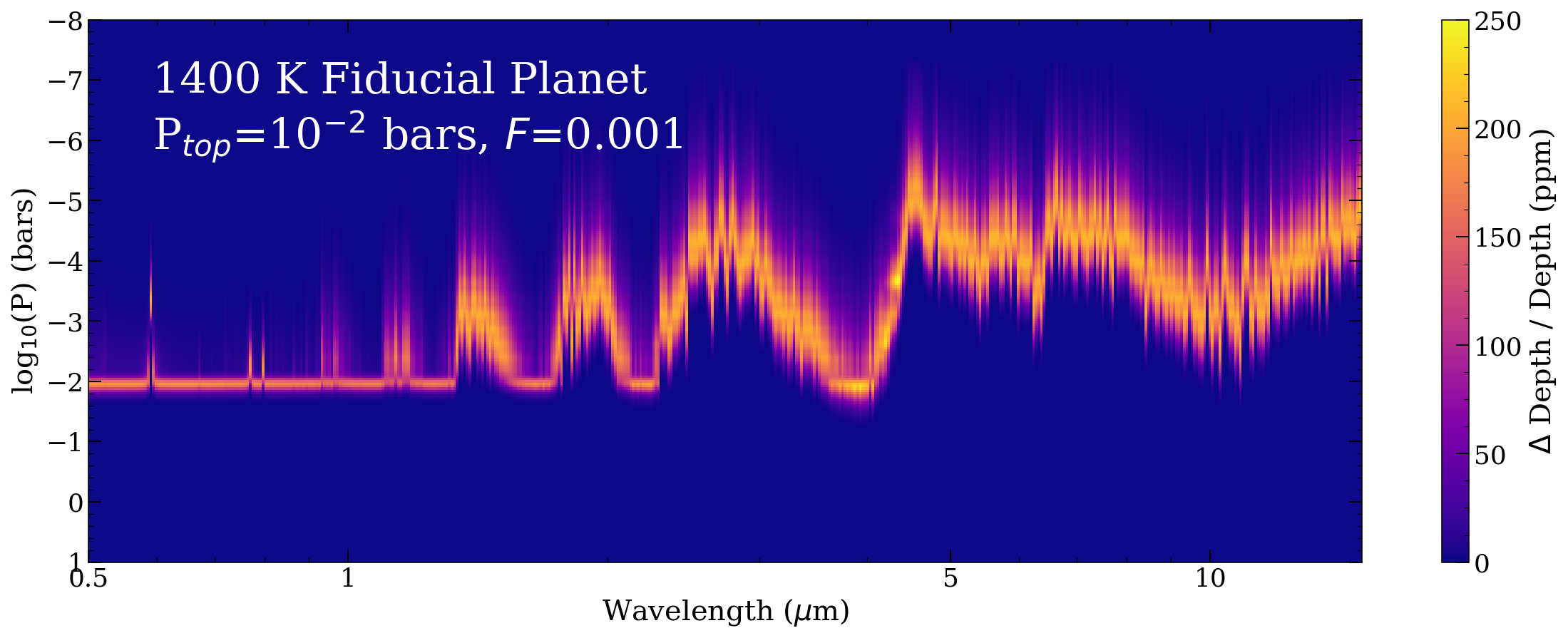}
    \includegraphics[width=0.499\textwidth]{smoothed_1400K_clear_cont.png}
    \includegraphics[width=0.45\textwidth]{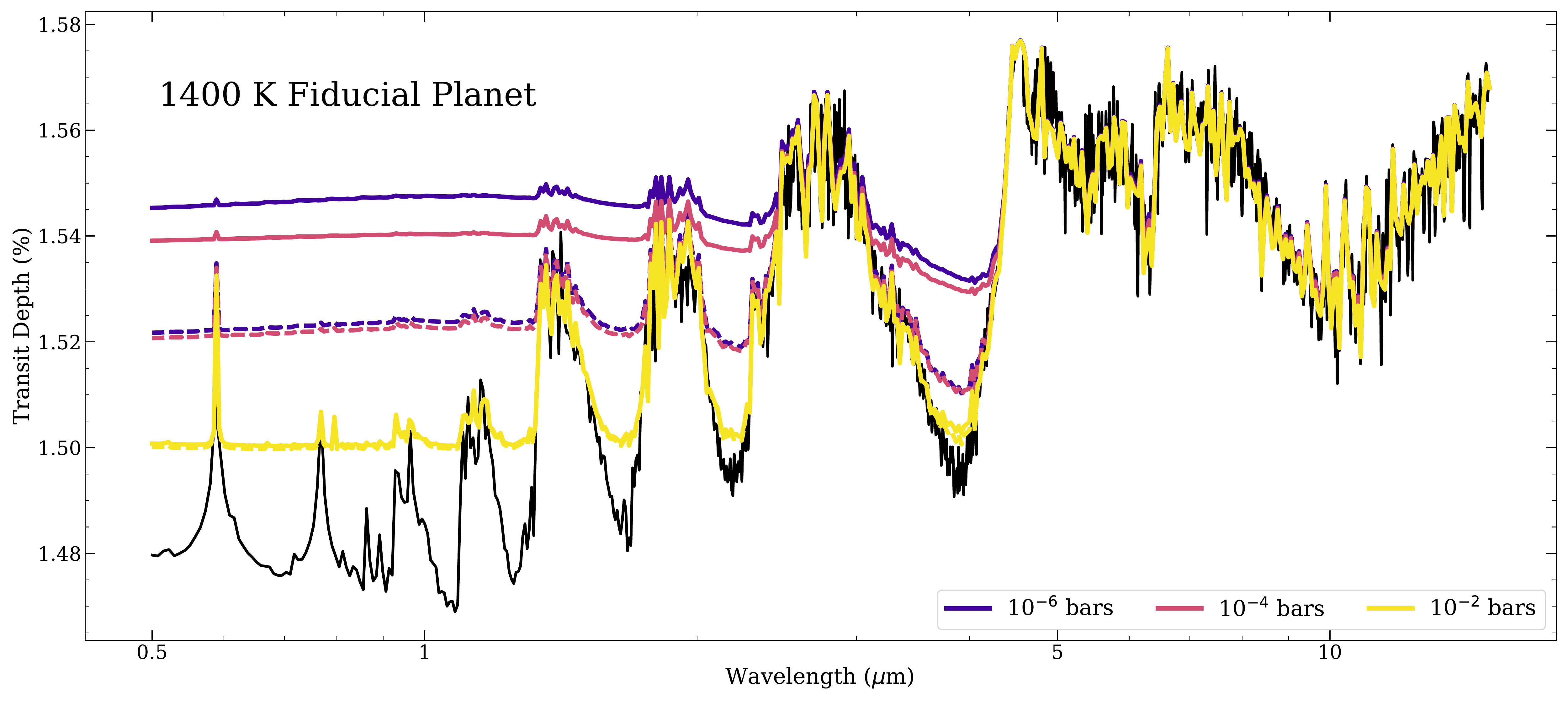}
    \caption{The top three panels show contribution functions for our 1400-K fiducial atmosphere with a tholin slab aerosol with different top-pressure cut-offs compared to the clear case in the fourth panel from the top. Lighter yellow and orange colors indicate which pressures are shaping the transit spectrum at a given wavelength. Darker purple and blue regions indicate pressures that do not shape transit spectra at a given wavelength. The lower pressures (higher altitudes) are purple because they are too low in density/opacity to block much light. The higher pressures (lower altitudes, deeper in the atmosphere) are purple because the atmosphere becomes optically thick to that wavelength at lower pressures. The bottom panel shows the corresponding transit spectra for these same cases. This Figure illustrates how a steep top-pressure cut-off results in gray behavior while a gradual drop in the amount of aerosol results in distinctive spectral signatures. $F$=0.01, Z=1,am=0.1,sigma=3.0 on left, $F$=0.001 on the right}
    \label{fig:hazeslab_cont}
\end{figure}

There are three regimes of transit spectra that can arise when a slab haze or cloud is included in an atmosphere. First, one can get transit spectra with a completely gradual cloud top exhibiting variation in aerosol extinction with wavelength (like we saw in the previous section). Second, one can have an aerosol that manifests as a purely gray opacity added onto the gaseous absorption. Finally, one can get transit spectra which exhibit the non-gray behavior described in regime 1 at some wavelengths (where the aerosols have a smaller total cross section), but which are also flattened by a top-pressure cut-off at other wavelengths (where the aerosols have a larger total cross section).

In the first regime, the top-pressure cut-off had no effect on the transit spectrum because it is at such a high pressure that the cloud or haze is optically thin to all wavelengths of light by the time it is reached. This enables the full variation of the aerosol's wavelength-dependent opacity to be imprinted on the transit spectrum. Alternatively to setting a very high P$_{top}$, one could achieve this behavior by setting $F$ and $Z$ to be very low, or by specifying a larger modal particle size such that less total particles form. If the aerosol opacity is stronger than the gaseous absorption across a range of wavelengths, then this will produce a transit spectrum with lots of information about species and particle-size distributions. An atmosphere can fall into the second regime if there is a steep top-pressure cut-off at an altitude where the haze or aerosol is optically thick across all wavelengths. It might also result if the aerosol is so optically thin that it just barely fills in troughs in between absorption peaks. If the particle-size distribution is simply very broad or has a large modal particle size then one will also see only gray effects on the transit spectrum, because large particles do not have as much variation in extinction efficiency with wavelength (see orange lines in Figure \ref{fig:efficiencies}). A gray aerosol can sometimes put a strong constraint on P$_{top}$, but will likely place only an upper limit on $F$. There can be strong degeneracies between modal particle size, $F$, and $Z$ depending on which species is included and what P$_{top}$ is. Transit spectra in the third regime can enable one to constrain P$_{top}$, particle-size distributions, and perhaps $F$ and $Z$, though this will depend on whether the aerosol is obscuring the gaseous signatures of metallicity or not. This third regime is thus the one most suitable for fitting with our full slab-aerosol model. The first regime would be better to leave out P$_{top}$, and the second regime would be better fit by only P$_{top}$ and no additional aerosol-related parameters.

The particle-size distribution, $F$, P$_{top}$, $Z$, and the species of aerosol dictate which of these three regimes manifests in the transit spectrum. The total available material to incorporate into aerosols depends on which species of aerosol is forming and the metallicity of the atmosphere. This means that the same combination of $F$, P$_{top}$ and particle size can fall into a different regime depending on which species and metallicity have been specified. Examples of this interplay are shown in Figure \ref{fig:hazeslab_cont}. Each panel in Figure \ref{fig:hazeslab_cont} shows the transit contribution function for the 1000-K fiducial planet with a tholin haze added with modal particle size 0.1 $\mu$m and $\sigma _a$=3. The first three rows have a different top pressure cut-off. The left column has $F$=0.01 and the right column has $F$=0.001. The bottom left panel shows the contribution function of the clear atmosphere, for comparison. The transit spectra corresponding to each of the contribution functions are shown in the bottom right. Dashed lines are for spectra with $F$=0.001, and solid lines are for spectra with $F$=0.01. 

The transit spectra corresponding to the top left panel, the top right panel, and the right panel in the second row fall into regime 1, in which there is a gradual top to the haze, allowing the full range of tholin spectral signatures to show up in the transit spectrum (see purple solid line, the purple dashed line, and the pink dashed line in the bottom right panel). Both panels in the third row, with top pressures of 10$^{-2}$ bars, contain a gray aerosol. These correspond to the transit spectra shown with yellow lines in the bottom right panel. The left panel in the second row falls into regime 3, and its transit spectrum is shown by the pink solid line in the bottom right panel. In this case, the wavelengths shorter than $\sim$4 $\mu$m are gray, evidence for the steep top-pressure cut-off at 10$^{-4}$ bars, but the wavelengths from 4-5 microns have a non-gray aerosol signature encoding information about the particle-size distribution. One can see how varying $F$ and P$_{top}$ shifts the transit spectrum about between regimes 1-3, even for a fixed metallicity, species of aerosol, and particle-size distribution. 

\begin{figure}
    \centering
    \includegraphics[width=\textwidth]{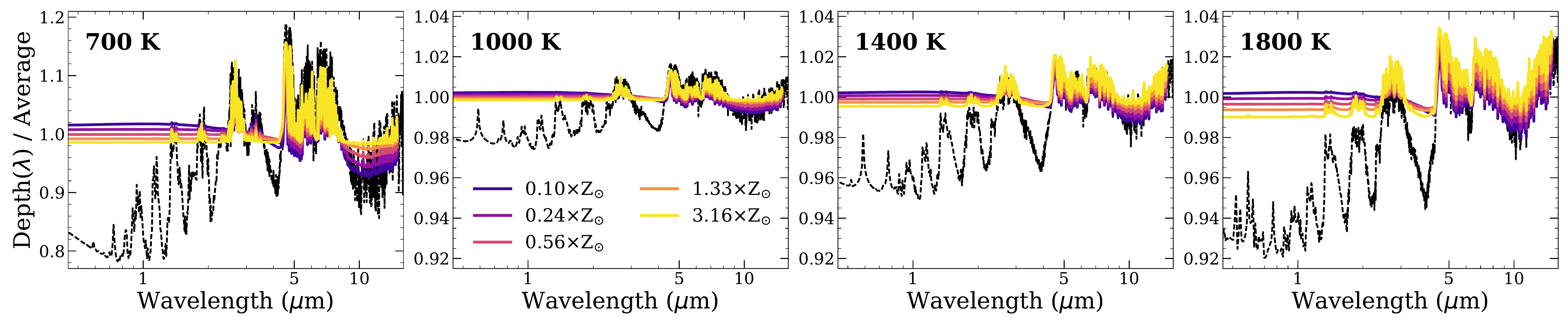}
    \caption{We show the transit depth for each wavelength divided by the average transit depth over all wavelengths as we vary metallicity, now including a Titan tholin slab aerosol like the one in Figure \ref{fig:hazeslab_cont}. Each panel shows a different one of our four fiducial atmospheres. The light gray line shows the same type of plot for a Z=1$\times$Z$_{\odot}$ clear atmosphere.}
    \label{fig:tholin_slab_z}
\end{figure}

What if we change the metallicity? We saw the effect of changing metallicity on a clear atmosphere in \S \ref{sec:clear_fiducials} where we learned that, to measure $Z$ well, we must compare wavelengths where Rayleigh scattering, CH$_4$ or CO absorption dominate to wavelengths where H$_2$O absorption dominates. 

If metallicity changes in an atmosphere with a slab aerosol, the values of $F$ and P$_{top}$ which set the bounds between regime 1-3 will also change. Changing the metallicity of a cloudy or hazy atmosphere changes the total amount of available material to incorporate into aerosols, so, for a given $F$, increasing $Z$ means more aerosols are present. Figure \ref{fig:tholin_slab_z} shows the transit depth at each wavelength divided by the average transit depth as we change the matallicity in a hazy version of our four fiducial atmospheres. $F$ is kept at 0.005, P$_{top}$ is kept at 10$^{-4}$ bars, $a_m$ is kept at 0.1 $\mu$m and $\sigma _a$ is kept at 2.5, while $Z$ varies. Varying $Z$ shifts the wavelengths longward of 2 $\mu$m from clearly showing tholin-like behavior at low metallicities to simply containing a gray opacity along with the gaseous absorption at high metallicities. Meanwhile, the transit spectra in the optical wavelength range up through 2 $\mu$m stay nearly constant, since the aerosols are always optically thick at the top-pressure cut off, regardless of $Z$. When $Z$=0.1$\times$ $Z_{\odot}$, the atmosphere falls into regime 3 with gray absorption at wavelengths blueward of 2 $\mu$m, but non-gray signatures red-ward of 2 $\mu$m. When $Z$=3.16$\times$ $Z_{\odot}$, there is only a gray aerosol opacity like regime 2. 

Comparing Figure \ref{fig:tholin_slab_z} to the equivalent figure for clear atmospheres (Figure \ref{fig:clear_1D_z_pstudy}), one can see that there is a larger in shape for the hazy transit spectra than for the clear transit spectra. But, since $F$ and $Z$ can be highly degenerate in forming the aerosol opacity, this sensitivity to metallicity will not always translate into a precise measurement. If we want to get a handle on the atmospheric metallicity, we need the cloud top to not obscure wavelengths which break the degeneracy between metallicity and reference pressure (see Figure \ref{fig:clear_1D_z_pstudy}), or to thoroughly understand how metallicity shapes the amount of aerosol present (see Figure \ref{fig:tholin_slab_z}). An example of this is the 700-K planet which shows the metallicity dependence of the 3.3-$\mu$m CH$_4$ feature even when aerosols are present at a pressure of 10$^{-4}$ bar (see left-most panel of Figure \ref{fig:tholin_slab_z}). Note that these examples looked specifically at Titan tholin slab hazes, but the qualitative trade-offs between P$_{top}$, $F$, $Z$, and particle size will occur for other species of clouds or hazes too. The quantitative details will be different since each species has a different limiting constituent with a different solar abundance and each species has a different density.

When we simulate data with slab-type aerosols and attempt to do retrievals, we will find results consistent with these model sensitivity studies. If nature presents us with the unfortunate reality of an exoplanet atmosphere in regime 2 with gray aersosols obscuring almost all the gaseous absorption, we could end up with transit spectra that contain very little information about what aerosols are present and very little information about the gaseous absorption. On the other hand, we could have something in regime 1 or 3. That is, a thinner cloud deck or haze layer situated such that it gradually tapers before its top-pressure cut-off, imprinting lots of non-gray behavior while still allowing some gaseous absorption to show through. To really warrant the full slab aerosol parameterization in our MCMC fits, we need an atmosphere to fall into regime 3, where some wavelengths are cut off by P$_{top}$ while others imprint some non-gray aerosol spectral signatures. The likelihood of one scenario over another is beyond the scope of our study, since it will depend on the details of vertical mixing, photochemistry and microphysics in an atmosphere. Such processes have long been explored on earth and other solar system bodies, and have begun to be applied to the study of exoplanet aerosols (\citealt{Ackerman2001}; \citealt{Lee2016}; \citealt{Gao2018a}; \citealt{Gao2018b}; \citealt{Ohno2018}; \citealt{Powell2018}; \citealt{Kawashima2019a}; \citealt{Helling2019a}; \citealt{Powell2019}; \citealt{Gao2020}). Future observations with JWST and ARIEL will hopefully reveal whether existing microphysical models are capturing the true behavior in exoplanet atmospheres.  

\subsection{Slab MCMC Experiment Results}\label{sec:aerosol_mcmc_results}

In order to assess how well JWST-like transit spectra will be able to distinguish which aerosol species are present in exoplanets and how well parameters can be retrieved from hazy transit spectra, we perform a series of experiments. For each of the four fiducial temperatures we simulated JWST-like transit spectra for atmospheres with candidate aerosols that either condense in that temperature range, or are formed through photochemical processes. We then use MCMC to fit all of the transit spectra with the correct species of aerosol and all of the other candidate aerosols in the same grouping. By examining the quality of the fits and the retrieved parameters, we can determine whether the true aerosol would be preferred by a blind retrieval. The posteriors from the fits done with the correct aerosol species for a given simulated data-set indicate how well measurements can be made for hazy transit spectra if you have identified the correct aerosol species.

We use the groupings of aerosols listed in Table \ref{tab:fiducials_uniform}. The slabs have a top-pressure cut-off of 10$^{-4.5}$ bars, $F$=1.0, and an overall atmospheric metallicity of Z=3$\times$Z$_{\odot}$. For each of these spatial distributions, we generate one spectra with a smaller modal particle size (a$_m$=0.05 $\mu$m) and one with a larger modal particle size (a$_m$=1.0-$\mu$m), always assuming a log-normal size distribution with dispersion $\sigma _a$=2.5. The noise for each simulated spectrum came from scaling Pandexo's\footnote{https://exoctk.stsci.edu/pandexo/calculation/new} noise calculation for HD209458b based based on the different transit depths of our fiducial planets and assuming 10 transits are observed with each JWST instrument/mode. This is then added in quadrature with the noise floor suggested by \citealt{Greene2016}. The procedure yields very high precision data, which means an extremely high $SNR$ for the 700-K and 1800-K planets with low surface gravities/large scale heights. The $SNR$ for the 1000-K and 1400-K planets are smaller.

\begin{figure}
    \centering
    \includegraphics[width=\textwidth]{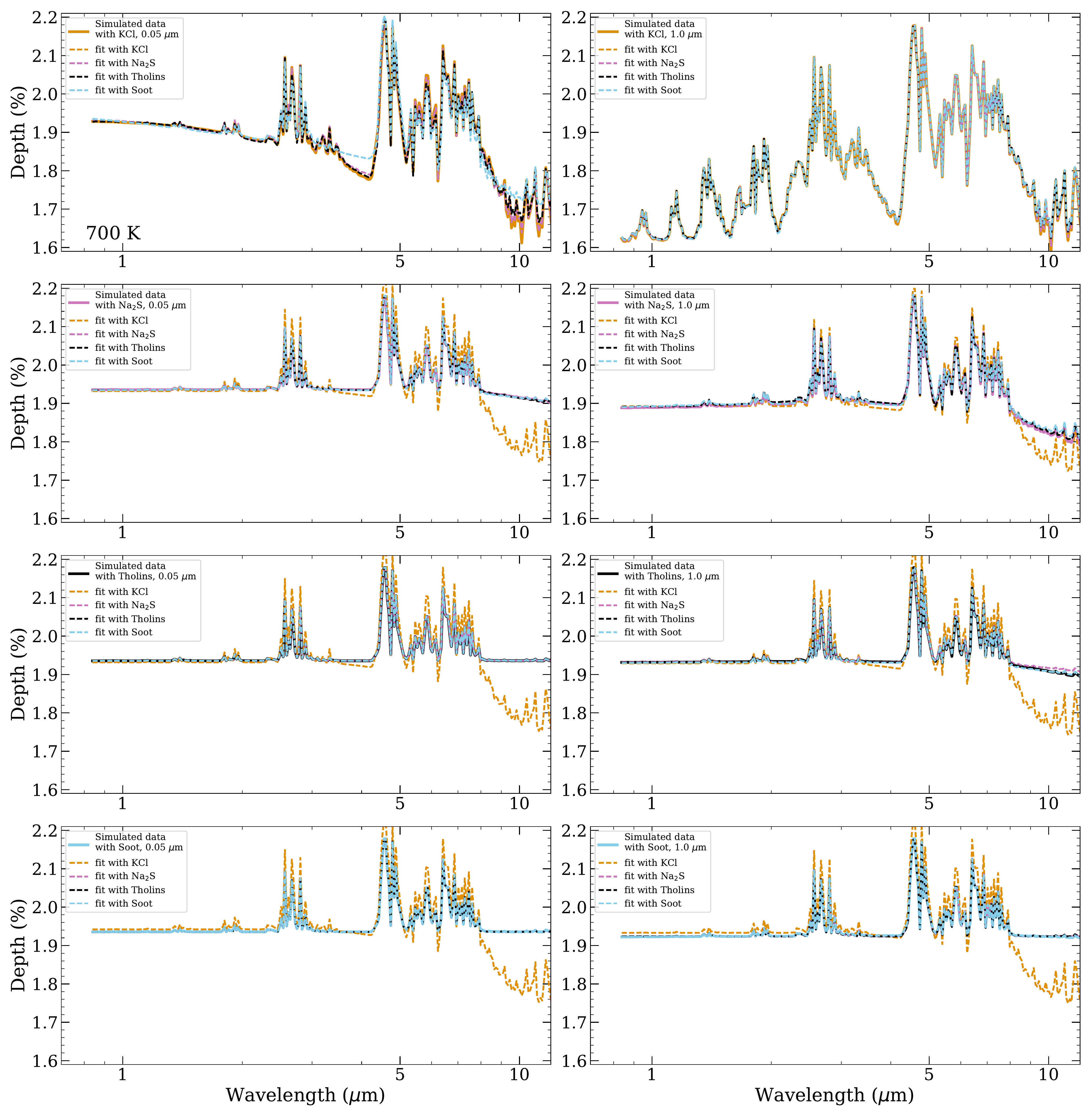}
    \caption{We show the transit spectra corresponding to median retrieved parameters in the MCMC experiments for the 700 K aerosol grouping: KCl, Na$_2$S, Tholins, and soot. Each row has a different true aerosol in the data, shown as the thick solid line with lighter shading indicating the error envelope. Dashed lines show the best fits with all four species. In the left column a modal particle size of 0.05 $\mu$m was used for the simulated data. In the right column a modal particle size of 1.0 $\mu$m was used for the simulated data. $F$ was kept at 1, P$_{top}$ at 10$^{-4.5}$ bars, and $\sigma _a$ at 2.5 when generating all the simulated data. }
    \label{fig:slab_700K_fits}
\end{figure}

\begin{figure}
    \centering
    \includegraphics[width=\textwidth]{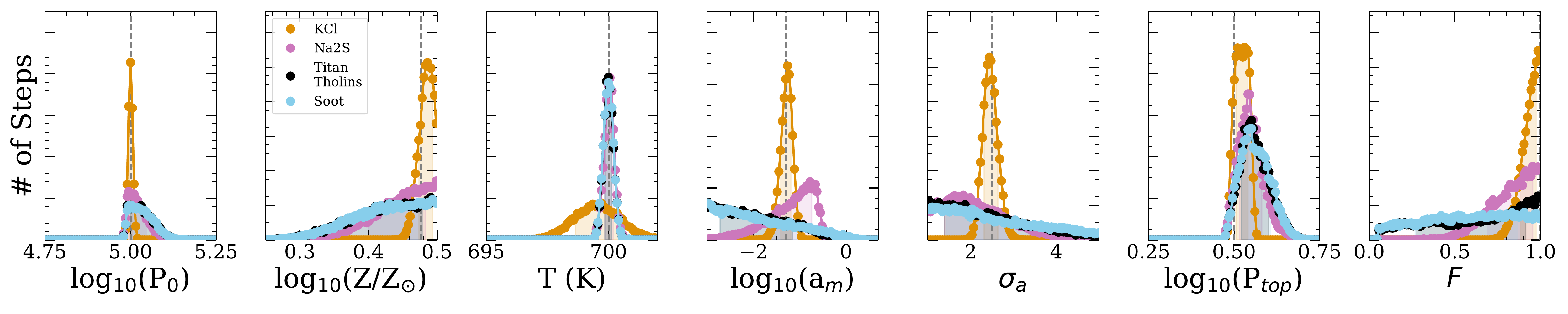}
    \includegraphics[width=\textwidth]{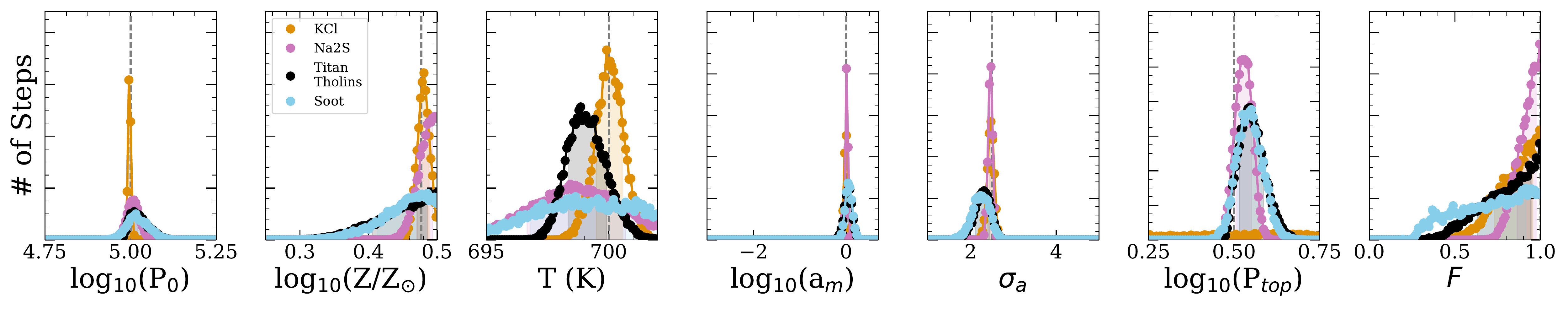}
    \caption{Posteriors for model parameters fit using the same aerosol species as the underlying simulated data. The top row shows results when the log-normal particle-size distribution had a mode of 0.5 $\mu$m, while the bottom row had a modal particle size of 1 $\mu$m. Each color corresponds to results for a different species. Vertical lines mark the true value of each parameter.}
    \label{fig:700_slab_hists}
\end{figure}

For the 700-K grouping, we test soot, tholins, KCl and Na$_2$S. The resulting best-fit transit spectra are shown in Figure \ref{fig:slab_700K_fits}, and histograms of the posteriors for fits with the correct aerosol species are shown in Figure \ref{fig:700_slab_hists}. With $F$=1.0 and metallicity Z=3.0$\times$Z$_{\odot}$, most of the slab aerosols show up as regime 2, essentially gray absorbers in the transit spectra. The exceptions are KCl which has so much less constituent material to form from and Na$_2$S which has a slight downward slope from 8 to 12 $\mu$m. It is thus unsurprising that it is only in the case of small KCl particles that a model with soot is not able to mimic the true aerosol species. If the size and spatial distributions of aerosols are such that we get gray absorption across all wavelengths, it will be impossible to unambiguously distinguish which species is present. As we saw in the previous section, a smaller value of $F$ or Z and a higher top pressure such that the aerosol layer is not optically thick all the way up to P$_{top}$ is necessary for species to look distinct. It is feasible that $F$ will be much less than 0.5. We chose such large values of $Z$ and $F$ in order to make sure KCl and Na$_2$S had non-negligible effects (and later NaCl, TiO$_2$ and Al$_2$O$_3$).

The posteriors and best fit transit spectra for all the other fiducial atmospheres are shown in Figures \ref{fig:1000_slab_hists} - \ref{fig:slab_1800K_fits} in the appendix. For the 1000-K grouping (Figure \ref{fig:1000_slab_hists} and Figure \ref{fig:slab_1000K_fits} in the appendix), we test soot, Tholins, NaCl, and Na$_2$S. Again, we see that the simulated data with soot and Tholins are nearly flat, while the Na$_2$S and NaCl which contain less common atomic constituents exhibit non-gray behavior. The soot nearly fits all the species within the error bars, but is certainly a weaker fit than the true species in all cases, aside from 1.0-$\mu$m NaCl and 0.05-$\mu$m Tholins. For the 1400-K grouping (Figure \ref{fig:1400_slab_hists} and Figure \ref{fig:slab_1400K_fits} in the appendix), we test soot, Tholins, iron, enstatite, and forsterite. All of the species are able to mimic each other very well when the modal particle size is 0.05 $\mu$m and they are essentially behaving as gray absorbers. However, when the particles are 1.0 $\mu$m the silicate species have a slight arch to them from 0.7 to 4 $\mu$m and have a hint of a bump at the 10-$\mu$m feature. Tholins are more opaque from 0.7 to 4 $\mu$m than in the 10-$\mu$m window. Since the data are very high SNR, these subtle changes are enough that the true species is generally a better fit than the mimicking species. For the 1800-K grouping (Figure \ref{fig:1800_slab_hists} and Figure \ref{fig:slab_1800K_fits} in the appendix), we test soot, Tholins, iron, Al$_2$O$_3$, and TiO$_2$. In this temperature range, each species is able to mimic all others, aside from when soot is used to fit TiO$_2$ with 0.05-$\mu$m modal particle size. Looking back at Figure \ref{fig:clear_contributions}, we can see that gaseous absorption in the 10-$\mu$m window extends up to 10$^{-4}$-10$^{-5}$ bars. This means that with our top pressure cut-off of 10$^{-4.5}$ bars none of them are really showing up in this wavelength range. That means that the aerosols only need to adapt to mimic wavelengths shorter than 4 $\mu$m. 

These MCMC experiments on the slab-aerosol model agree well with our expectations from the parameter sensitivity studies. If aerosols are present in such a way that the transit spectrum falls into regime 2, they can be adequately treated as a gray opacity source. We only get an upper bound on particle size and the breadth of the size-distribution, and it will be difficult to determine which species of aerosol is dominant. If aerosols fall into regime 1 or 3 then we can learn what species and size particles are present. Whether or not we can also gauge the overall atmospheric metallicity and temperature will depend on whether the aerosol is fully overshadowing the gaseous absorption at key wavelengths. Alternatively, if we can accurately couple microphysics to gas-phase chemistry within the retrieval framework, then the aerosols themselves can be highly sensitive to the metallicity. 

\section{Phase Equilibrium Clouds}\label{sec:equilibrium}

Now we will move on to explore the range of behavior that can result when we use a phase equilibrium cloud model. Recall that this model incorporates some assumptions based on the thermodynamic properties of each species to set how much material will be incorporated into the cloud and where the base of the cloud forms, so the only additional free parameters are the relative scale height between the aerosol particles and the gas (set by $\alpha$), and the parameters used to describe the particle-size distribution. This rather simplistic treatment of clouds is intended to give a taste of how physically-motivated assumptions based upon our understanding of cloud formation in the Solar System can help us learn more from transit spectra than we get from using empirical models alone. 

\begin{figure}
    \centering
    \includegraphics[width=0.5\textwidth]{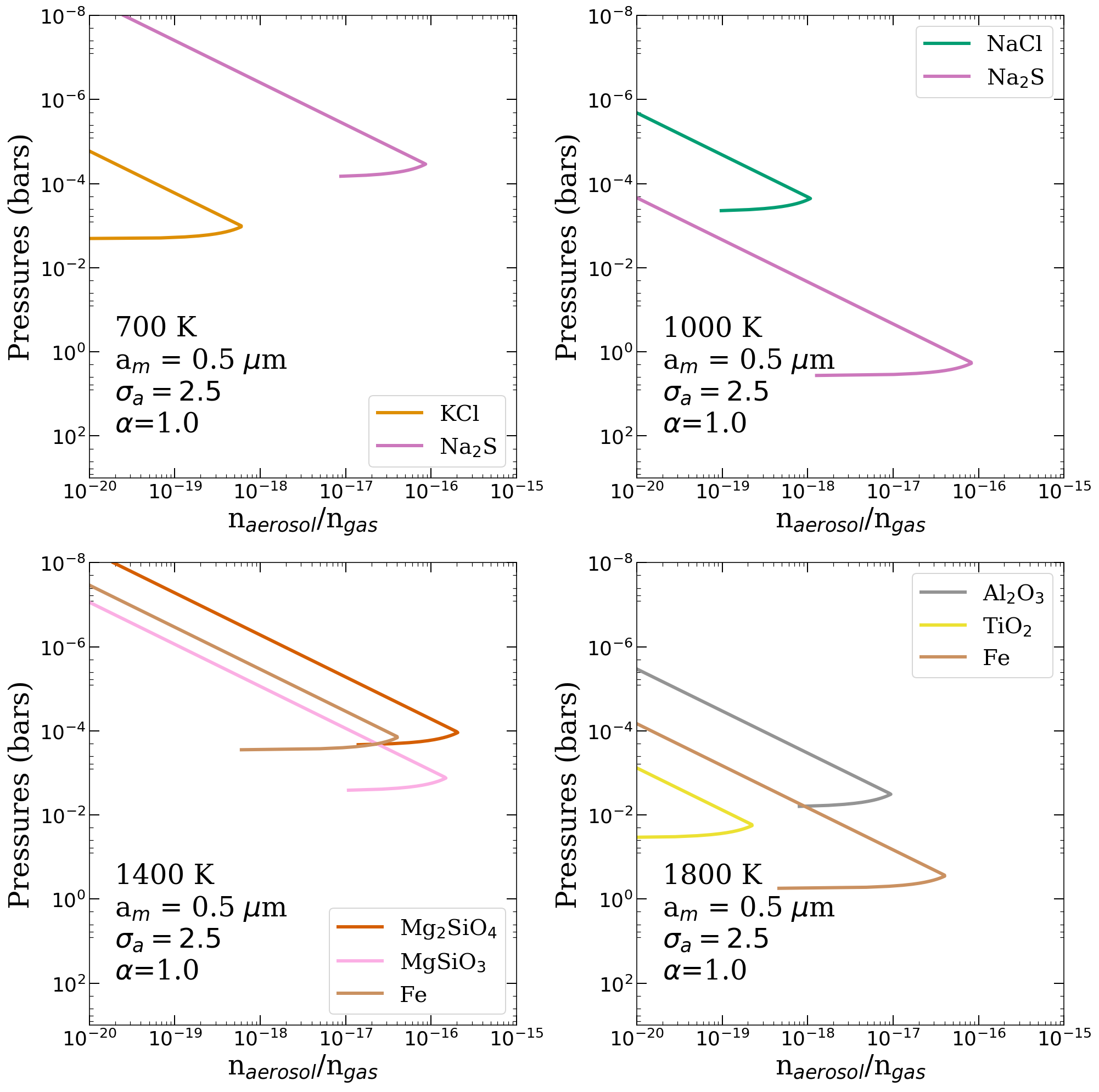}
    \caption{The profiles of the ratio of aerosol particle number density ($n_{aerosol}$) over gas number density ($n_{gas}$) for species that might condense at 700 K, 1000 K, 1400 K, and 1800 K. The base of the cloud is assumed to occur where the T-P profile intersects with the Claussius-Clapeyron line for each respective species. We have used a log-normal particle-size distribution with a modal size of 0.5 $\mu$m and a dispersion of 2.5, and we have assumed that the ratio of $n_{aerosol}$/$n_{gas}$ falls off with pressure as (P/P$_{base}$)$^{-1}$, and the atmosphere has a metallicity of Z=1$\times$Z$_{\odot}$.}
    \label{fig:Fiducial_Eq_Clouds}
\end{figure}

Figure \ref{fig:Fiducial_Eq_Clouds} shows the approximate ratio of cloud particle number density to gas number density versus pressure for the possible condensates in each of our four fiducial atmospheres (assuming a 0.5-$\mu$m modal particle size with a log-normal dispersion of 2.5 and a fall-off of $\alpha$=1.0). Different species will condense at different levels in the atmosphere for a given temperature, and some species form more particles than others. For example, Na$_2$S condenses much higher up in the 700-K atmosphere than in the 1000-K atmosphere, and Fe condenses higher up in the 1400-K atmosphere than in the 1800-K atmosphere. Recall that the number of aerosol particles is limited by the solar mixing ratio of the aerosol species' least abundant constituent atomic species. The solar mixing ratios of Cl, K, and Ti to H are all of order 10$^{-7}$, so the number densities of aerosol particles of KCl, NaCl, and TiO$_2$ are much smaller than those of the other possible condensate species, even assuming the same particle-size distribution.   

\begin{figure}
    \includegraphics[width=0.49\textwidth]{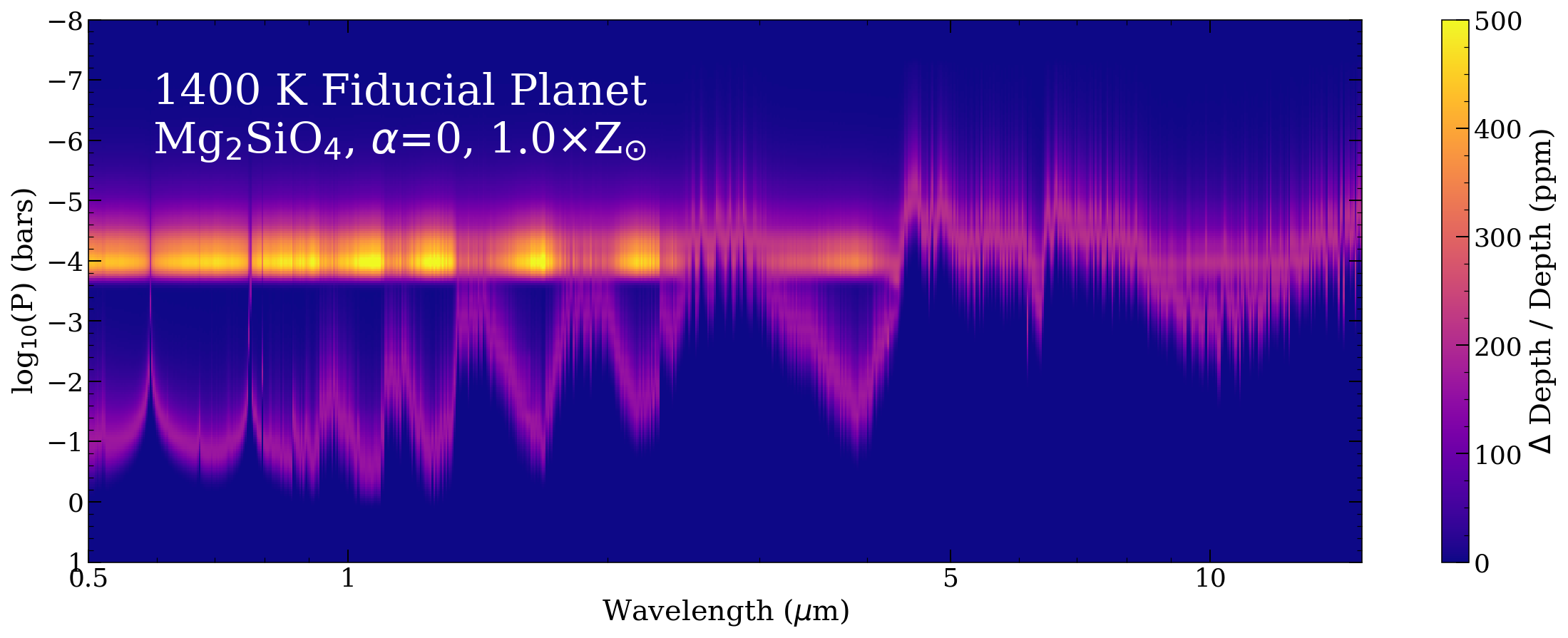}
    \includegraphics[width=0.49\textwidth]{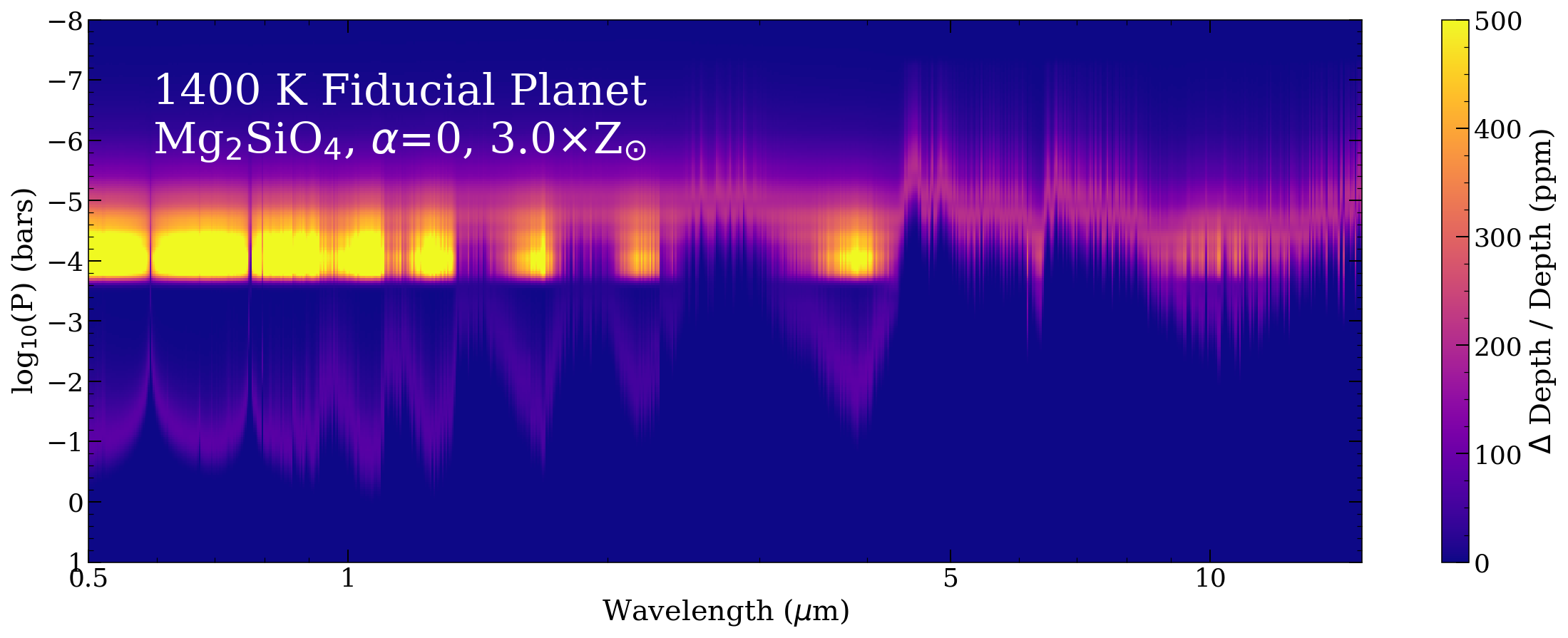}
    \includegraphics[width=0.49\textwidth]{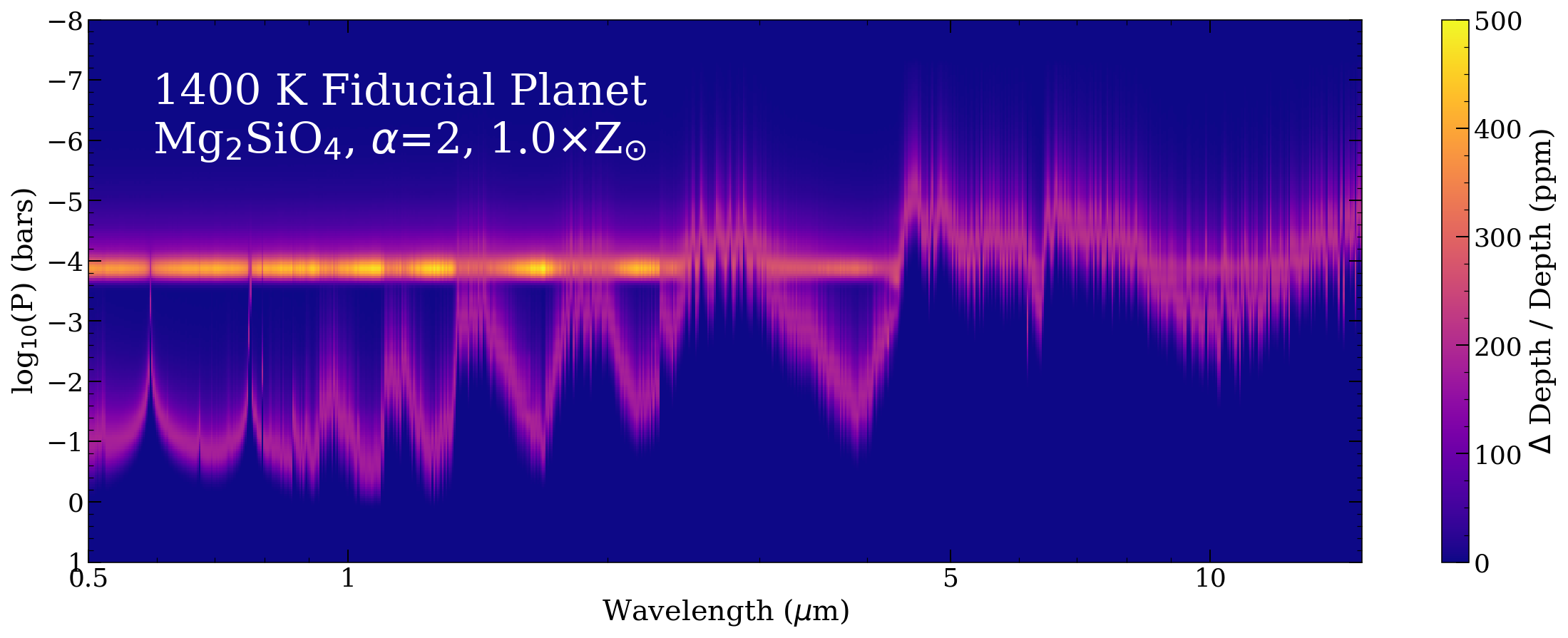}
    \includegraphics[width=0.49\textwidth]{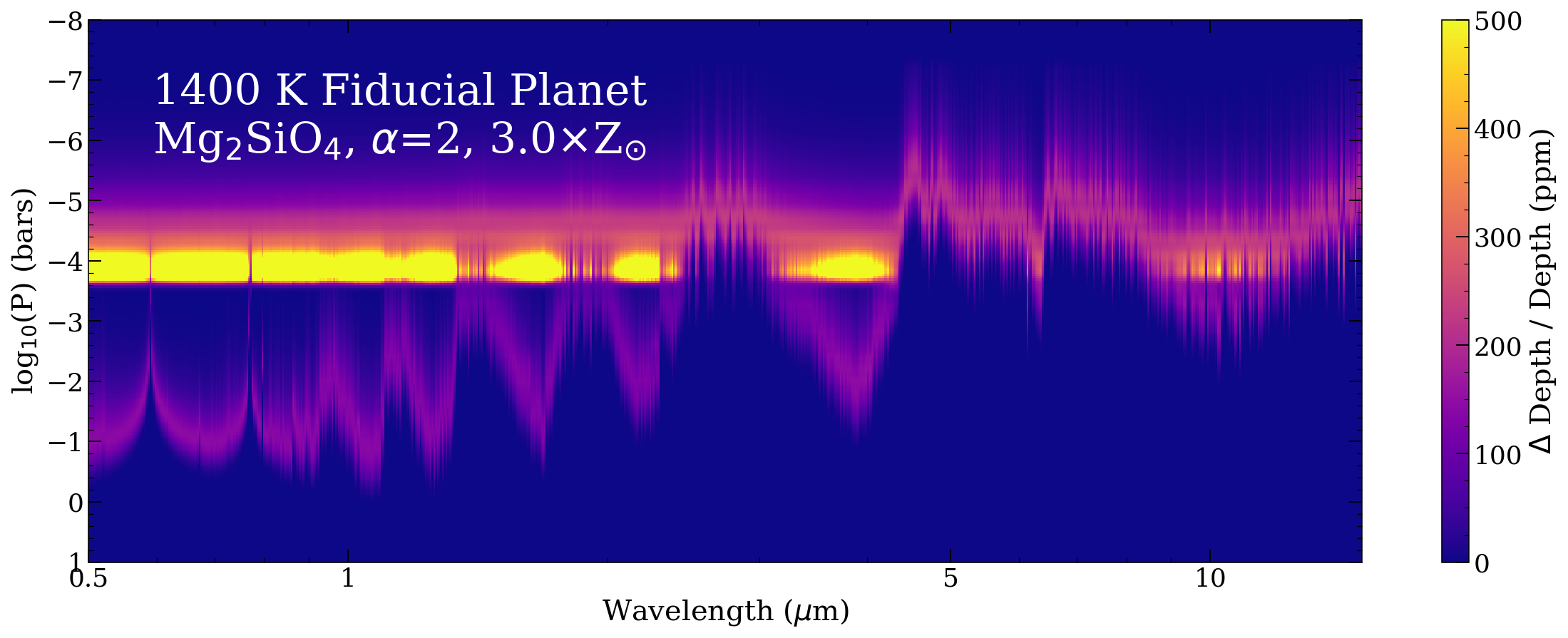}
    \includegraphics[width=0.49\textwidth]{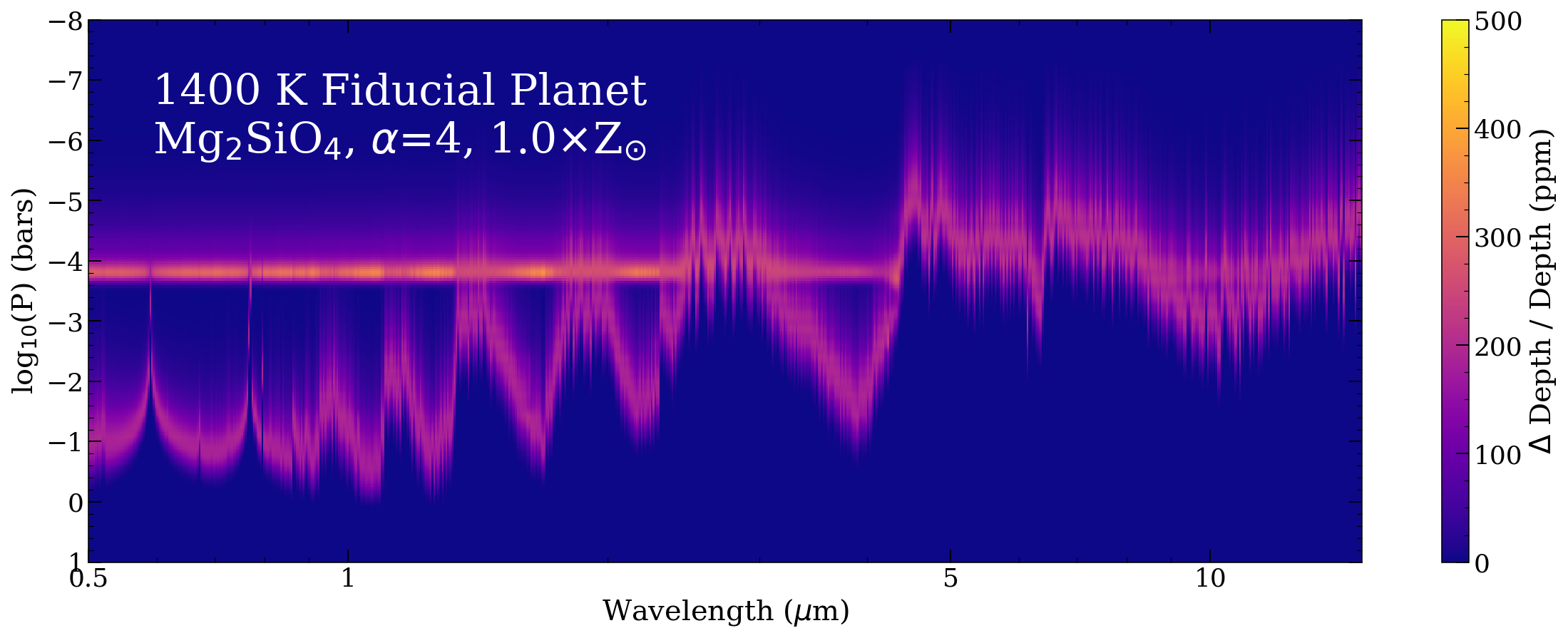}
    \includegraphics[width=0.49\textwidth]{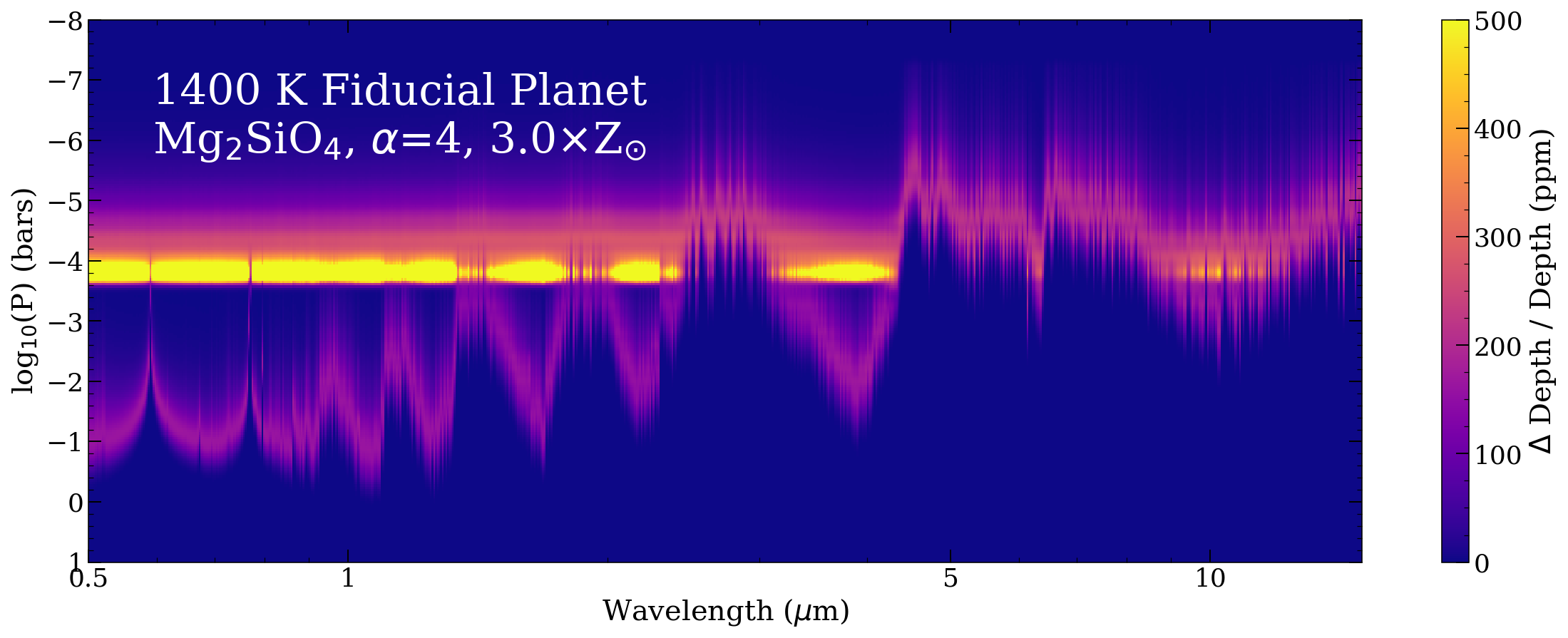}
    \includegraphics[width=0.49\textwidth]{smoothed_1400K_clear_cont.png}
    \includegraphics[width=0.45\textwidth]{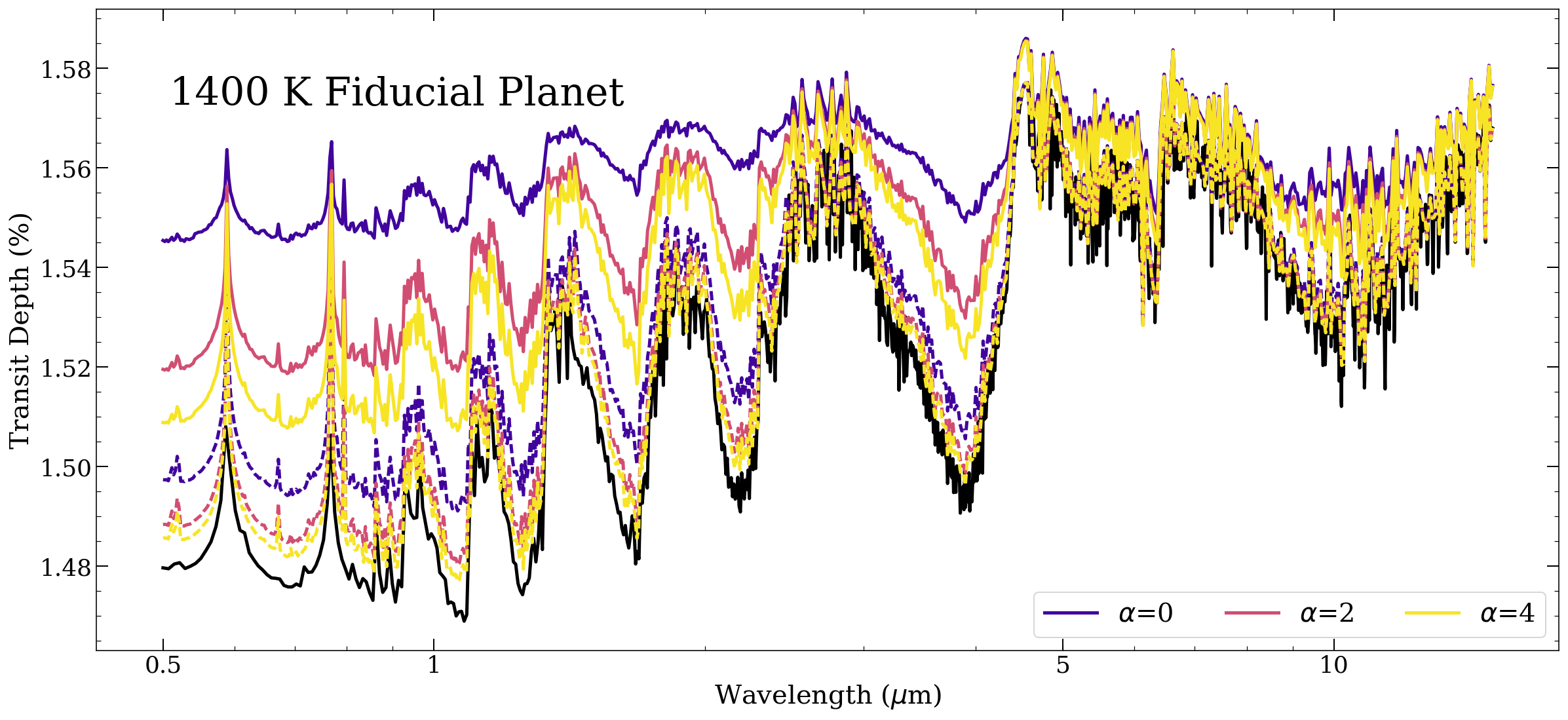}
    \caption{Transit Contribution functions for the 1400 K fiducial planet with a phase equilibrium forsterite cloud for varying metallicities and $\alpha$. In all cases the log-normal particle-size distribution has a modal particle size of 1 $\mu$m and a size dispersion of $\sigma _a$ = 2. In the left column the atmospheres have a metallicity of Z=1.0$\times$Z$_{\odot}$ and in the right column the atmosphere Z=3.0$\times$Z$_{\odot}$.  In the top row $\alpha$=0, in the second row $\alpha$=2, in the third row $\alpha$=4, and in the bottom left panel we show the transit contribution function for a clear atmosphere. In the bottom right row we show the transit spectra that correspond to the transit contribution functions. Solid lines indicate transit spectra for the higher metallicity atmospheres and dashed lines indicate the transit spectra for lower metallicity atmospheres.}
    \label{fig:forsterite_contributions}
\end{figure}

First, we will look at transit contribution functions for a Mg$_2$SiO$_4$ cloud in the 1400-K fiducial atmosphere to get a sense for what's going on as we change $\alpha$ (Figure \ref{fig:forsterite_contributions}). In the left column, the metallicity is $Z$=1$\times Z_{\odot}$ and in the right column the metallicity is $Z$=3$\times Z_{\odot}$. The first row has $\alpha$=0, then the second has $\alpha$=2, and the third has $\alpha$=4. The bottom left panel shows the contribution function for the clear 1400-K atmosphere, and the bottom right panel shows the transit spectra for all the other panels. Spectra shown by solid lines correspond to atmospheres with $Z$=3$\times Z_{\odot}$ and spectra shown by dashed lines correspond to $Z$=1$\times Z_{\odot}$. The particle-size distribution has a modal particle size of 1.0 $\mu$m and a dispersion of $\sigma _a$=2.5. The cloud base always forms a little deeper than 10$^{-4}$ bars. For lower values of alpha, there are cloud particles contributing opacity higher up in the atmosphere, while for higher values of alpha the cloud only contributes within a small range of pressures. As metallicity increases the cloud opacity makes up a larger portion of the contribution function than the gaseous opacity, but it never totally dominates, even for $\alpha$=0 and $Z=3\times Z_{\odot}$. We can see the cloud base in the contribution functions around 10$^{-4}$ bars, then a gap in contribution, and then the patterns of the gaseous absorption at higher pressures picks up. This is why, in the corresponding transit spectra, we can always see the gaseous absorption peaks.

Transit spectra for all the candidate cloud species with a range of $Z$ and $\alpha$ are shown in Figure \ref{fig:1400K_eq_pstudy} and in Figures \ref{fig:700K_eq_pstudy} - \ref{fig:1800K_eq_pstudy} in the appendix, grouped by temperature. $Z$, cloud species, and the particle-size distribution work together to modulate how optically thin or thick the cloud is. In the examples show in Figure \ref{fig:forsterite_contributions}, the cloud is always optically thin enough that the gaseous absorption is still affecting the transit spectrum. However, for many species, clouds can overpower gaseous extinction at many wavelengths (eg. Na$_2$S in the 1000-K atmosphere, MgSiO$_3$ in the 1400-K atmosphere if metallicity is high, Fe in the 1800-K atmosphere, and Al$_2$O$_3$ in the 1800-K atmosphere if metallicity is high). Once one chooses a cloud species and a temperature, the cloud's base pressure is set. If the cloud base is at a middling pressure or deep in the atmosphere (say deeper than 10$^{-2}$ bars), then $\alpha$ shapes whether the cloud is extended or resides only at depth. If the cloud-base forms very high up in the atmosphere, then varying $\alpha$ will have less effect. 

\begin{figure}
    \centering
    \includegraphics[width=\textwidth]{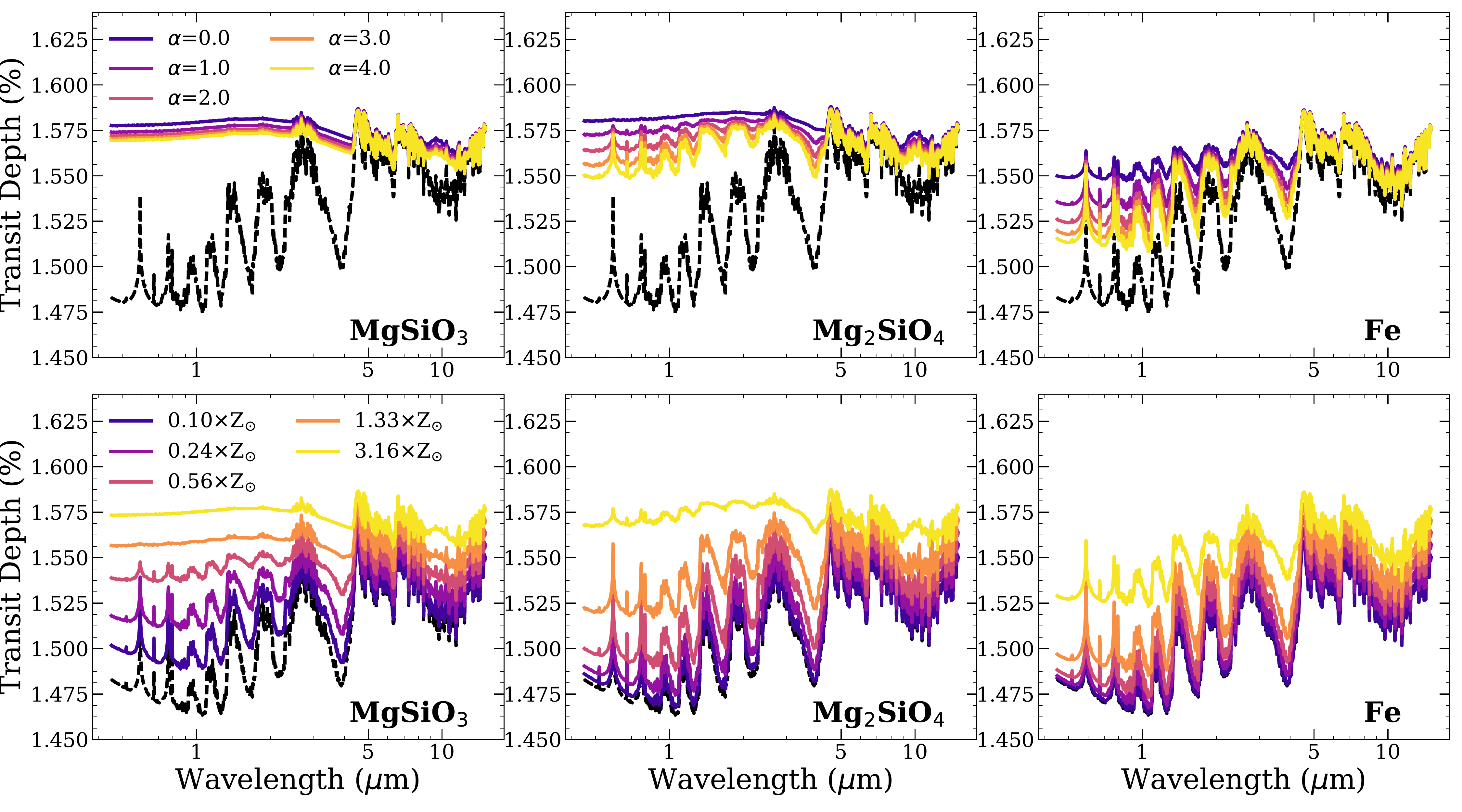}
    \caption{A demonstration of the 1400-K planet's sensitivity to metallicity and $\alpha$ when an equilibrium cloud of MgSiO$_3$ (left column), Mg$_2$SiO$_4$ (center column), or Fe (right column) is present. The top row shows transit spectra with varying $\alpha$ and the bottom row shows transit spectra with varying Z. In each panel a light gray dashed line shows the transit spectra for the 1400-K atmosphere when clear.}
    \label{fig:1400K_eq_pstudy}
\end{figure}

A key point we wish to emphasize with Figure \ref{fig:1400K_eq_pstudy} and Figures \ref{fig:700K_eq_pstudy}-\ref{fig:1800K_eq_pstudy} in the appendix, is that changing Z and $\alpha$ can only alter the transit spectra within a small range of behaviors in some cases (e.g. KCl, NaCl, and TiO$_2$), while in others it can move the atmosphere from appearing nearly clear to appearing very clouded (eg. Fe and Mg$_2$SiO$_4$). When a large number of particles form, then the atmosphere looks very cloudy regardless of $\alpha$ (e.g. Na$_2$S in the 1000-K atmosphere, MgSiO$_3$ in the 1400-K atmosphere, or Fe in the 1800-K atmosphere). In these cases, changing Z can make a large difference to the balance between gaseous opacity and cloud opacity. When a very small number of particles form (e.g. KCl at 700 K, NaCl at 1000 K, TiO$_2$ at 1800 K), then the transit spectra can never appear much different from a clear atmosphere, and changing the metallicity is mostly changing the gaseous absorption features. 

Clearly, assuming the cloud base must form where the Clausius-Clapeyron line intersects with the T-P profile and using phase equilibrium to setting the amount of available material that goes into condensates places severe constraints on the patterns a cloud species can impart onto the transit spectrum of a given atmosphere. If these assumptions are a good approximation for actual cloud formation on exoplanets, then this is good news. One should be able to break degeneracies and obtain much tighter constraints on atmosphere and aerosol properties. If these assumptions are wrong, then one will struggle to fit the data at all or will end up with erroneous results. We will see the consequences of this in the MCMC experiments of the next section. Compared to the results for the slab aerosols, it is much easier to differentiate between the candidate species. This is partly because of the model and partly because the sample data tend to exhibit more non-gray behavior than the sample data for the hazy experiments. When the correct species is used in the fit, we also get tighter constraints on $Z$ than for the slab aerosol.

\subsection{Equilibrium Cloud MCMC Experiment Results}

Is one likely to be able to unambiguously identify condensing species with JWST transit spectra? To address this question, we now consider the results of MCMC retrieval experiments for transit spectra with phase equilibrium clouds. In the experiment we simulated data for each possible cloud species with a tapering shape parameter of $\alpha$=2, and an overall atmospheric metallicity of Z=3$\times$Z$_{\odot}$. Like the slab-MCMC experiments, we compare one spectrum with a smaller (0.05-$\mu$m) modal particle size and one with a larger (1.0-$\mu$m) modal particle size, and we always used a log-normal size distribution with dispersion 2.5. 

\begin{figure}
    \centering
    \includegraphics[width=\textwidth]{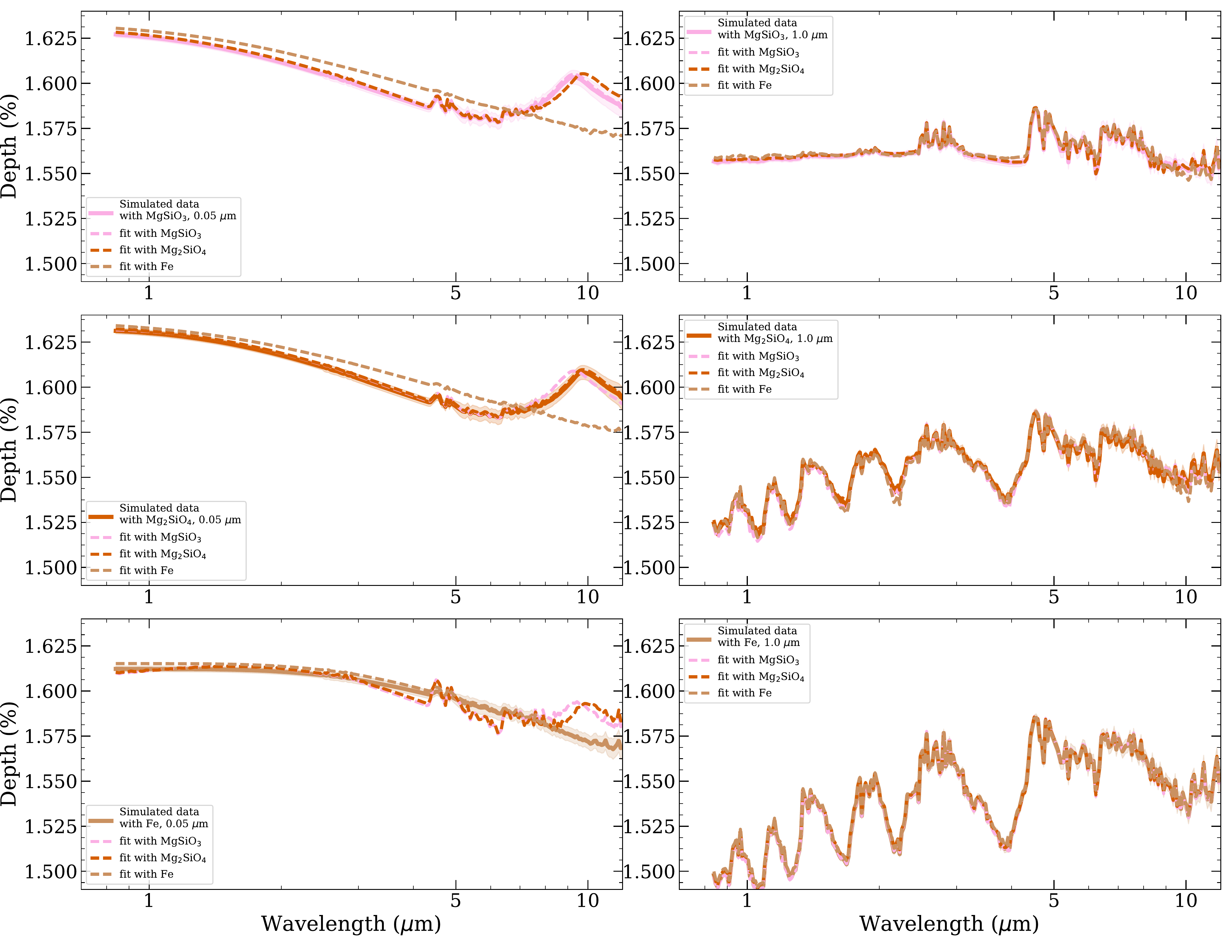}
    \caption{Results for MCMC experiments with the 1400-K atmosphere and phase equilibrium clouds. The solid lines with shaded error envelopes indicate the simulated data. Dashed lines show the best fit spectra with MgSiO$_3$ clouds, Mg$_2$SiO$_4$ clouds, and Fe clouds respectively. In the top row the true cloud species is MgSiO$_3$, in the middle row it is Mg$_2$SiO$_4$, and in the bottom row it is Fe. On the left hand side the modal particle size is 0.05 $\mu$m and on the right side the modal particle size is 1 $\mu$m.}
    \label{fig:eq_1400K_fits}
\end{figure}

\begin{figure}
    \centering
    \includegraphics[width=\textwidth]{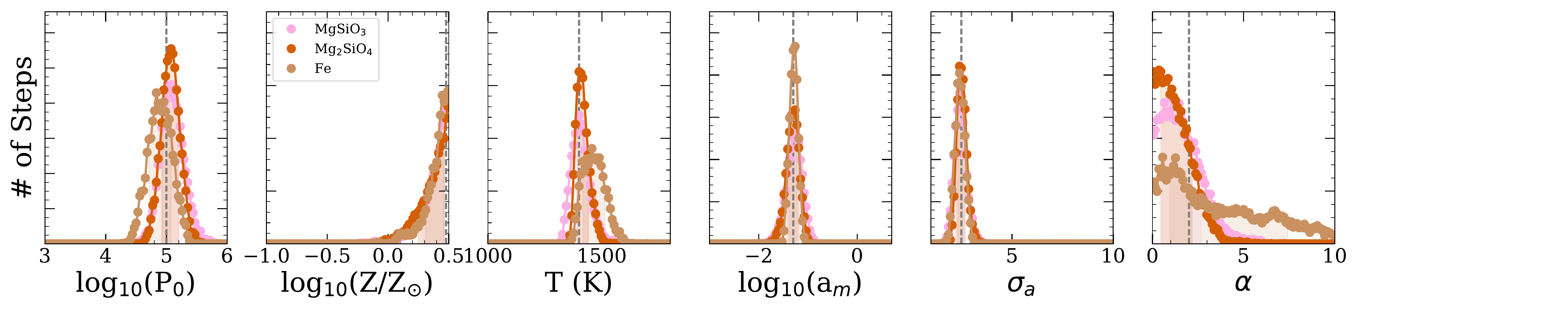}
    \includegraphics[width=\textwidth]{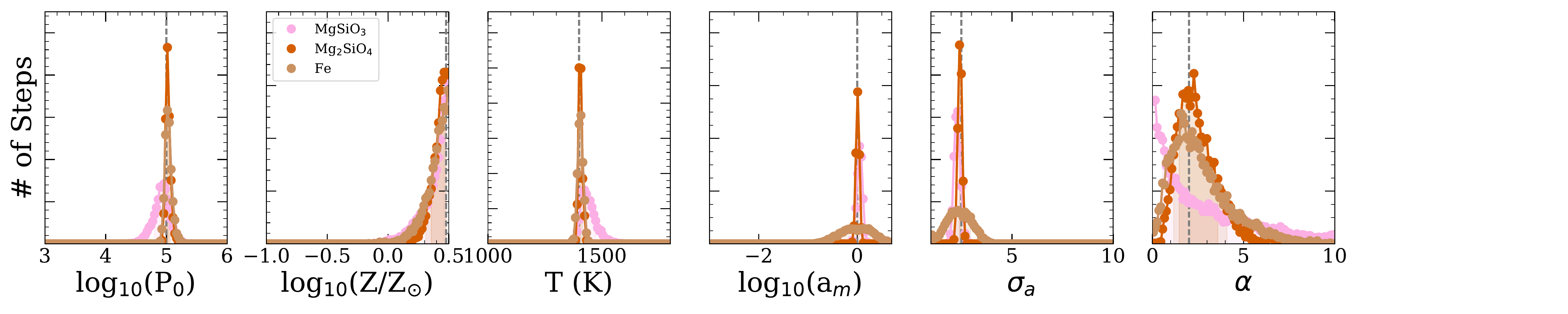}
    \caption{Posteriors for all parameters when the correct cloud species is used to fit data simulated with the phase equilibrium cloud model in the 1400-K fiducial atmosphere. The top row used data simulated with a modal cloud particle size of 0.05 $\mu$m, and the bottom row was simulated with a larger modal cloud particle size of 1.0 $\mu$m. Each color represents a different condensate species. Vertical lines indicate the true values of parameters.}
    \label{fig:1400_eq_hists}
\end{figure}

The best-fit spectra for the 1400-K atmospheres are shown in Figure \ref{fig:eq_1400K_fits}. Fe and the two types of silicates look very different for the small modal particle size. Distinguishing between MgSiO$_3$ and Mg$_2$SiO$_4$ is not as easy, but it does look feasible provided sufficient SNR and spectral resolution to locate the peak and shape of the 10-$\mu$m resonance feature. When the particle-size distribution has a mode of 1 $\mu$m, all three species are able to mimic each other quite well. If $\alpha$ were lower or the metallicity were higher, then the 10-$\mu$m feature might rise above the gas absorption, again making iron and the two types of silicates distinguishable. 

The posteriors for each MCMC fit done with the correct species are shown in Figure \ref{fig:1400_eq_hists}. The modal particle size and the dispersion of the size distribution are accurately retrieved in all cases. The tapering parameter $\alpha$ is constrained for the 1-$\mu$m Fe and forsterite, but otherwise just places an upper bound. For the smaller Fe particles in particular, $\alpha$ is poorly constrained. The reference pressure and temperature are more tightly constrained for the large particle sizes. The temperature for small iron particles and large enstatite particles skews a bit higher than the true value, accompanied by a slight shift towards lower reference pressures. Within our prior of 0.1-3.16 $\times$ solar, the metallicity is only able to place a lower bound. Note that this lower bound is much tighter than those placed on spectra which had slab aerosols. 

We also show results for all the other fiducial atmospheres in the appendix, Figures \ref{fig:700_eq_hists}-
\ref{fig:eq_1800K_fits}. 

For the 700-K atmospheres, if the size distribution has 0.05-$\mu$m modal particles, the fit with the true species is stronger than the fit with the wrong species. For the size distributions with 1.0-$\mu$m modal particles the wrong species provides a suitable fit to the data. With this modal particle size, the amount of material available to make KCl or Na$_2$S particles given our equilibrium assumptions is simply not sufficient to create a significant cloud opacity compared to the gaseous opacity. For the 1000-K atmospheres, in all cases where the underlying data have strong effects due to the aerosols, Na$_2$S and NaCl are very distinguishable. It is only for the 1.0-$\mu$m  NaCl cloud that the two species fit the data equally well. In the 1000-K atmosphere, the Na$_2$S cloud base forms deeper in the atmosphere, around 1 bar, where there is enough material to make a reasonable number of particles. In contrast, in the 700-K atmosphere the Na$_2$S cloud base formed up around 10$^{-5}$ bars. For the 1800 K grouping we tested Fe, TiO$_2$, and Al$_2$O$_3$. Similar to the other temperatures, when particle sizes are smaller, then the three species have trouble mimicking each other. When particles sizes are 1.0 $\mu$m, they are almost indistinguishable. However, with very high SNR, Fe may be unable to mimic the Al$_2$O$_3$ cloud.

The posteriors for the 1000-K and 1800-K atmospheres mostly show similar results to the 1400-K atmosphere. The metallicity is only a lower bound and $\alpha$ is only an upper bound. The lower bounds on metallicity are again much tighter than those for the slab aerosols. The posteriors for the 1.0-$\mu$m NaCl clouds in the 1000-K atmosphere show that the particle-size distribution and $\alpha$ are totally unconstrained. This reflects the fact that the NaCl cloud has only an extremely week effect on the transit spectrum. In the 700-K atmosphere, the metallicity was actually retrieved for all combinations of species and particle sizes, not just a lower bound. The correct value of $\alpha$ is also retrieved. The constraints for the 700-K are tightest, then the 1800-K planet and finally the 1400-K and 1000-K planets have looser constraints. This reflects the varying scale heights of the planets compared to the precision of the depth measurements.

Overall, it is very clear from the parameter sensitivity studies and MCMC experiment results, that incorporating some assumptions based on thermodynamics into your model can lead to fewer degeneracies between model parameters and makes aerosol species appear more distinct. This is no surprise. However, the benefits of the equilibrium cloud model depend heavily on whether you have made good assumptions. If condensed aerosols in exoplanet atmospheres form roughly where the T-P profile crosses the Claussius-Clapeyron line, then fall off in density in a manner proportional to the gaseous pressure scale height, and if they have a small modal particle size (less than 1.0 $\mu$m), then they tend to embed a lot of information about the species and size-distribution of the aerosol in the transit spectra. These are strong assumptions, but consistent with the behavior of condensed clouds in the solar system. 

\section{Summary and Conclusions}\label{sec:conclusions}

First, we demonstrated the sensitivity of JWST-like transit spectra to atmospheric properties, such as temperature and metallicity, and to aerosol properties, such as particle-size distribution, aerosol species, and spatial extent of aerosols. In our explorations, we considered spectra with 15 different aerosol species. We did our calculations using METIS and two forms of aerosol parameterization. One type of aerosol was a ``slab" specified by a fraction of available material incorporated into the aerosol, $F$, and an arbitrary top-pressure cut-off, P$_{top}$. The second type was a ``phase equilibrium" cloud which assigns the base of a condensing species to form where the T-P profile intersects with the Claussius-Clapeyron line, and then uses a free parameter, $\alpha$, to describe how quickly the number density of aerosol particles falls off relative to the number density of gas particles. We paired these two types of aerosol spatial parameterization with log-normal size distributions. With this context of parameter sensitivity laid out, we then presented results from a wide array of retrieval experiments to look in depth at the prospects for inferring aerosol and planetary properties from a variety of hazy and cloudy simulated JWST transit spectra. We focused on fiducial atmospheres with temperatures of 700 K, 1000 K, 1400 K, and 1800 K in order to include all the candidate condensing aerosol species in our set of 15 outside of NH$_3$ and H$_2$O clouds. 

Before we review the results of our retrieval experiments and their implications for JWST, we pause to remind the reader of a few caveats. As mentioned throughout the methods and results sections, our forward model for retrievals and simulated data assume isothermal atmospheres, uniform aerosol coverage, and equilibrium chemistry with a solar C/O ratio. This forward model would need modifications to also allow the variation of C/O ratio, patchy clouds, and a parameterized T-P profile rather than an isothermal profile in order to be applied to real data. Several works have already presented the problem with assuming uniform aerosol-coverage and found successful remedies (\citealt{Line2016}; \citealt{MacDonald2017}). \citealt{Rocchetto2016} point out biases that will arise from assuming an isothermal atmosphere in fitting real observations of transit spectra, and \citealt{Kempton2012} compare results from assuming equilibrium chemistry versus allowing non-equilibrium mixing ratios when fitting the transit spectrum of GJ1214b. With this in mind, our findings should be interpreted as indicative of the information content about aerosols and other atmospheric properties that can feasibly be encoded in JWST-like transit spectra of cloudy, hazy exoplanets. The retrieval approach used here is not intended to be applied to real data, rather it is meant to provide an upper-limit on the precision of constraints that can be placed on model parameters and demonstrate two options for incorporating aerosol parameterizations with greater physical significance in future retrieval efforts. 

We now summarize the questions we investigated in this paper and the answers we found.
\begin{enumerate}
    \item \textbf{Which JWST wavelengths contain the most information about aerosol properties and which provide information  about  gaseous  absorption? } Looking at Jacobians and transit spectra for a representative array of particle-size distributions and aerosols species, we found that it is the combination of JWST's longest (8+ $\mu$m) and shortest (less than 2 $\mu$m) wavelength coverage which provide the most information about 0.1-1.0 $\mu$m aerosols, while the middle IR wavelengths usually provide information about gaseous absorption (unless aerosols are present at very high altitudes). In other words, transit spectra which appear very flat at shorter wavelengths, could still exhibit recognizable spectral features from gaseous absorption, or from the aerosols themselves in the longer wavelength range accessible by JWST. This trend arises both because there are strong gaseous absorption features in the near-mid IR and because of the optical properties of many of the leading candidate aerosol species. 
    
    \item \textbf{How well can we recover atmospheric metallicities and temperatures, even when aerosols are present as we extend the wavelength coverage of transit spectra?} The metallicity was often difficult to retrieve when aerosols are present at high altitudes. The peaks of the broad groupings of absorption features at 2.5-3 $\mu$m and 4.5-8.5 $\mu$m are usually recognizable in transit spectra, even with aerosols present. However, the shapes of the edges and troughs of these gaseous absorption features relative to the peaks are needed to show a change due to metallicity that is not degenerate with simply changing the reference pressure or the amount of available material that is incorporated into aerosols. On the other hand, if the amount and/or species of aerosol present could be accurately tied to the bulk metallicity of the atmosphere, then spectra with aerosols are actually very sensitive to the metallicity of the atmosphere (often much more sensitive than the gas alone). This result emphasizes the importance of developing relevant microphysical models and using them to place reasonable priors on how much available material is likely to be incorporated into aerosols. That will only be possible if we know what species are present.
    
    \item \textbf{Can we uniquely identify which dominant aerosol species are present in atmospheres using JWST transit spectroscopy?} \textbf{Can we constrain the size-distribution of aerosols?} \textbf{How do these tasks differ for condensed clouds and photochemical hazes?} We found that log-normal size distributions of different aerosol species could often be distinguished, so long as modal particle sizes and spatial positions are such that the aerosols do not just behave as a gray opacity source relative to the gaseous contributions to the transit spectra. Aerosols can present themselves as a gray opacity source when particles are large, when the aerosol opacity is negligible compared to gas opacity so it only raises the bottoms of absorption windows slightly, or when there is a steep top-pressure cut-off to the physical location of aerosols at a height such that the aerosol is optically thick for all wavelengths of light. The good news is that this type of aerosol can often be marginalized over to retrieve unbiased temperatures. The bad news is that it doesn't allow us to identify what species the aerosol is. In particular, different types of hydrocarbon haze (various soots, Tholins, poly-HCN) tend to look very similar. They can mimic each other and other aerosols quite well if the other aerosol either behaves as a gray opacity source or is just slightly different from a gray opacity source (eg. with a slight downward slope from 8 to 10 $\mu$m). However, in many cases the aerosol type, modal particle size, and spread in particle-size distribution can be recovered. The slab and phase equilibrium aerosol formulations of NaCl and KCl with small particles look quite distinctive. The Na$_2$S cloud with equilibrium base in the 1000-K atmosphere for both 0.05- and 1.0-$\mu$m particles, and in the 700-K atmosphere for just 0.05-$\mu$m particles. For 0.05-$\mu$m particles silicates and iron formed as equilibrium base in 1400-K look very distinct. At 1800 K equilibrium base Al$_2$O$_3$ TiO$_2$ and Fe look different when the modal particle size is 0.05 $\mu$m. Different types of silicates (enstatite vs forsterite and different iron percentages) may even be distinguishable if the 10-$\mu$m feature is observable, and the observations have sufficient $SNR$ and spectral resolution. This requires the presence of 0.1-1 $\mu$m size particles lofted up in the air above 10$^{-4}$-10$^{-5}$ bars for a strong 10-$\mu$m feature to be visible above water absorption. The question thus becomes, will aerosols form with size and spatial distributions such that they appear gray, or will they form in such as way as to leave distinctive features indicating their nature in transit spectroscopy? Answering this is beyond the scope of our paper, as it relies on detailed microphysical models and GCM's. This is the purview of coupled dynamical cores and microphysical models and is still an open question.
\end{enumerate}

 Overall, our results support the community's wide-spread optimism for JWST and ARIEL transit spectroscopy. Through a coincidence of stronger gaseous absorption and weaker aerosol extinction, the 4-9 $\mu$m range is consistently most likely to be dominated by gaseous absorption, even when the shorter and longer wavelengths tend to be shaped by aerosols across a wide variety of temperatures and aerosol species. This means that, despite the fact that some transit spectra lack gaseous absorption features in current observations, JWST could still detect signatures of gaseous absorption in the near through mid-IR. For both a slab-type aerosol and a phase equilibrium cloud, our ability to distinguish between some of the leading candidate aerosol species and recover particle-size distributions will depend on what the ground truth is in the exoplanet atmospheres, even with the full wavelength coverage of JWST. If nature is kind, and aerosols form such that we can unambiguously determine what the dominant aerosol species are and make empirical measurements of particle-size distributions and spatial extent, then this information would provide a valuable test of whether theories developed with GCM's and detailed microphysical modeling are correctly capturing the behavior of cloudy hazy exoplanet atmospheres. Our results point to the importance of continuing efforts to accurately link enrichment and depletion by aerosols to gas-phase chemistry within retrievals. This will ultimately be the key to inferring metallicities of cloudy hazy atmospheres.  

\acknowledgments
The authors would like to acknowledge support for this research under NASA 
WFIRST-SIT award \# NNG16PJ24C and NASA Grant NNX15AE19G. This research has made use of the NASA Exoplanet Archive, which is operated by the California Institute of Technology, under contract with the National Aeronautics and Space Administration under the Exoplanet Exploration Program.
\vspace{5mm}
\software{astropy \citep{astropy}, numpy (\citealt{numpy1}; \citealt{numpy2}), scipy (\citealt{scipy1}; \citealt{scipy2}), matplotlib \citep{matplotlib}, emcee (\citealt{emcee1}; \citealt{Goodman2010}), corner (\citealt{corner}), mpi4py (\citealt{Dalcin2011})}

\appendix

\begin{figure}
    \centering
    \includegraphics[width=\textwidth]{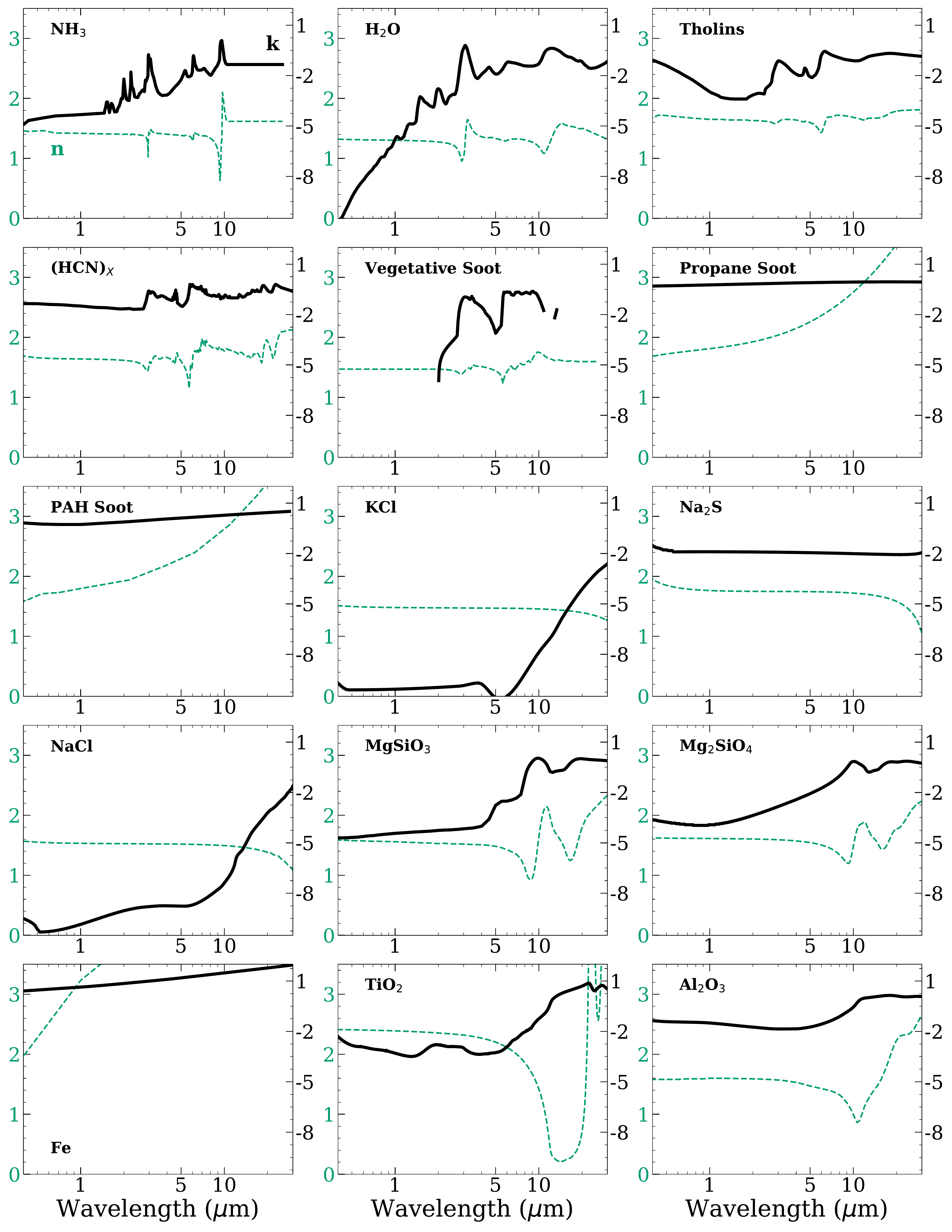}
    \caption{Complex indices of refraction used to incorporate aerosols into our models. References are in Table \ref{tab:aerosol_properties}. }
    \label{fig:indices}
\end{figure}

\begin{figure}
    \centering
    \includegraphics[width=\textwidth]{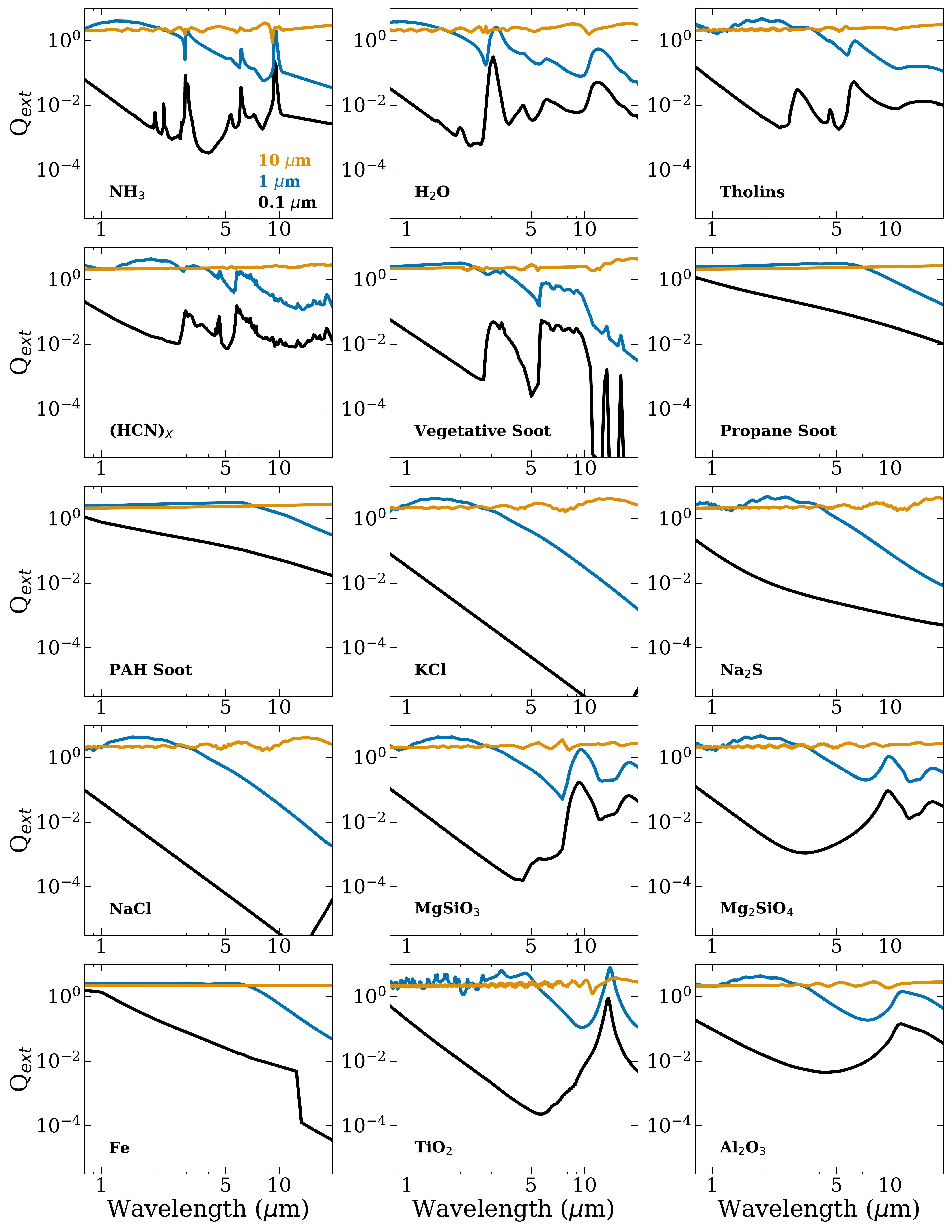}
    \caption{Extinction efficiency factors for several particle sizes. Recall, $\sigma _{ext}$ = $Q_{ext}\times \pi a^2$, where $a$ is the radius of the particle.}
    \label{fig:efficiencies}
\end{figure}

\begin{figure}
    \centering
    \includegraphics[width=\textwidth]{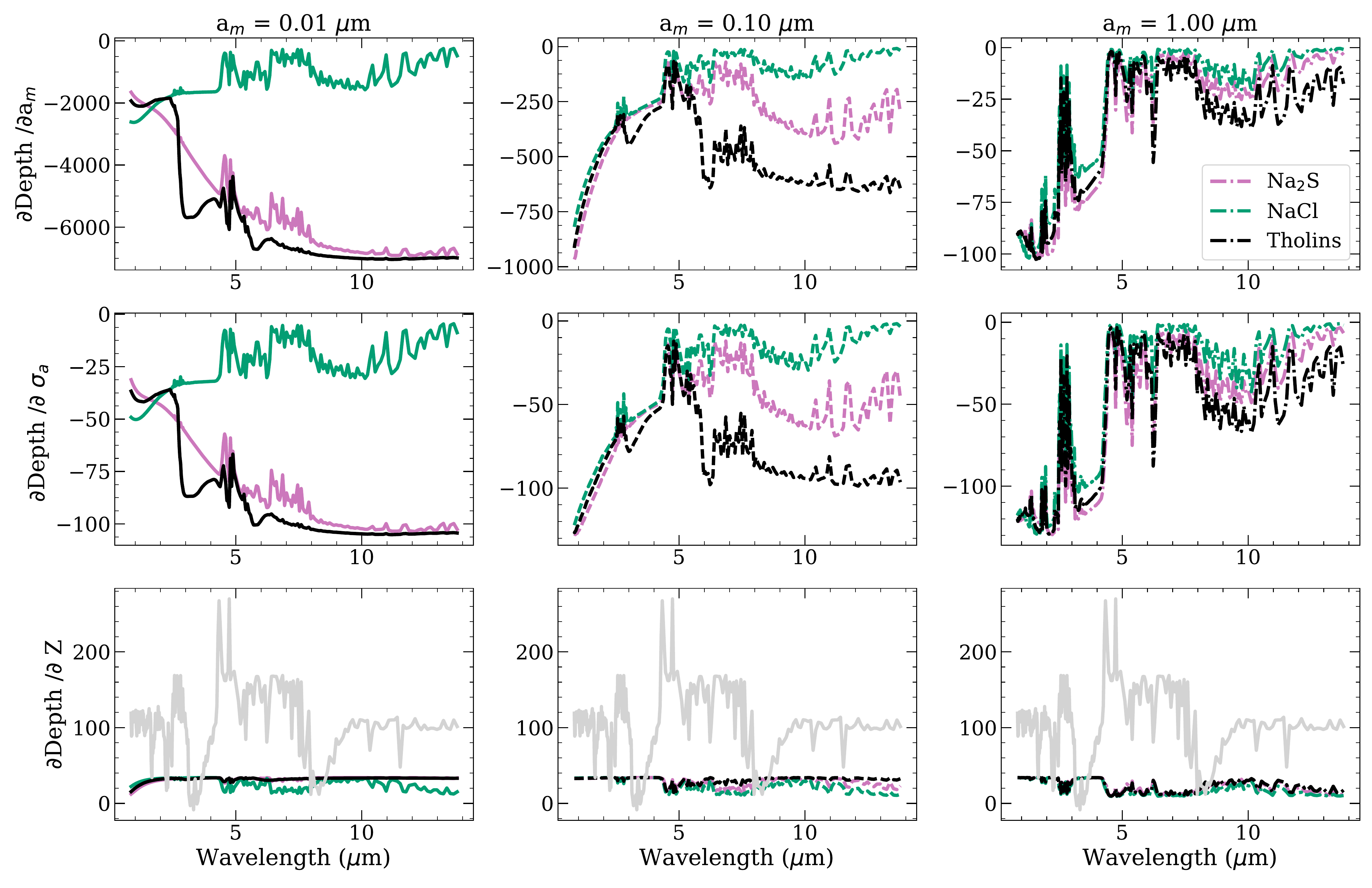}
    \caption{Transit depth Jacobians for our 1000-K fiducial atmosphere when Na$_2$S (pink), NaCl (teal), and Titan tholins (black) are included as slab aerosols. We have made the same assumptions about the slab aerosol as in Figure \ref{fig:700K_jacobian}: varying $F$ such that $F\times$the solar abundance of the limiting atomic species equal to 3$\times$10$^{-6}$, setting the top pressure cut-off too high up to make a difference, and using a size dispersion of 2.5 on the log-normal size distribution. Each column shows results for a different modal particle size, as labeled. The top row shows the partial derivative with modal particle size, the middle row shows the partial derivative with size dispersion, and the bottom row shows the partial derivative with metallicity. Again, the light gray line indicates the partial derivative with metallicity for the clear atmosphere.}
    \label{fig:1000K_jacobian}
\end{figure}

\begin{figure}
    \centering
    \includegraphics[width=\textwidth]{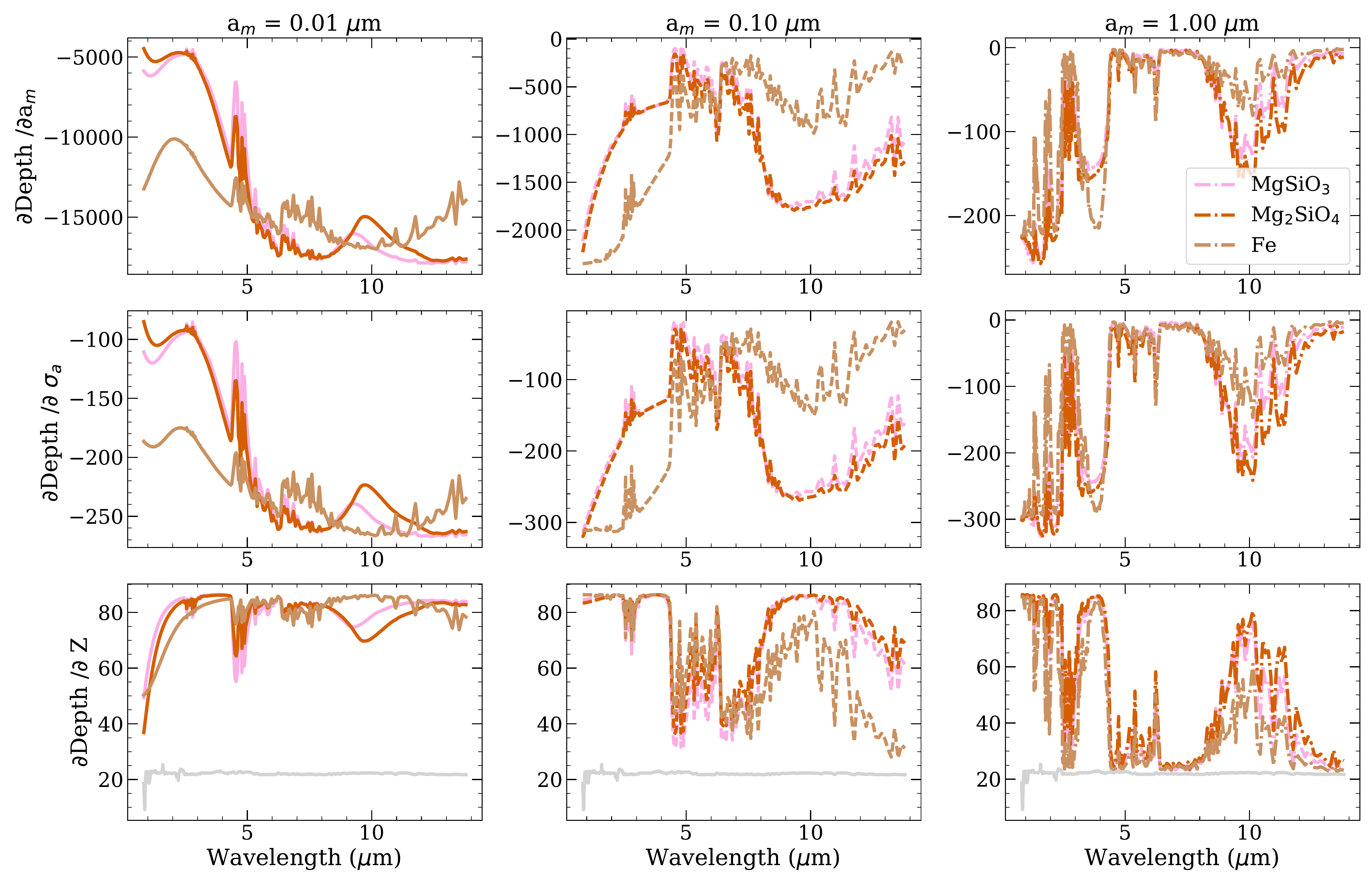}
    \caption{Transit depth Jacobians for our 1400-K fiducial atmosphere when MgSiO$_3$ (pink), Mg$_2$SiO$_4$ (orange), and Fe (buff) are included as slab aerosols. Each column sets a different modal particle size and each row is the partial derivative with respect to a different parameter. In the bottom panels showing the partial derivative with respect to Z, we include a light gray line showing the result for a clear atmosphere.}
    \label{fig:1400K_jacobian}
\end{figure}

\begin{figure}
    \centering
    \includegraphics[width=\textwidth]{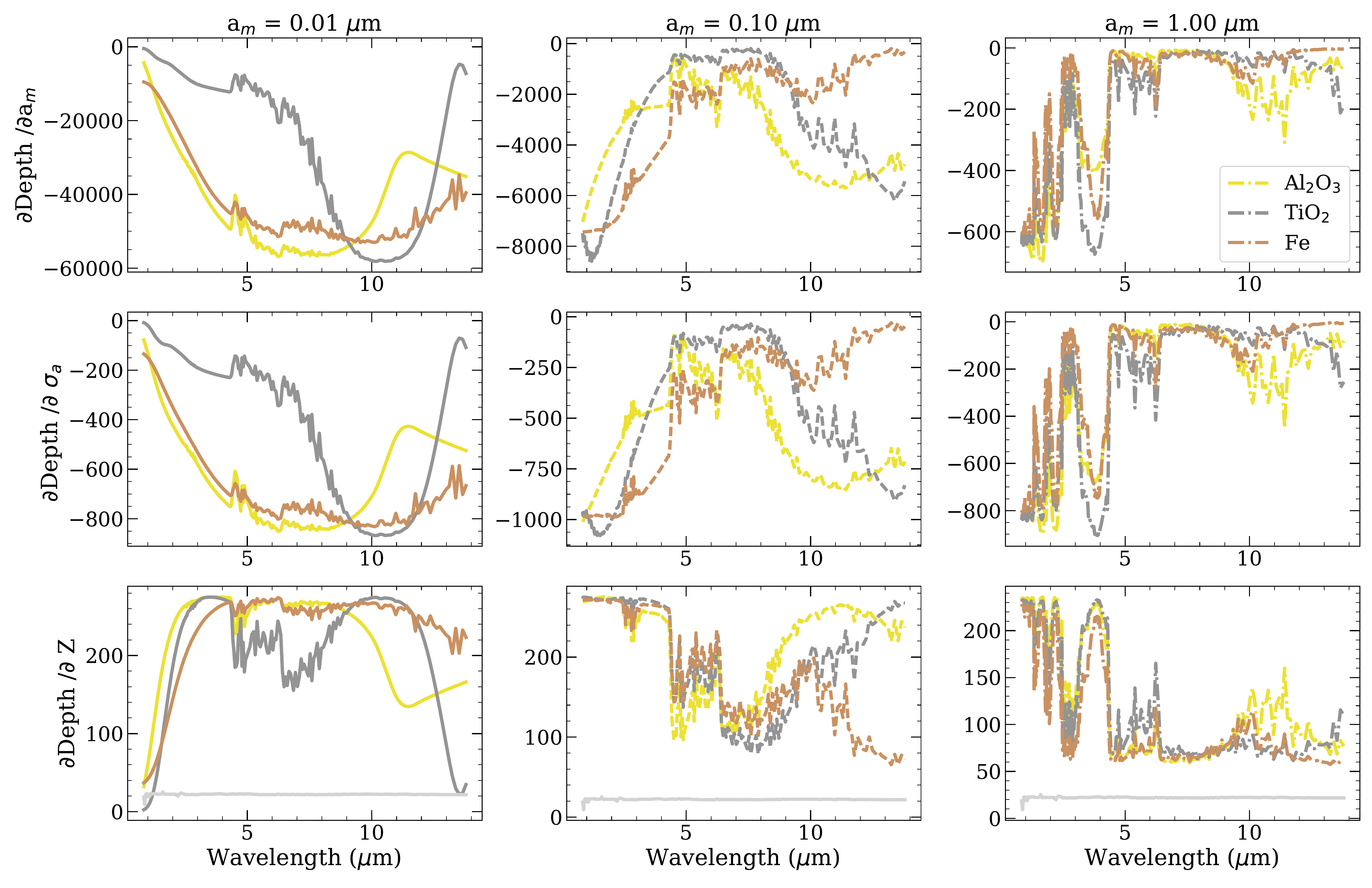}
    \caption{Transit depth Jacobians for our 1800-K fiducial atmosphere when Al$_2$O$_3$ (dark gray), TiO$_2$ (yellow), and Fe (buff) are included as slab aerosols. The top row shows the partial derivative with respect to modal particle size, the center row shows the partial derivative with respect to the dispersion of the log-normal particle-size distribution, and the bottom shows the partial derivative with respect to metallicity. These panels include a light gray line with the results for a clear atmosphere. Each column has a different modal particle size. The same values of P$_{top}$ and $\sigma _a$ are used throughout all the panels for all species. $F$ is chosen for each species such that $F\times$the solar abundance of the limiting atomic species equals 3$\times$10$^{-6}$.}
    \label{fig:1800K_jacobian}
\end{figure}

\begin{figure}
    \centering
    \includegraphics[width=\textwidth]{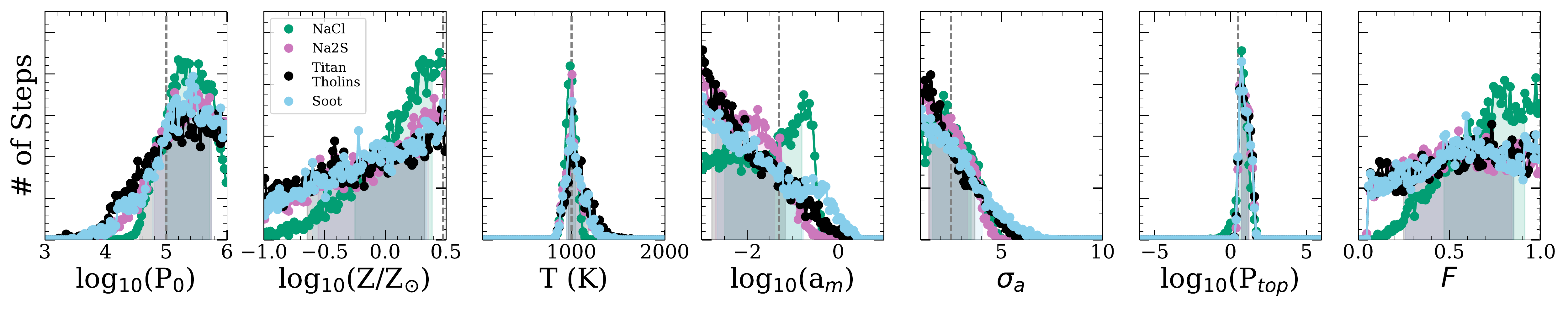}
    \includegraphics[width=\textwidth]{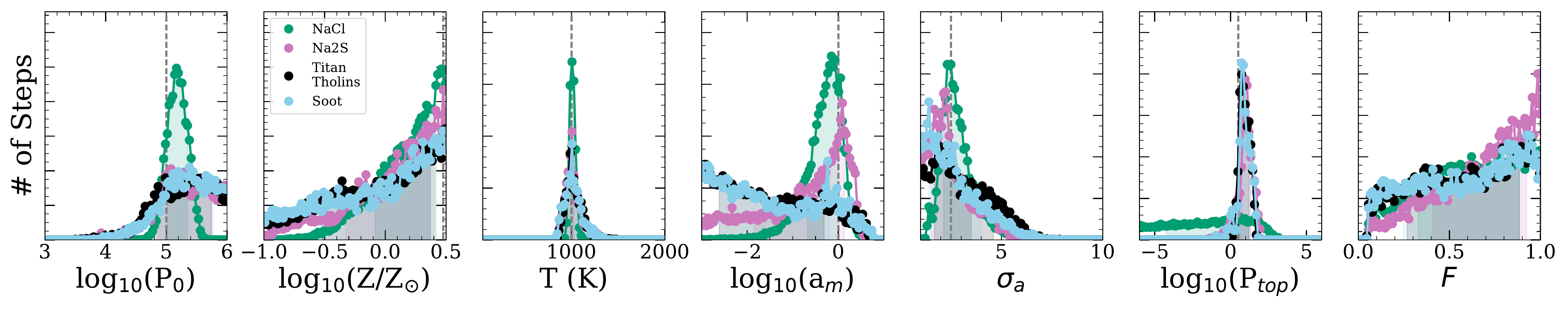}
    \caption{Histograms of parameter posteriors for the 1000-K fiducial atmosphere with slab aerosols. Colors show different species of areosols and vertical lines show true values of parameters. In the top row, data was simulated with a modal particle size of 0.05$\mu$m. In the bottom row, data was simulated with a modal particle size of 1$\mu$m.}
    \label{fig:1000_slab_hists}
\end{figure}

\begin{figure}
    \centering
    \includegraphics[width=\textwidth]{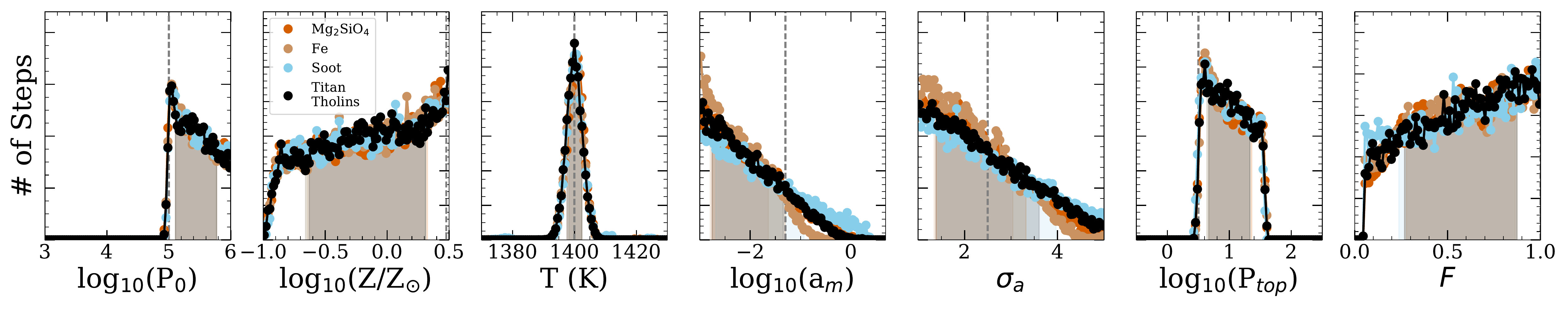}
    \includegraphics[width=\textwidth]{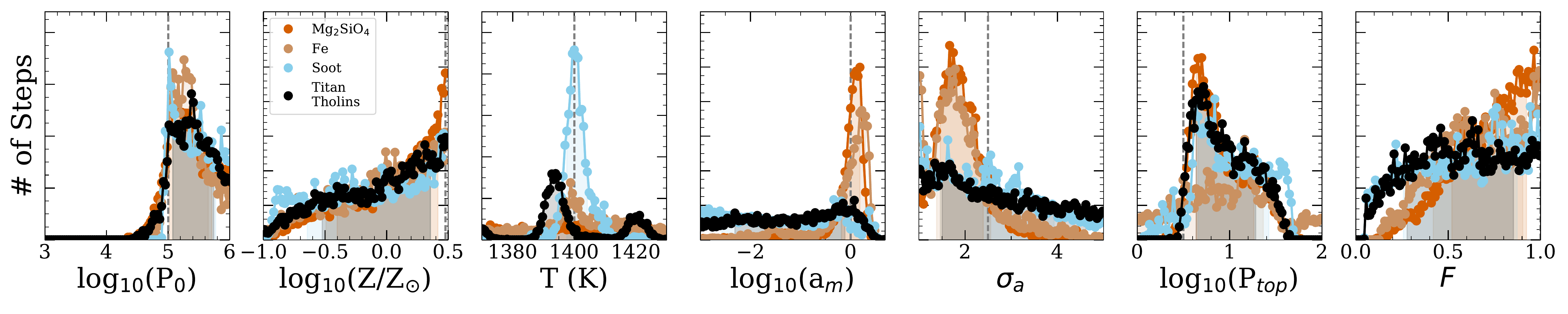}
    \caption{Histograms of parameter posteriors for the 1400-K fiducial atmosphere with slab aerosols. Each color is a retrieval for simulated data with a different species of aerosol. In the top row the modal particle size was 0.05$\mu$m, while in the bottom row the modal particle size was 1$\mu$m. Vertical lines indicate the true parameter values.}
    \label{fig:1400_slab_hists}
\end{figure}

\begin{figure}
    \centering
    \includegraphics[width=\textwidth]{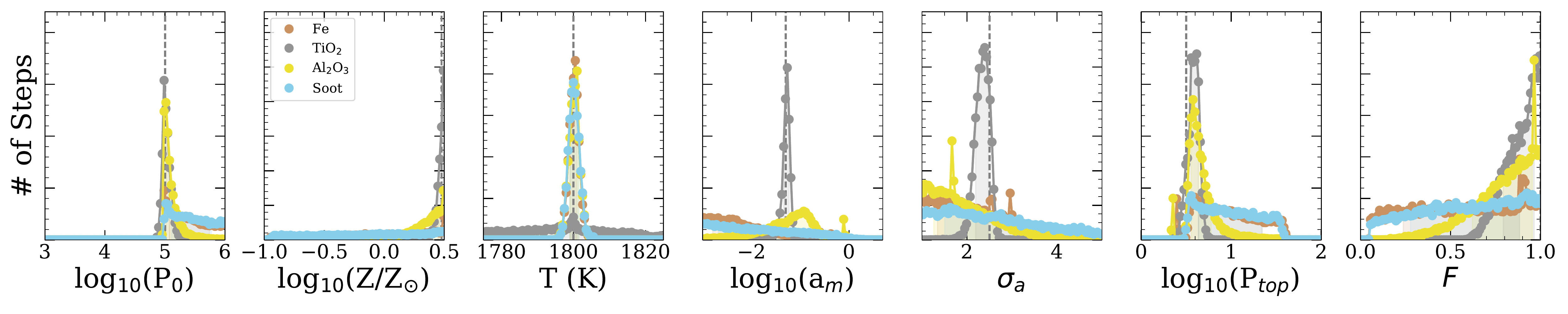}
    \includegraphics[width=\textwidth]{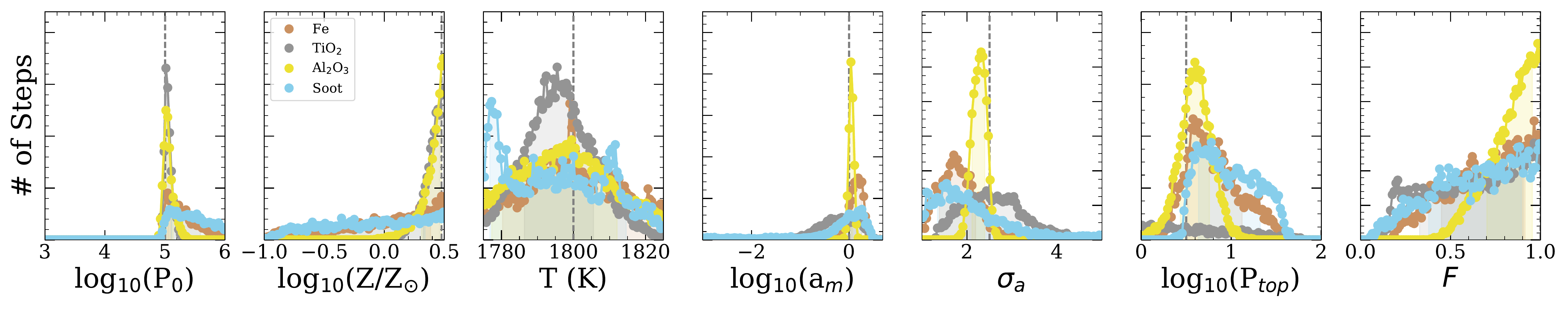}
    \caption{Histograms of parameter posteriors for the 1800-K fiducial atmosphere with slab aerosols. Colors show results for data simulated with different species of aerosols. The vertical lines mark the true values of parameters used when simulating data. In the top row the modal particle size was 0.05$\mu$m, and in the bottom row the modal particle size was 1$\mu$m.}
    \label{fig:1800_slab_hists}
\end{figure}

\begin{figure}
    \centering
    \includegraphics[width=\textwidth]{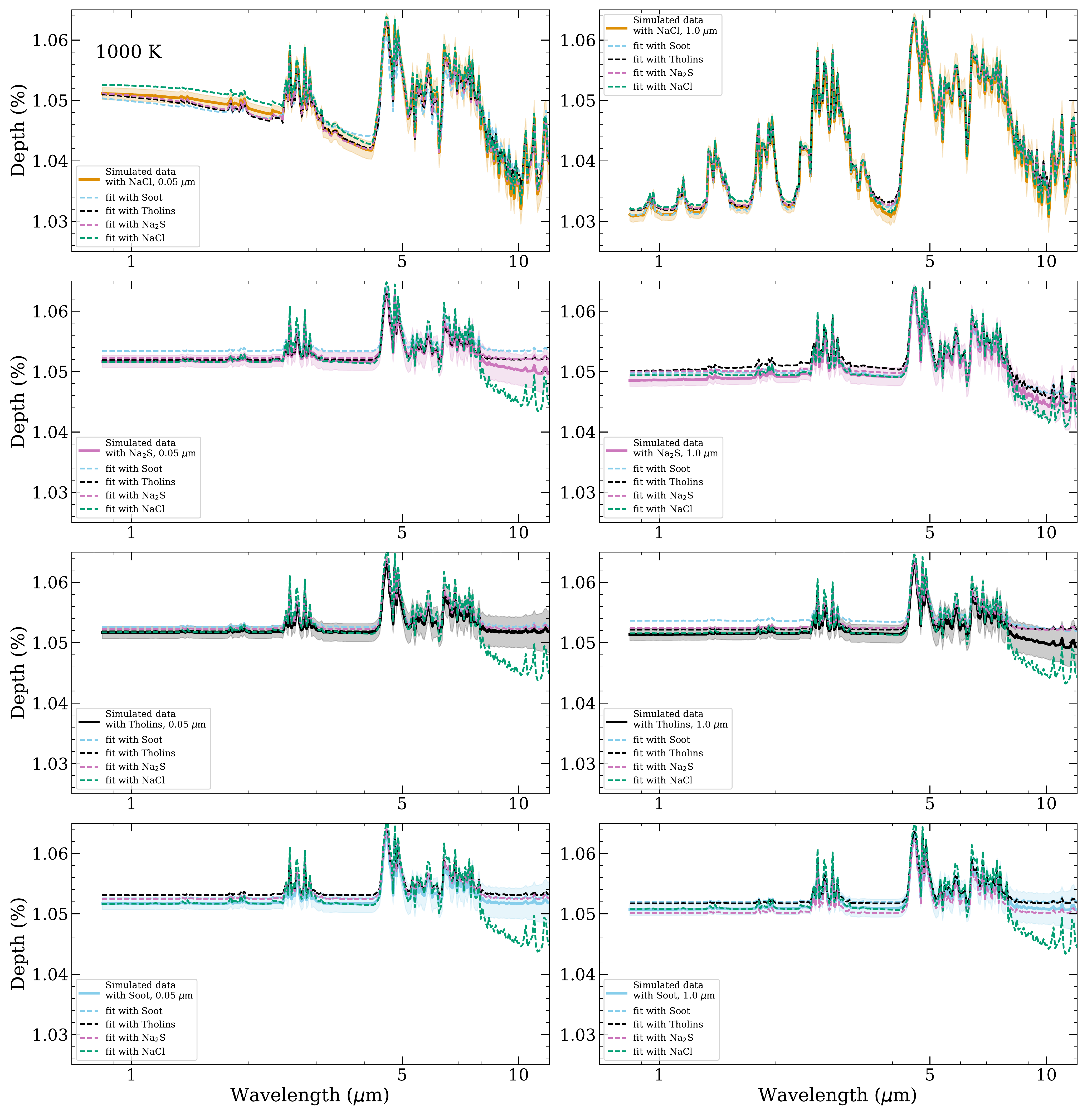}
    \caption{Results for the MCMC experiments with aerosols in the 1000-K fiducial atmosphere. The solid line in each panel shows the simulated data and the surrounding light shading indicates the error envelope. In the top row the true species is NaCl, second row Na$_2$S, third row Tholins, and finally soot in the bottom row. Dashed lines show the spectra corresponding to the median parameter values retrieved with all the different aerosol species. In the left column, the data was simulated with a log-normal size distribution with modal particle size of 0.05 $\mu$m. The right-side had a modal particle size of 1.0 $\mu$m. The true value of $F$ was always 0.5, P$_{top}$ was 10$^{-4.5}$ bars, and Z was 3 $\times$ Z$_{\odot}$.}
    \label{fig:slab_1000K_fits}
\end{figure}

\begin{figure}
    \centering
    \includegraphics[width=0.8\textwidth]{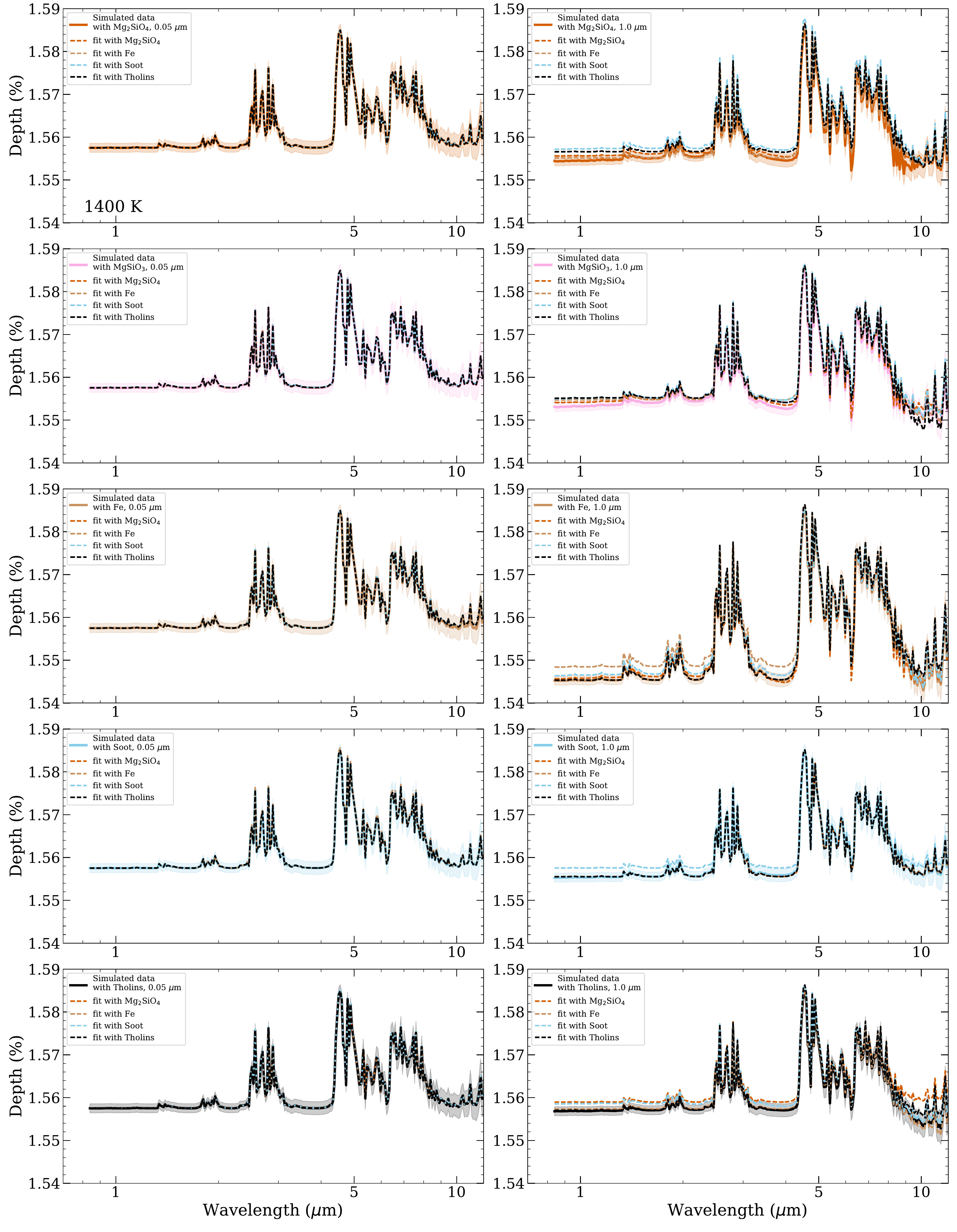}
    \caption{The results of our mixed aerosol MCMC experiments for the 1400-K atmosphere. We simulated data with a slab of Mg$_2$SiO$_4$ (top row), MgSiO$_3$ (second row), Fe (third row), Soot (fourth row), and Tholins (bottom row). Simulated data are shown with a solid line, the errors are shown with a shaded envelope, and transit spectra from MCMC retrievals with all different species are shown with dashed lines. The slab parameterization used to simulate data always had $F$=1.0, P$_{top}$=10$^{-4.5}$ bars, and a log-normal size distribution with $\sigma _a$=2.5. The modal particle size was 0.05 $\mu$m in the left column and 1.0 $\mu$m on the right column. The overall atmospheric metallicity was Z=3$\times$Z$_{\odot}$.}
    \label{fig:slab_1400K_fits}
\end{figure}

\begin{figure}
    \centering
    \includegraphics[width=0.8\textwidth]{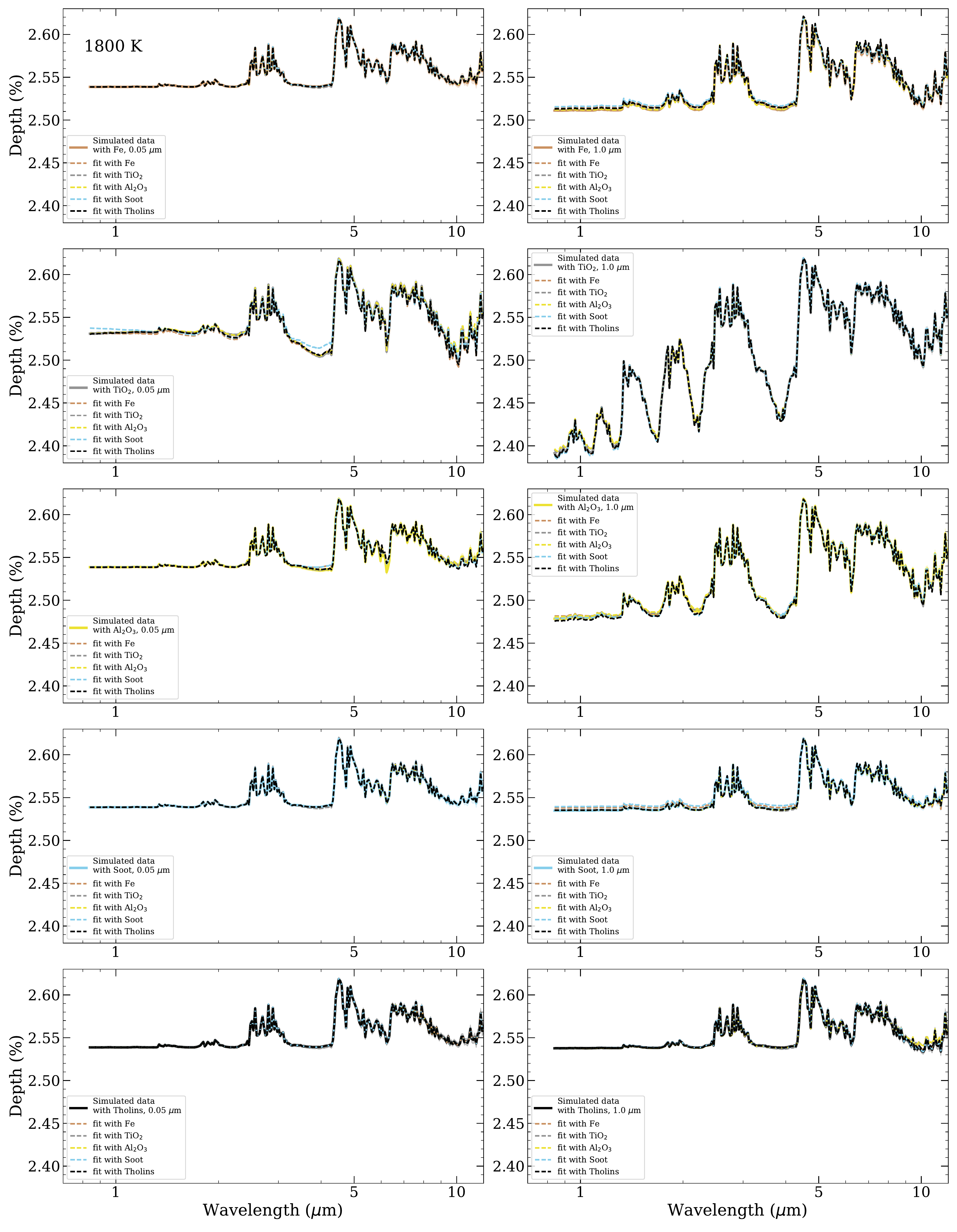}
    \caption{The results of our mixed aerosol MCMC experiments for the 1800-K atmosphere. We simulated data with a slab aerosol of Fe (top row), TiO$_2$ (second row), Al$_2$O$_3$ (third row), Soot (fourth row), and Tholins (bottom row). The slab aerosol used to simulate data always had $F$=1.0, P$_{top}$=10$^{-4.5}$ bars, and a log-normal size distribution with $\sigma _a$=2.5. The modal particle size was 0.05 $\mu$m in the left column and 1.0 $\mu$m on the right column. In each panel, the solid line represents the simulated data and the lightly shaded region represents the error envelope. The dashed lines show the transit spectra corresponding to the median parameter values of the posteriors mapped by the MCMC chains.}
    \label{fig:slab_1800K_fits}
\end{figure}

\begin{figure}
    \centering
    \includegraphics[width=0.6666\textwidth]{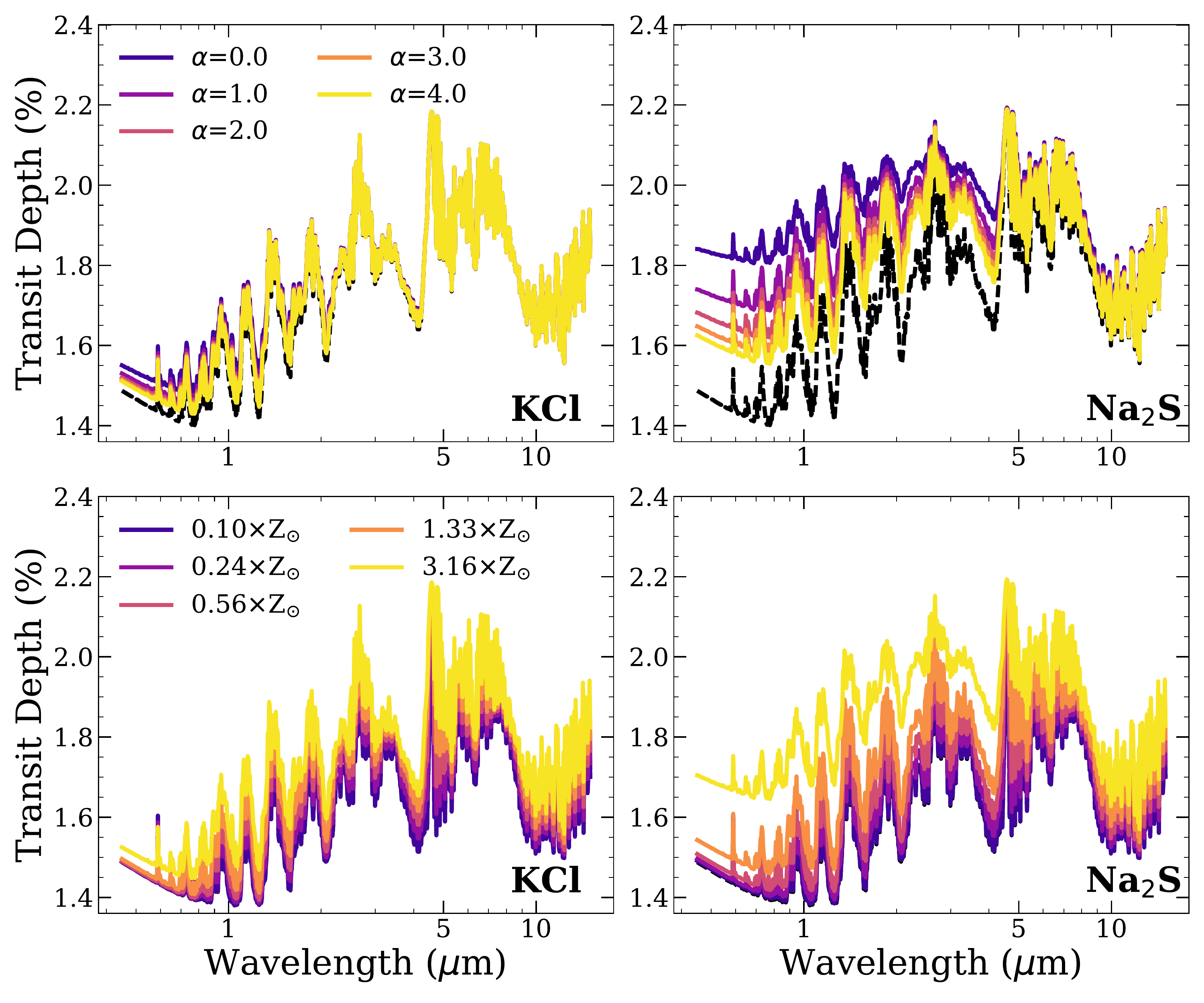}
    \caption{A demonstration of the 700-K planet's sensitivity to metallicity and $\alpha$ when an equilibrium cloud of KCl (left column) or Na$_2$S (right column) is present. The top row shows transit spectra with varying $\alpha$ and the bottom row shows transit spectra with varying Z. In each panel, a light gray dashed line shows the transit spectra for the 700-K atmosphere when clear.}
    \label{fig:700K_eq_pstudy}
\end{figure}

\begin{figure}
    \centering
    \includegraphics[width=0.6666\textwidth]{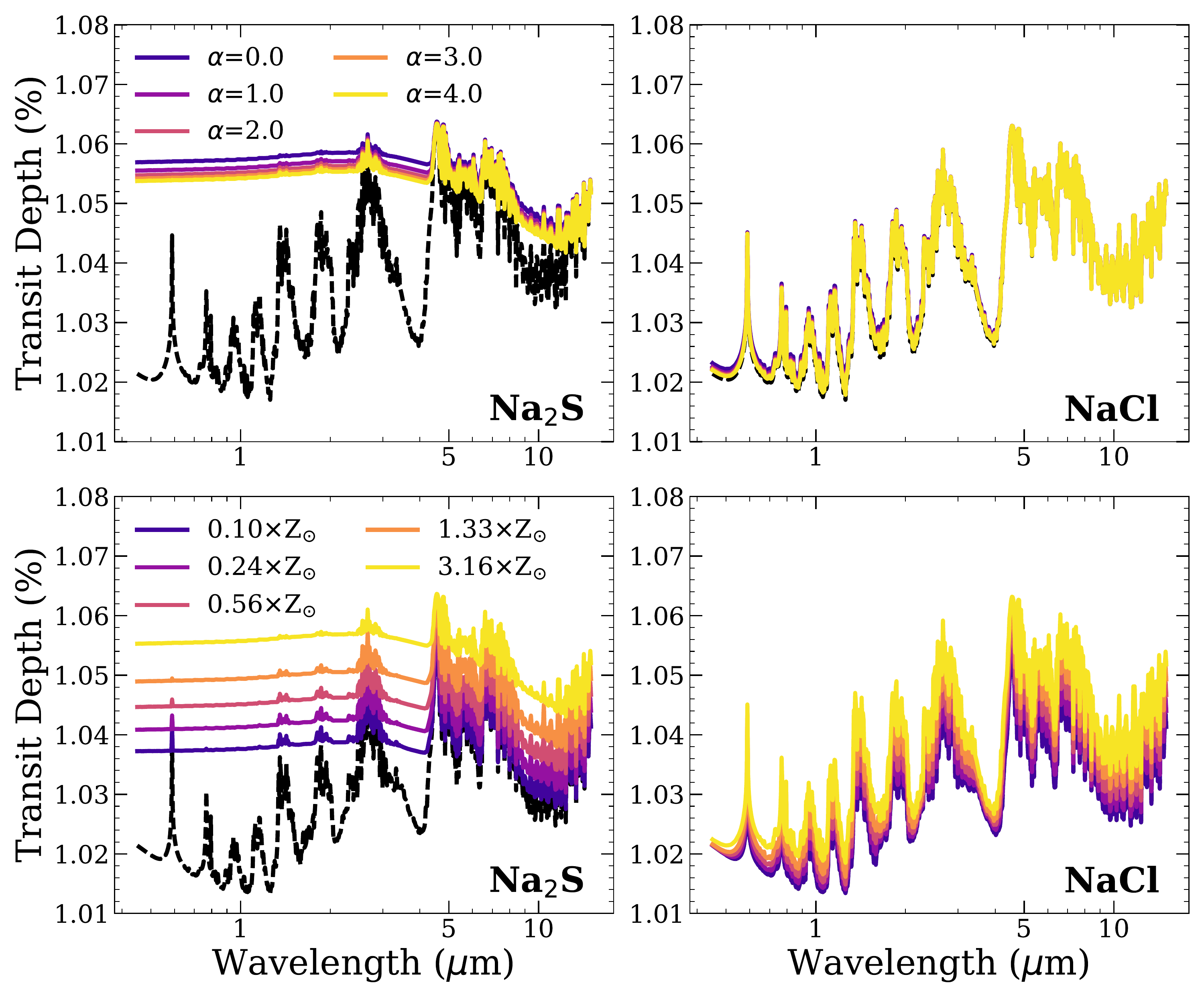}
    \caption{A demonstration of the 1000-K planet's sensitivity to metallicity and $\alpha$ when an equilibrium cloud of Na$_2$S (left column) or NaCl (right column) is present. The top row shows transit spectra with varying $\alpha$ and the bottom row shows transit spectra with varying Z. In each panel a light gray dashed line shows the transit spectra for the 1000-K atmosphere when clear.}
    \label{fig:1000K_eq_pstudy}
\end{figure}

\begin{figure}
    \centering
    \includegraphics[width=\textwidth]{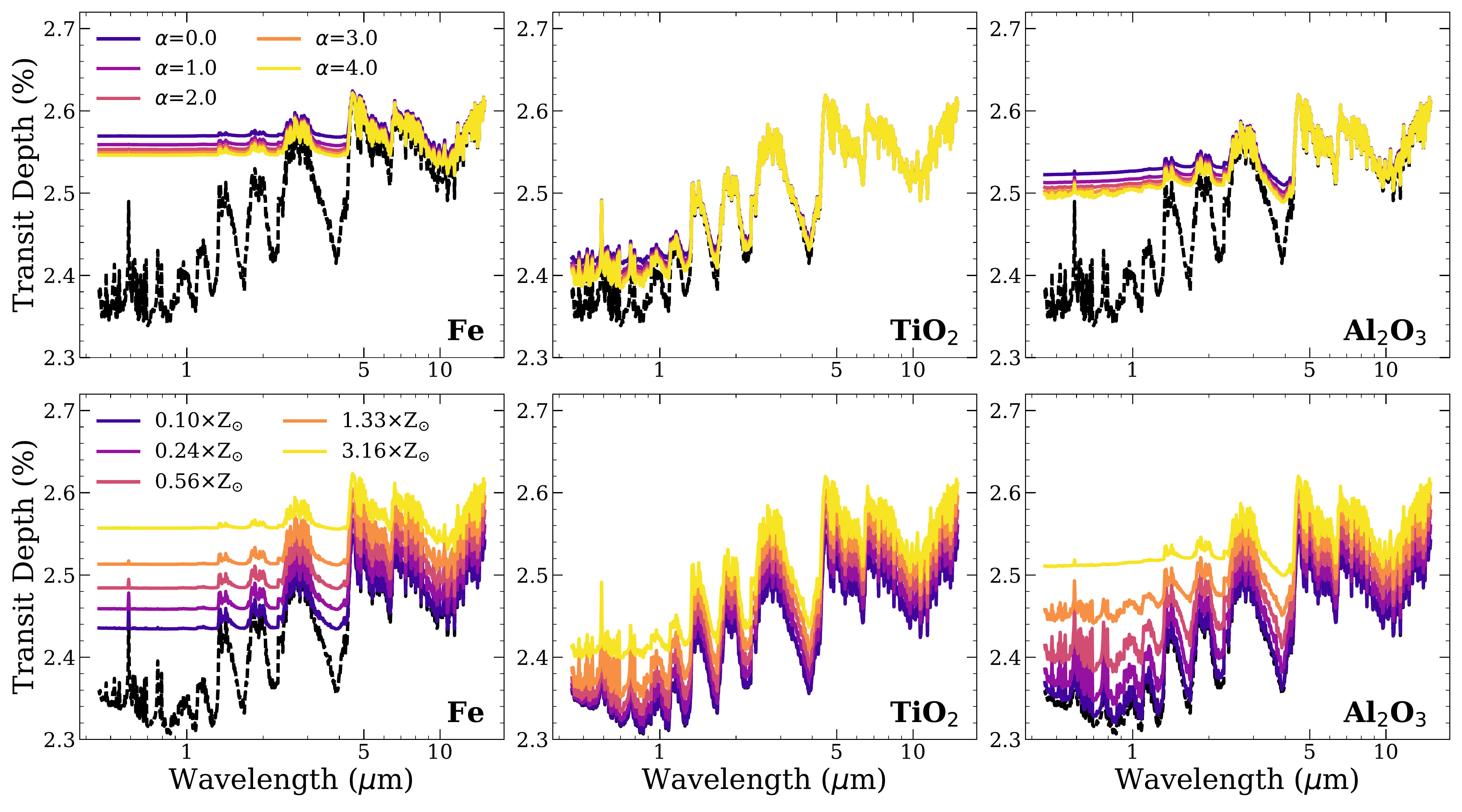}
    \caption{A demonstration of the 1800-K planet's sensitivity to metallicity and $\alpha$ when an equilibrium cloud of Fe (left column), TiO$_2$ (center column), or Al$_2$O$_3$ (right column) is present. The top row shows transit spectra with varying $\alpha$ and the bottom row shows transit spectra with varying Z. In each panel a light gray dashed line shows the transit spectra for the 1800-K atmosphere when clear.}
    \label{fig:1800K_eq_pstudy}
\end{figure}

\begin{figure}
    \centering
    \includegraphics[width=\textwidth]{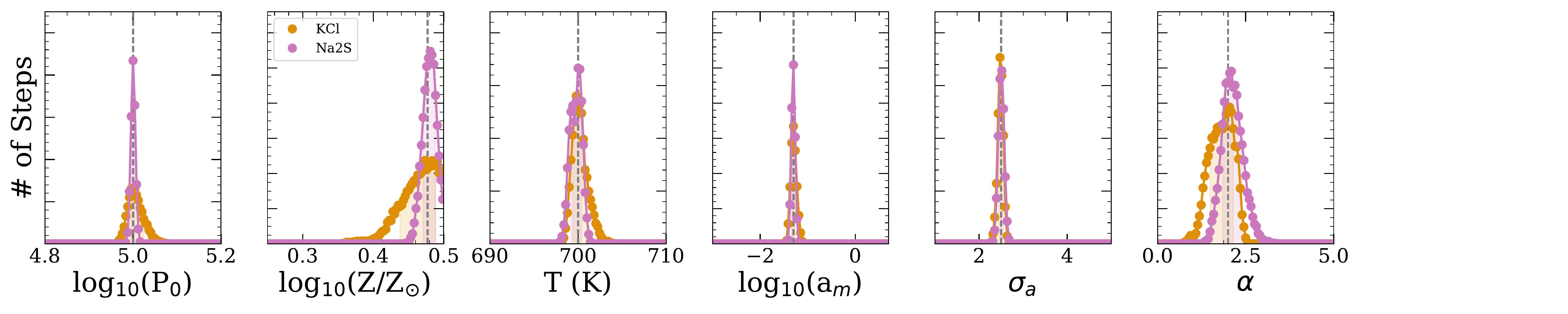}
    \includegraphics[width=\textwidth]{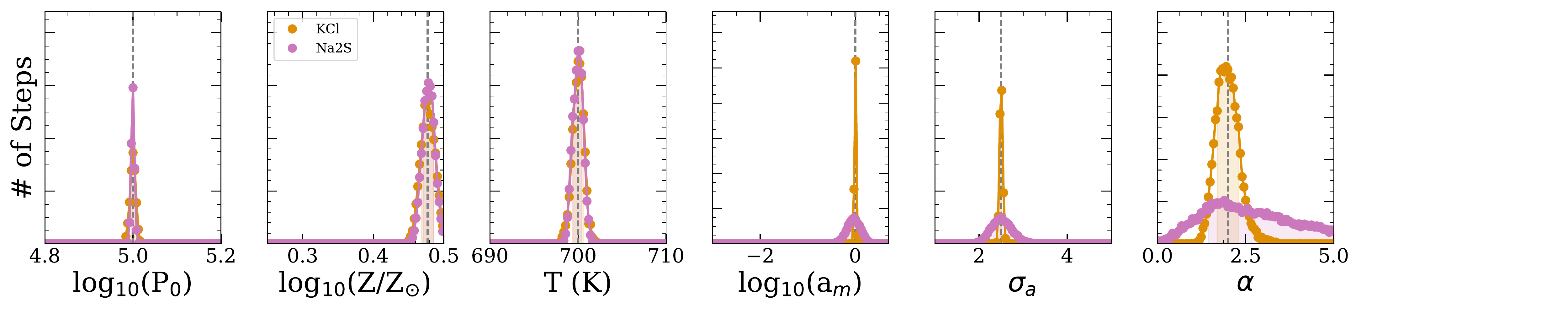}
    \caption{Histograms of posteriors for parameters from retrievals using the 700-K fiducial atmosphere with phase-equilibrium clouds. The top row shows results for a cloud with a log-normal particle-size distribution with modal particle size a$_m$=0.05$\mu$m, and the bottom row shows results for log-normal particle-size distribution with modal particle size a$_m$=1$\mu$m. Each color shows a different species of cloud. Vertical lines mark the true underlying values of parameters.}
    \label{fig:700_eq_hists}
\end{figure}

\begin{figure}
    \centering
    \includegraphics[width=\textwidth]{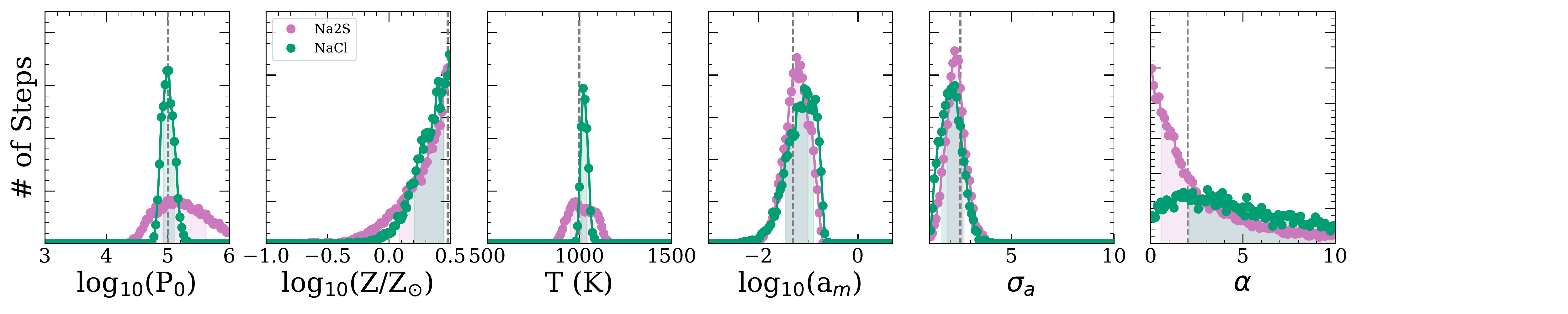}
    \includegraphics[width=\textwidth]{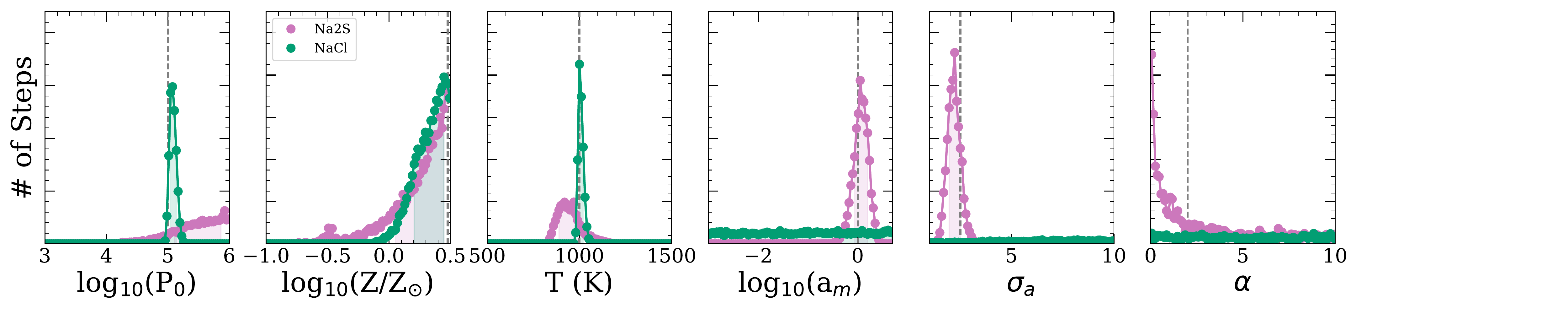}
    \caption{Histograms of posteriors for parameters from retrievals using the 1000-K fiducial atmosphere with phase-equilibrium clouds. The top row has a log-normal particle-size distribution with modal particle size a$_m$=0.05$\mu$m, while the bottom row has a log-normal particle-size distribution with a$_m$=1$\mu$m. Each color indicates a different species of cloud, and vertical dashed lines indicates the true values of parameters.}
    \label{fig:1000_eq_hists}
\end{figure}

\begin{figure}
    \centering
    \includegraphics[width=\textwidth]{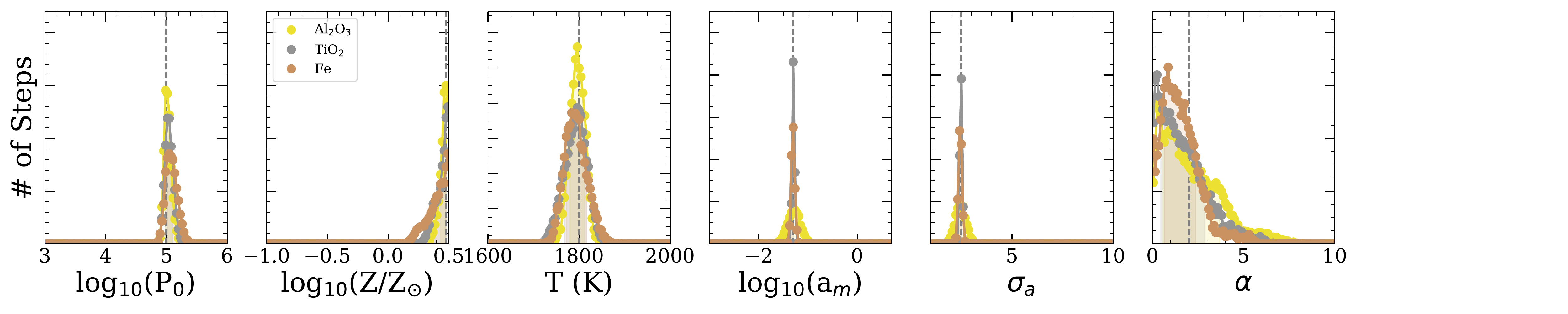}
    \includegraphics[width=\textwidth]{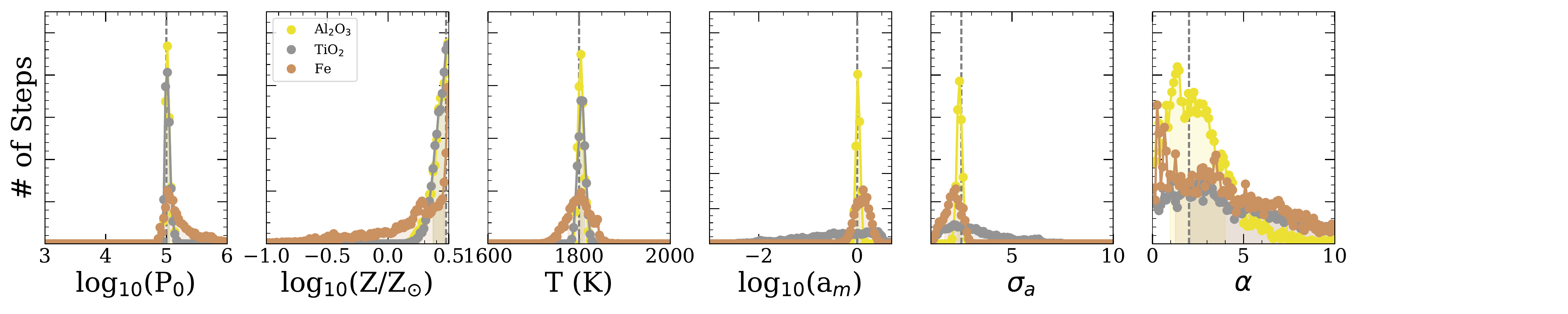}
    \caption{Histograms of posteriors for parameters from retrievals using the 1800-K fiducial atmosphere with phase equilibrium clouds. The top row has a log-normal particle-size distribution with modal particle size a$_m$=0.05$\mu$m, and the bottom row has a log-normal particle-size distribution with a modal particle size a$_m$=1$\mu$m. Colors show different cloud species, and vertical lines denote the true values of parameters used to simulate data.}
    \label{fig:1800_eq_hists}
\end{figure}

\begin{figure}
    \centering
    \includegraphics[width=\textwidth]{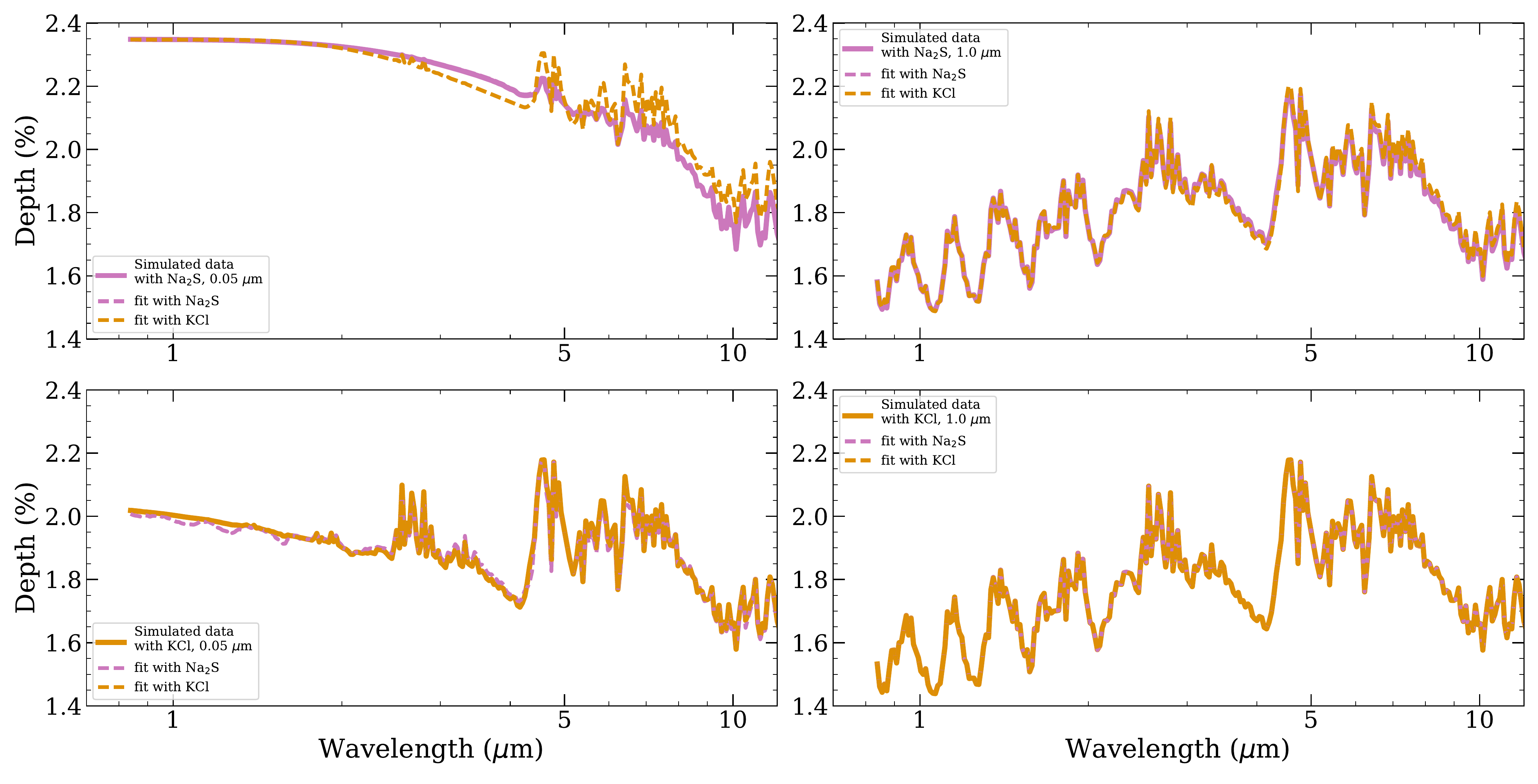}
    \caption{Results for the MCMC experiments with a temperature of 700 K and phase equilibrium clouds. The top row had simulated spectra with Na$_2$S and the bottom row had simulated spectra with KCl. The left column has a modal particle size of 0.05 $\mu$m and the right column has a modal particle size of 1.0 $\mu$m. The metallicity is always Z=3$\times$Z$_{\odot}$ and the dispersion for the log-normal size distribution is always $\sigma _a$=2.5. In each panel, the solid line indicates the simulated data and the shaded region shows the corresponding error envelope. The dashed lines show the transit spectra for retrieved parameters with different species of aerosols.}
    \label{fig:eq_700K_fits}
\end{figure}

\begin{figure}
    \centering
    \includegraphics[width=\textwidth]{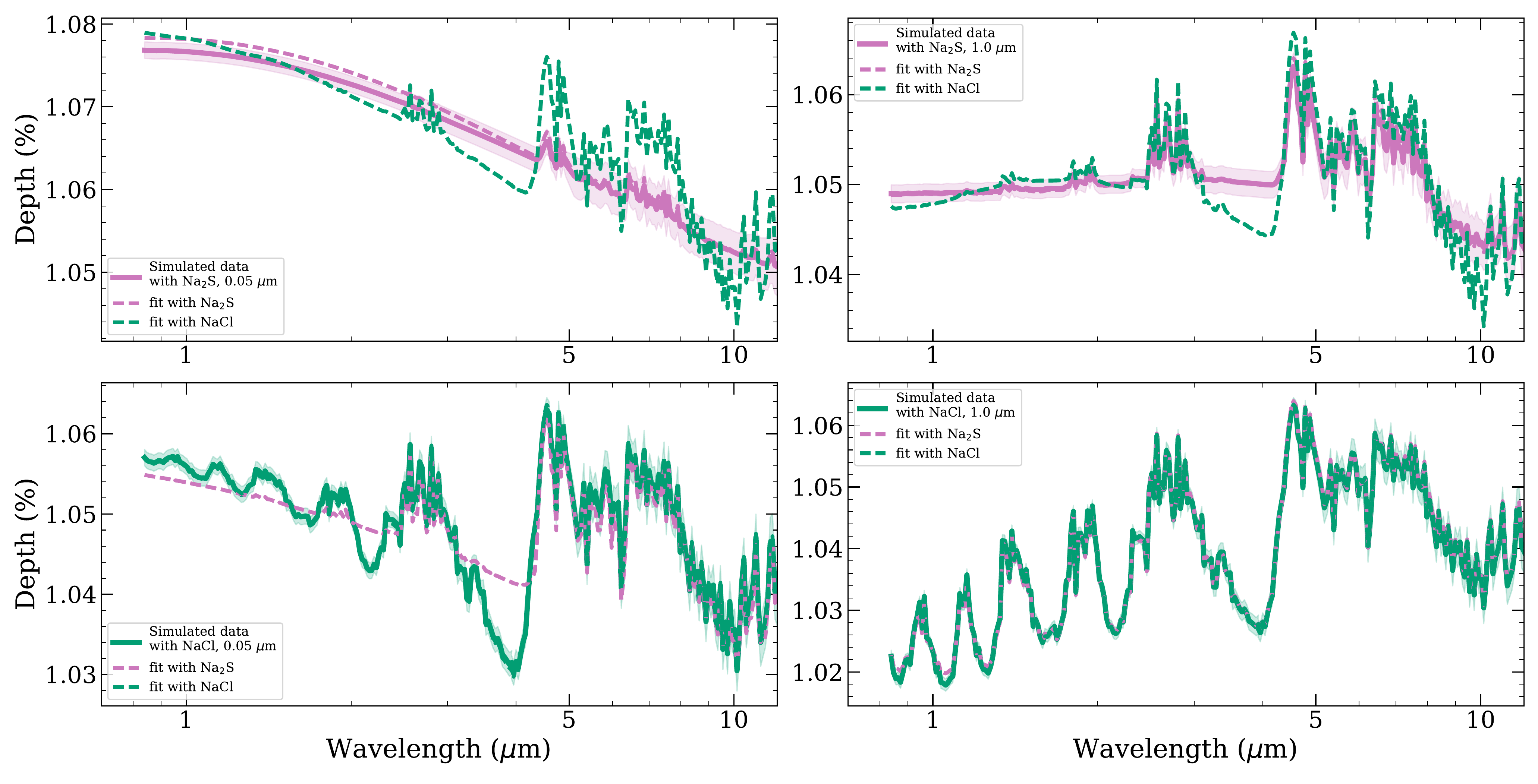}
    \caption{Results for MCMC experiments with the 1000-K atmosphere and phase equilibrium clouds. The solid lines with shaded error envelopes indicate the simulated data. Dashed lines show the best fit spectra with Na$_2$S clouds, and NaCl clouds. In the top row the true cloud species is Na$_2$S and in the bottom row it is NaCl. On the left hand side the modal particle size is 0.05 $\mu$m and on the right side the modal particle size is 1 $\mu$m.}
    \label{fig:eq_1000K_fits}
\end{figure}

\begin{figure}
    \centering
    \includegraphics[width=\textwidth]{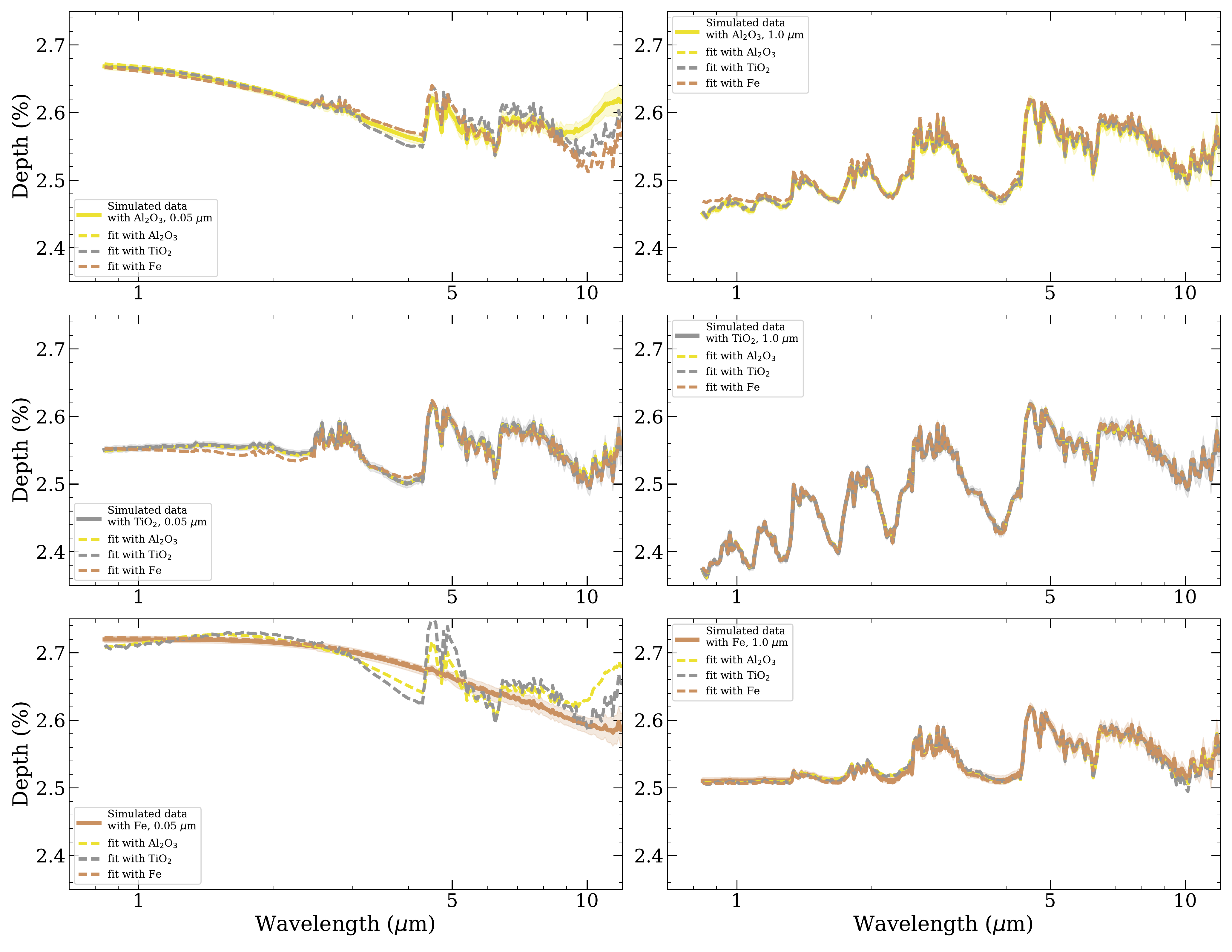}
    \caption{Results for MCMC experiments on the 1800-K atmosphere with phase equilibrium clouds. The solid lines with shaded error envelopes indicate the simulated data. Dashed lines show the best fit spectra with different condensing species. In the top row the true cloud species is Al$_2$O$_3$, in the middle row the true cloud species is TiO$_2$, and in the bottom row it is Fe. On the left hand side the modal particle size is 0.05 $\mu$m and on the right side the modal particle size is 1 $\mu$m.}
    \label{fig:eq_1800K_fits}
\end{figure}

\end{document}